\documentclass[a4paper,12pt]{article}

\PassOptionsToPackage{table}{xcolor}
\usepackage{jheppub} 

\usepackage[T1]{fontenc} 
\usepackage{lmodern} 

\usepackage{orcidlink}

\usepackage[all]{xy}
\usepackage[percent]{overpic}
\usepackage{slashed}
\usepackage{wrapfig}
\usepackage{diagbox}
\usepackage{mathrsfs,amsmath,amssymb,amsthm,amsfonts,tikz,graphicx,accents,hyperref, color}
\usepackage{dsfont,epiolmec, latexsym, stmaryrd, comment}
\usepackage{slashed,ccaption}
\usepackage{mathrsfs, calligra}
\usepackage{leftidx}
\usepackage{import}
\usepackage{multirow}
\usepackage{amsfonts}
\usepackage{pifont}
\usepackage{tabularx}
\usepackage[utf8]{inputenc}
\usetikzlibrary{intersections,calc}
\usepackage{tikz-3dplot}
\usepackage{ifthen}
\usetikzlibrary{arrows}
\usepackage{amsmath}
\usepackage{xcolor}

\usepackage{tensor}
\usepackage{mathtools}

\usepackage{caption}

\usepackage{array}

\usepackage{booktabs}
\usepackage{tabularx}
\usepackage{float} 

\usepackage[percent]{overpic}
\usepackage{wrapfig}
\usepackage{bbm}
\usepackage{slashed}
\usepackage{fancyhdr} 
\usepackage{amsmath}
\usepackage{amsfonts}
\usepackage{amssymb}
\usepackage{diagbox}

\usepackage{multirow}
\usepackage{pifont}

\hypersetup{ linktoc=all,
	colorlinks, linkcolor={blue}, 
	citecolor={red}, urlcolor={darkpink}
}

\graphicspath{{Images/}}

\definecolor{light gray}{RGB}{220,220,220}
\definecolor{dark purple}{RGB}{108,0,217}
\definecolor{pink}{RGB}{190,20,100}
\definecolor{orang}{RGB}{193,63,0}
\definecolor{green}{RGB}{11,98,17}
\definecolor{darkpink}{RGB}{153,0,76}
\definecolor{bluegreen}{RGB}{0,102,102}
\definecolor{greenlagan}{RGB}{0,102,0}
\definecolor{redgreen}{RGB}{102,102,0}
\definecolor{Redgreen}{RGB}{153,76,0}
\definecolor{vividviolet}{rgb}{0.62, 0.0, 1.0}
\definecolor{amaranth}{rgb}{0.9, 0.17, 0.31}
\definecolor{palatinateblue}{rgb}{0.15, 0.23, 0.89}
\definecolor{brightpink}{rgb}{1.0, 0.0, 0.5}
\definecolor{cornflowerblue}{rgb}{0.39, 0.58, 0.93}
\definecolor{deepcarminepink}{rgb}{0.94, 0.19, 0.22}
\definecolor{radicalred}{rgb}{1.0, 0.21, 0.37}
\usepackage{graphicx}
\usepackage{tikz}
\usetikzlibrary{shapes.geometric, arrows.meta, positioning, calc}

\pgfdeclarelayer{background}
\pgfsetlayers{background,main}

\usetikzlibrary{decorations.markings}
\tikzstyle arrowstyle=[scale=1]
\tikzstyle directed=[postaction={decorate,decoration={markings,
		mark=at position .65 with {\arrow[arrowstyle]{stealth}}}}]
\tikzstyle reverse directed=[postaction={decorate,decoration={markings,
		mark=at position .65 with {\arrowreversed[arrowstyle]{stealth};}}}]
\usepackage{import}
\usepackage{accents}
\usepackage{mathrsfs,amsmath,amssymb,slashed}
\usepackage{multirow,multicol}
\usepackage{enumitem}
\usepackage[percent]{overpic}
\usepackage{slashed}

\usepackage{wrapfig}
\usepackage{diagbox}
\usepackage{comment}
\usepackage{tikz}
\usepackage{mathrsfs,amsmath,amssymb,amsthm,amsfonts,graphicx,accents,hyperref,color}
\usepackage{leftidx}
\usepackage{import}
\usetikzlibrary{decorations.pathmorphing}
\DeclareFontFamily{OT1}{rsfs}{}

\DeclareFontShape{OT1}{rsfs}{m}{n}{ <-7> rsfs5 <7-10> rsfs7 <10->rsfs10}{} 

\DeclareMathAlphabet{\mycal}{OT1}{rsfs}{m}{n}

\newcommand{\be}{\begin{equation}}
	\newcommand{\ee}{\end{equation}}

\begin{document}
	\newcommand{\mytitle}{{\textbf{\centerline{\Large{Trinity of Carrollian Gravity:}} \\ \vskip 2mm \centerline{\large{Dynamically Equivalent Formulations of Magnetic Carrollian Gravity}}
	}}}
	
	\title{{\mytitle}}
	\author[a,b]{\, Mehdi~Ahmadi-Jahmani \orcidlink{0009-0008-0637-8764}}
	\author[a]{, Aliasghar~Parvizi \orcidlink{0000-0002-2581-767X} }
	\affiliation{$^a$ School of Physics, Institute for Research in Fundamental Sciences (IPM),\\ P.O.Box 19395-5531, Tehran, Iran }
	\affiliation{$^b$ Department of Physics, Faculty of Science, Ferdowsi University of Mashhad, Mashhad, Iran}
	
	\emailAdd{mehdiahmadijahmani@gmail.com,
		a.parvizi@ipm.ir,
	}
	
	\abstract{In this paper, we extend the paradigm of the Geometrical Trinity of Gravity to the Carrollian geometry to construct a framework of dynamically equivalent theories for Magnetic Carrollian gravity. Utilizing the inherent ambiguities and freedoms of affine connections on degenerate metrics, we relax the metric-compatibility and zero-torsion constraints of the rigid magnetic sector. By imposing global flatness on these modified connections, we show that the curvature of a Magnetic Carrollian universe can be dynamically traded for isolated kinetic energy densities of torsion and non-metricity. Unlike the Lorentzian Trinity which maps gravity to alternative gravitational frameworks, the Carrollian equivalences can also map spacetime geometry to non-gravitational theories. Specifically, we demonstrate that intrinsic torsion recovers Electric Carrollian gravity as the Teleparallel equivalent of Magnetic Carroll gravity; purely spatial torsion reproduces kinetic terms stracturally similar to the magnetic Hamiltonian of scalar-charge Fracton gauge theory; and temporal torsion sectors map independently to $U(1)$ Maxwell electric fields and Carroll-Weyl conformal dilaton theories. Finally, by formulating a Symmetric Teleparallel Carroll connection with minimal non-metricity, we show that the Magnetic Carrollian Lagrangian is dynamically equivalent to an ultralocal, isotropic Electric Carrollian gravity upon exploiting the present freedoms in connection. These results establish a geometric dictionary translating Carrollian gravitational dynamics into the language of tensor gauge theories, electromagnetism, and dilaton theory.}

	\maketitle
	
	\section{Introduction}
	The ultra-local limit of Lorentzian spacetime, first formalized by L\'evy-Leblond \cite{levy1965nouvelle}, yields Carrollian geometry: a degenerate spacetime structure in which light cones collapse onto time-like lines. In recent years, Carrollian geometry has experienced a significant resurgence, owing to its central role in flat-space holography \cite{Ruzziconi:2026bix,Nguyen:2025zhg,Bagchi:2023cen}, the fluid/gravity correspondence \cite{Ciambelli:2018wre, Adami:2023fbm}, the physics of the BMS algebra at null infinity \cite{Duval:2014uva}, fracton phases of matter \cite{Figueroa-OFarrill:2023vbj,Ahmadi-Jahmani:2025iqc}, Carrollian hydrodynamics and fluid \cite{Petkou:2022bmz,Freidel:2022bai,Bagchi:2023ysc,Kolekar:2024cfg,Shukla:2026xig} \footnote{For a broad review of Carrollian physics, including hydrodynamics and condensed-matter connections, see \cite{Bagchi:2025vri}.}, Carrollian gravity \cite{Hansen:2021fxi,Musaeus:2023oyp,Ciambelli:2025unn,Ahmadi-Jahmani:2025bls,Adami:2023wbe},  and tensionless string theory \cite{Bagchi:2019cay,Sheikh-Jabbari:2026vqh,Sheikh-Jabbari:2026tpf,Sheikh-Jabbari:2026kwr}. 

    In the formulation of Carrollian gravity, the dynamics are traditionally split into ``magnetic'' and ``electric'' sectors \cite{Bekaert:2015xua,Hansen:2021fxi,Hartong:2015xda}. Magnetic Carrollian gravity is characterized by the intrinsic curvature and a rigid spatial geometry, whereas Electric Carrollian gravity is governed by the extrinsic curvature, which encodes the spatial shearing and expansion along the flow of time. 
	
	In the study of General Relativity, the ``Gravity Trinity'' has proven that standard curvature-based gravity (Einstein-Hilbert) can be reformulated into dynamically equivalent geometric theories using Torsion (Teleparallel Equivalent of GR, TEGR) or Non-Metricity (Symmetric Teleparallel Equivalent of GR, STEGR) \cite{BeltranJimenez:2019esp,Aldrovandi:2013wha,Koivisto:2018loq,Nester:1998mp,BeltranJimenez:2017tkd,Maluf:2013gaa,Heisenberg:2018vsk,Heisenberg:2023lru}. 
	
	In this article, we merge these two paradigms to establish an equivalent dictionary for Magnetic Carrollian gravity, within the strict framework of metric compatibility and controlled torsion conditions. By systematically relaxing the torsion-free condition of the Carrollian connection, we show that the magnetic curvature can be exchanged for distinct manifestations of torsion and non-metricity. 

    The problem of extending the geometric trinity of gravity to non‑Lorentzian settings has been considered previously. In \cite{Wolf:2023rad}, the authors construct a non‑relativistic analog of the geometric trinity by taking the $c \to\infty$ limit. The ultra‑relativistic counterpart was recently investigated in \cite{Vigneron:2025vgn}, where the analysis is based on the $c \to 0$ limit of relativistic Lorentzian metrics. In that approach, the geometric constraints arise from identifying the structures that survive the limiting procedure. By contrast, our construction in this article is entirely intrinsic to Carrollian geometry. Rather than relying on a relativistic parent theory and an ultra‑relativistic limit, we derive the equivalences directly from the Carrollian fiber‑bundle structure by fixing the ambiguities of Carrollian connection according to physically motivated principles. Our goal is to establish their explicit dynamics by deriving the corresponding equivalent actions.

	In Carrollian geometry, the spatial metric \(h_{\mu\nu}\) is degenerate: the time-flow vector \(v^\mu\) lies in its kernel, so that \(h_{\mu\nu}v^\nu = 0\), which means \(h_{\mu\nu}\) possesses a zero eigenvalue. Because \(h_{\mu\nu}\) has no true inverse, the standard algebraic procedure that uniquely determines the Levi-Civita connection in Riemannian geometry no longer applies \cite{Ciambelli:2025unn, Henneaux:2021yzg, Musaeus:2023oyp}. This degeneracy opens a mathematical gap: the conditions of metric compatibility and vanishing torsion no longer single out a unique connection. Instead, they admit a family of solutions parametrized by arbitrary tensor fields, the connection ambiguities that we can parametrize based on the physical considerations.
	
	Throughout our analysis, we treat these connection ambiguities not as pathological defects of the degenerate metric, but as fundamental degrees of freedom to be systematically exploited. 
    By fixing these geometric ambiguities according to physical considerations, one can isolate and emphasize distinct kinematical sectors of the theory. For instance, fixing the arbitrary clock gradient through \eqref{eq:dilatonsolution} yields a Newtonian‑like scalar theory as one of the geometric manifestations of Magnetic Carrollian Gravity (MCG), while relating its divergence to the extrinsic curvature of the universe via \eqref{eq:Xchoice}, manifests its equivalence to Electric Carrollian gravity (ECG). In our search for equivalent theories, we leverage these gauge-like geometric choices to map the curvature of the magnetic sector into entirely different, isolated phenomena.
	
	We construct a framework of equivalent theories for Magnetic Carrollian gravity, focusing on the different connections and their physical meanings. Following the logic of Geometric Trinity of Gravity, we demonstrate that the alternative geometric formulation of Magnetic Carrollian gravity (governed by a torsion-free curvature) can be recast into dynamically equivalent theories by shifting the curvature into distinct, isolated torsion sectors. We derive the exact mappings between the torsion-free scalar curvature and a class of theories characterized by intrinsic torsion (Electric Carrollian gravity), spatial torsion (a teleparallel formulation closely related to tensor gauge theories), and temporal torsion (a dilatonic scalar sector). These equivalences provide a dictionary connecting Carrollian gravitational fields with the scalar and tensor gauge theories.

    In analogy with general relativity, which admits three dynamically equivalent formulations \cite{BeltranJimenez:2019esp,Vigneron:2025vgn,Wolf:2023rad}, we will demonstrate in the following sections that Magnetic Carrollian gravity likewise exhibits a structural sub-trinity. This arises from the different choices one can make regarding intrinsic, spatial, and temporal torsion, as well as the way in which non-metricity is defined within the Carrollian manifold. These structural choices for non-metricity and torsion determine whether magnetic Carroll curvature is encoded in intrinsic Ricci scalars or redistributed into torsion and non-metricity sectors, establishing a Carrollian sub-trinity analogous to the GR trinity. The resulting equivalent theories demonstrate that Carrollian gravity is not a single rigid framework, but rather admits a family of dynamically equivalent formulations, each representing a distinct geometric manifestation of Magnetic Carroll Gravity and highlighting different features.

    The paper is \textbf{organized} as follows. In Sec.~\ref{sec:GTG}, we provide a brief review of the Geometric Trinity of General Relativity in the standard Lorentzian manifold to establish the methodological foundation for constructing equivalent theories. Sec.~\ref{sec:CarrollGandG} introduces the essential mathematical structures of Carrollian geometry and outlines the standard formulation of Carrollian gravity. In Sec.~\ref{sec:metric-sector}, we investigate the metric-compatible sector to construct the Teleparallel Equivalent of Magnetic Carroll Gravity. After deriving the most general metric-compatible Carrollian connection, we establish Magnetic Carrollian Gravity as our baseline theory. By systematically isolating distinct torsion sectors and imposing a global flatness constraint, we introduce four dynamical equivalent theories: Electric Carrollian Gravity (\ref{section:ETI-ECG}) (via intrinsic torsion), Tensor Gauge Theory (\ref{section:ETII-FractonGT}) (via spatial torsion), a Dilatonic scalar theory (\ref{section:ETII-DilatonicS})(via temporal vector torsion), and Maxwell Electric Theory (\ref{section:ETIV-MaxwellElectricT})(via Carrollian clock torsion). Sec.~\ref{sec:STEMCG} transitions to the torsionless, non-metric sector to construct the Symmetric Teleparallel Equivalent. We derive a specific non-metric Carroll connection and demonstrate that fixing the geometry to a minimal non-metricity yields a fifth equivalent theory, mapping the magnetic curvature to an Isotropic Electric Carrollian gravity. Finally, Sec.~\ref{sec:conclusion} summarizes our findings, discusses the broader physical implications of these dualities, and outlines potential avenues for future research. For completeness, App.~\ref{app:gauging} provides a supplementary review of gauging the Carroll algebra within the first-order formalism.

    Given the variety of notations and connections employed throughout this work, each motivated by distinct physical considerations, Table \ref{tab:connections} provides a summary of the conventions, connections, and corresponding equivalent theories to facilitate navigation of the formalism and results.
    
	\begin{table}[H]
\centering
\renewcommand{\arraystretch}{1.5}
\small
\caption{Summary of Carrollian connections, their geometric constraints, and the corresponding dynamically equivalent theories. The first block lists the metric-compatible connections; the second block lists the non-metric extensions.}
\label{tab:connections}
\begin{tabularx}{\textwidth}{@{} l >{\raggedright\arraybackslash}X >{\raggedright\arraybackslash}X @{}}
\toprule
\textbf{Connection} & \textbf{Geometric Constraints} & \textbf{Equivalent Theory} \\
\midrule
\multicolumn{3}{@{}l}{\textit{(a) Teleparallel sector}} \\
\midrule
$\Gamma^\lambda_{\mu\nu}$, $\nabla_\mu$ &
\textbf{General metric-compatible.}\par
$\nabla h = 0$, $\nabla v = 0$.\par
Arbitrary torsion: $K_{\mu\nu}$, $A^\lambda_{\mu\nu}$, $X_{\mu\nu}$, $F_{\mu\nu}$. &
\textit{Base framework.} \\
\addlinespace[3pt]
$\tilde{\Gamma}^\lambda_{\mu\nu}$, $\tilde{\nabla}_\mu$ &
\textbf{Metric and torsion-free.}\par
$T = 0 \Rightarrow K_{\mu\nu} = 0$ (rigid space),\par
$A^\lambda_{\mu\nu} = 0$, $X_{[\mu\nu]} = -\tfrac{1}{2}F_{\mu\nu}$. &
\textbf{Magnetic Carrollian gravity (MCG)}\par
$S = \int e\,\tilde{R}$ \\
\addlinespace[3pt]
$\check{\Gamma}^\lambda_{\mu\nu}$, $\check{\nabla}_\mu$ &
\textbf{Intrinsic torsion.}\par
$A^\lambda_{\mu\nu} = 0$, $K_{\mu\nu} \neq 0$,\par
Clock gradient linked to $K$:$X_{(\mu\nu)} \propto K_{\mu\nu}$. &
\textbf{Electric Carrollian gravity (ECG)}\par
$S = \int e\,\alpha(K_{\mu\nu}K^{\mu\nu} - K^2)$ \\
\addlinespace[3pt]
$\bar{\Gamma}^\lambda_{\mu\nu}$, $\bar{\nabla}_\mu$ &
\textbf{Spatial torsion.}\par
$K_{\mu\nu} = 0$, $X_{[\mu\nu]} = -\tfrac{1}{2}F_{\mu\nu}$,\par
$A^\lambda_{\mu\nu} \neq 0$. &
\textbf{Tensor gauge theories}\par
$S = \int e\,A_{\alpha\beta\gamma}A^{\gamma\alpha\beta}$ \\
\addlinespace[3pt]
$\mathring{\Gamma}^\lambda_{\mu\nu}$, $\mathring{\nabla}_\mu$ &
\textbf{Temporal vector torsion.}\par
$K = 0$, $A = 0$,\par
$s_\mu \equiv v^\lambda \Upsilon_{[\lambda\mu]} \neq 0$. &
\textbf{Dilaton / Weyl scaling}\par
$S = \int e\,(-s_\mu s^\mu)$ \\
\addlinespace[3pt]
$\breve{\Gamma}^\lambda_{\mu\nu}$, $\breve{\nabla}_\mu$ &
\textbf{Clock 2-form torsion.}\par
$K = 0$, $A = 0$, $X_{[\mu\nu]} = 0$,\par
torsion sourced by $F_{\mu\nu} = d\tau$. &
\textbf{Electric Maxwell theory}\par
$S = \int e\,(-\tfrac{1}{4}E_\mu E^\mu)$ \\
\midrule
\multicolumn{3}{@{}l}{\textit{(b) Symmetric Teleparallel sector}} \\
\midrule
$\mathring{\bar{\Gamma}}^\lambda_{\mu\nu}$, $\mathring{\bar{\nabla}}_\mu$ &
\textbf{General non-metric, torsion-free.}\par
$T = 0$, $\mathring{\bar{\nabla}} v \neq 0$, $\mathring{\bar{\nabla}} h \neq 0$. &
\textit{General STECG framework.} \\
\addlinespace[3pt]
$\mathring{\tilde{\Gamma}}^\lambda_{\mu\nu}$, $\mathring{\tilde{\nabla}}_\mu$ &
\textbf{Minimal non-metric (STECG).}\par
$a_\mu = 0$ (no excess non-metricity),\par
gauge fixing $\Upsilon = \lambda K$. &
\textbf{Isotropic Electric Carroll gravity}\par
$S = \int e\,(-\lambda K^2)$ \\
\bottomrule
\end{tabularx}
\end{table}

	
	\section{A Brief Review of the Geometric Trinity of Gravity}\label{sec:GTG}
	
	General relativity (GR) provides a geometric description of gravitation in which the gravitational field is identified with the dynamical structure of spacetime itself. The natural setting of the theory is a four-dimensional differentiable manifold $\mathcal{M}$ endowed with a Lorentzian metric $g_{\mu\nu}$. The metric determines the causal structure of spacetime, distinguishing timelike, null, and spacelike directions, while simultaneously providing the notion of proper time, spacetime intervals, and the metric volume element. In this sense, the Lorentzian metric is not merely an auxiliary field but constitutes the fundamental geometrical structure through which the physical content of GR is formulated. The Einstein field equations then relate this geometry to the distribution of matter and energy, so that spacetime geometry and gravitational dynamics are inseparably intertwined.
	
	The metric alone, however, does not determine how tensors are differentiated or parallel transported. These operations require an affine connection. In GR, the connection is chosen to be the unique Levi--Civita connection compatible with the metric and free of torsion,
	\begin{equation}
		{}^{\rm GR}\nabla_{\alpha}g_{\mu\nu}=0,
		\qquad
		{}^{\rm GR}T^{\alpha}{}_{\mu\nu}=0.
	\end{equation}
	Consequently, the connection will be completely determined by the metric.
	The gravitational field can therefore be characterized geometrically by the curvature associated with this connection. Explicitly,
	\begin{equation}
		{}^{\rm GR}R^{\alpha}{}_{\beta\mu\nu}
		:=
		\partial_{\mu}{}^{\rm GR}\Gamma^{\alpha}{}_{\nu\beta}
		-
		\partial_{\nu}{}^{\rm GR}\Gamma^{\alpha}{}_{\mu\beta}
		+
		{}^{\rm GR}\Gamma^{\alpha}{}_{\mu\lambda}
		{}^{\rm GR}\Gamma^{\lambda}{}_{\nu\beta}
		-
		{}^{\rm GR}\Gamma^{\alpha}{}_{\nu\lambda}
		{}^{\rm GR}\Gamma^{\lambda}{}_{\mu\beta}.
	\end{equation}
	The Riemann tensor measures the failure of covariant derivatives to commute and, equivalently, the failure of a vector to return unchanged after parallel transport around an infinitesimal closed loop. Contracting the Riemann tensor yields the Ricci tensor ${}^{\rm GR}R_{\mu\nu}$ and Ricci scalar ${}^{\rm GR}R$.
	The Einstein-Hilbert Lagrangian will then be constructed from this scalar curvature $\mathcal{L}_{\rm GR}=\sqrt{-g} \, {}^{\rm GR}R$, up to the cosmological and matter contributions.
	
	The geometrical description of gravity in terms of curvature is not, however, the only way in which the gravitational degrees of freedom of GR can be represented. More generally, one may regard the metric and affine connection as independent geometrical structures and relax the conditions of metric compatibility and vanishing torsion. The three fundamental geometrical quantities associated with a general metric-affine spacetime are then curvature, torsion, and non-metricity. For a general affine connection
	$\overset{=}{\Gamma}{^\alpha}{}_{\mu\nu}$, these are defined respectively by
	\begin{subequations}
		\begin{align}
			\overset{=}R{^\alpha}{}_{\beta\mu\nu}
			&:=
			2\partial_{[\mu}
			\overset{=}{\Gamma}{^\alpha}{}_{\nu]\beta}
			+
			2\overset{=}{\Gamma}{^\alpha}{}_{[\mu|\lambda|}
			\overset{=}{\Gamma}{^\lambda}{}_{\nu]\beta},
			\\
			\overset{=}T{^\alpha}{}_{\mu\nu}
			&:=
			2\overset{=}{\Gamma}{^\alpha}{}_{[\mu\nu]},
			\\
			\overset{=}Q_{\alpha\mu\nu}
			&:=
			\overset{=}{\nabla}_{\alpha}g_{\mu\nu}.
		\end{align}
	\end{subequations}
	Curvature characterizes the non-integrability of parallel transport, torsion measures the antisymmetric part of the affine connection and the associated failure of infinitesimal parallelograms to close, while non-metricity measures the failure of parallel transport to preserve the metric and hence lengths and angles.
	
	These three geometrical structures are not independent in the usual formulation of GR. The Levi--Civita connection imposes
	\begin{equation}
		{}^{\rm GR}T^{\alpha}{}_{\mu\nu}=0,
		\qquad
		{}^{\rm GR}Q_{\alpha\mu\nu}=0,
	\end{equation}
	leaving curvature as the geometrical carrier of gravitational information. Nevertheless, one may redistribute the geometrical content of GR among curvature, torsion, and non-metricity by choosing different affine connections. This observation leads to what is now commonly referred to as the \emph{geometric trinity of gravity} \cite{BeltranJimenez:2019esp}: three formulations of GR in which the gravitational dynamics can be represented, respectively, through spacetime curvature, spacetime torsion, or spacetime non-metricity. The three vertices are the standard formulation of GR, the teleparallel equivalent of general relativity (TEGR) \cite{Aldrovandi:2013wha,Koivisto:2018loq,Heisenberg:2023lru}, and the symmetric teleparallel equivalent of general relativity (STEGR) \cite{Nester:1998mp, BeltranJimenez:2017tkd, Maluf:2013gaa,Heisenberg:2023lru}.
	
	The existence of this trinity can be understood particularly clearly by considering the relation between a general affine connection and the Levi--Civita connection. Their difference is a tensor and can be decomposed into contortion and disformation (also called distorsion in parts of the literature),
	\begin{equation}
		\overset{=}{\Gamma}{^\alpha}{}_{\mu\nu}
		=
		{}^{\rm GR}\Gamma{^\alpha}{}_{\mu\nu}
		+
		K{^\alpha}{}_{\mu\nu}
		+
		L{^\alpha}{}_{\mu\nu},
		\label{eq:connection_decomposition}
	\end{equation}
	where $K{^\alpha}{}_{\mu\nu}$ is the contortion associated with torsion and
	$L^{\alpha}{}_{\mu\nu}$ is the disformation associated with non-metricity. Defining
	\begin{equation}
		M{^\alpha}{}_{\mu\nu}
		:=
		K{^\alpha}{}_{\mu\nu}
		+
		L{^\alpha}{}_{\mu\nu},
	\end{equation}
	one obtains a direct relation between the curvature tensors of the two connections. In particular,
	\begin{align}
		\overset{=}R{^\alpha}{}_{\beta\mu\nu}
		={}&
		{}^{\rm GR}R{^\alpha}{}_{\beta\mu\nu}
		+{}^{\rm GR}\nabla_{\mu}M{^\alpha}{}_{\nu\beta}
		-{}^{\rm GR}\nabla_{\nu}M{^\alpha}{}_{\mu\beta} +
		M{^\gamma}{}_{\nu\beta}M{^\alpha}{}_{\mu\gamma}
		-
		M{^\gamma}{}_{\mu\beta}M{^\alpha}{}_{\nu\gamma}
		+
		\overset{=}T{^\gamma}{}_{\mu\nu}
		M{^\alpha}{}_{\gamma\beta},
		\label{eq:curvature_relation}
	\end{align}
	where the covariant derivatives on the right-hand side are constructed with the Levi--Civita connection. This identity provides the geometrical mechanism underlying the different formulations of GR \cite{Wolf:2023rad}.
	
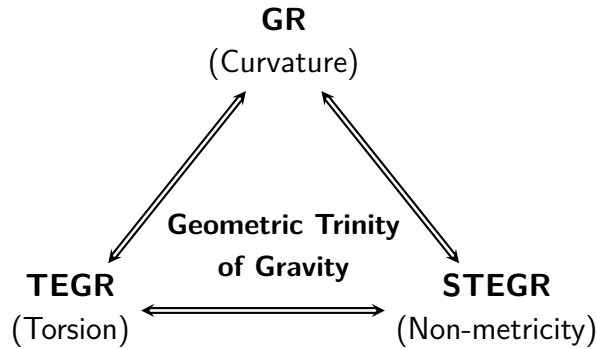
\begin{figure}[htbp]
		\centering
		\begin{tikzpicture}[node distance=2.5cm, every node/.style={align=center, font=\sffamily}]
			
			\node (gr) at (90:2.5) {\textbf{GR} \\ (Curvature)};
			\node (tegr) at (200:3) {\textbf{TEGR} \\ (Torsion)};
			\node (stegr) at (340:3) {\textbf{STEGR} \\ (Non-metricity)};
			
			\draw[thick, <->, >=stealth, double, double distance=1pt] (gr) -- (tegr);
			\draw[thick, <->, >=stealth, double, double distance=1pt] (tegr) -- (stegr);
			\draw[thick, <->, >=stealth, double, double distance=1pt] (stegr) -- (gr);
			
			\node at (0,-0.2) {\small\textbf{Geometric Trinity}\\\small\textbf{of Gravity}};
			
		\end{tikzpicture}
		\caption{Schematic representation of the Geometric Trinity of Gravity.}
		\label{fig:trinity_gravity}
	\end{figure}

	The three formulations correspond to different restrictions on the affine connection.     In standard GR, the connection is metric-compatible and torsion-free but possesses non-vanishing curvature in general,
	\begin{equation}
		\overset{=}Q_{\alpha\mu\nu}=0,
		\qquad
		\overset{=}T{^\alpha}{}_{\mu\nu}=0,
		\qquad
		\overset{=}R{^\alpha}{}_{\beta\mu\nu}\neq0.
	\end{equation}
	In TEGR, by contrast, one employs a metric-compatible and curvature-free connection,
	\begin{equation}\label{eq:TEGR}
		\overset{=}Q_{\alpha\mu\nu}=0,
		\qquad
		\overset{=}R{^\alpha}{}_{\beta\mu\nu}=0,
		\qquad
		\overset{=}T{^\alpha}{}_{\mu\nu}\neq0,
	\end{equation}
	so that gravity is encoded entirely in torsion. In STEGR, one instead imposes vanishing curvature and torsion,
	\begin{equation}\label{eq:STEGR}
		\overset{=}R{^\alpha}{}_{\beta\mu\nu}=0,
		\qquad
		\overset{=}T{^\alpha}{}_{\mu\nu}=0,
		\qquad
		\overset{=}Q_{\alpha\mu\nu}\neq0,
	\end{equation}
	and the gravitational interaction is described through non-metricity \cite{Wolf:2023rad,BeltranJimenez:2017tkd}. Thus, schematically we have Fig.~\ref{fig:trinity_gravity}
	
	The equivalence among these formulations is most transparent at the level of their actions. In TEGR \eqref{eq:TEGR}, the Ricci scalar constructed from the Levi--Civita connection can be expressed in terms of the quadratic torsion scalar $\overset{=}T$ together with a total divergence,
	\begin{equation}
		\overset{=}R =0 \Rightarrow {}^{\rm GR}R
		=
		-\overset{=}T
		+
		2{}^{\rm GR}\nabla_{\alpha}
		\overset{=}T{^{\beta\alpha}}{}_{\beta},
		\label{eq:GR_TEGR_identity}
	\end{equation}
	where
	\begin{equation}
		\overset{=}T
		=
		\frac{1}{4}
		\overset{=}T_{\alpha\mu\nu}\overset{=}T{^{\alpha\mu\nu}}
		+
		\frac{1}{2}
		\overset{=}T_{\alpha\mu\nu}\overset{=}T{^{\mu\alpha\nu}}
		-
		\overset{=}T{^\alpha}{}_{\alpha\mu}
		\overset{=}T{^\beta}{}_{\beta\nu}g^{\mu\nu}.
	\end{equation}
	Similarly, in STEGR formulation \eqref{eq:STEGR} one may write
	\begin{equation}
		\overset{=}R =0 \Rightarrow {}^{\rm GR}R
		=
		-\overset{=}Q
		+
		{}^{\rm GR}\nabla_{\alpha}
		\left(
		\overset{=}Q{^\alpha}{}_{\lambda}{}^{\lambda}
		-
		\overset{=}Q{^\lambda}{}_{\lambda}{}^{\alpha}
		\right),
		\label{eq:GR_STEGR_identity}
	\end{equation}
	where $\overset{=}Q$ denotes the appropriate quadratic non-metricity scalar
	\begin{equation}
		\overset{=}Q = g^{\mu\nu} \left(L^\alpha{}_{\mu\beta}L^\beta{}_{\nu\alpha} - L^\alpha{}_{\alpha\beta}L^\beta{}_{\mu\nu}\right) \, .
	\end{equation}
	The differences between the corresponding gravitational Lagrangians are therefore boundary terms. Under suitable boundary conditions, these terms do not alter the bulk field equations, establishing the dynamical equivalence of GR, TEGR, and STEGR \cite{BeltranJimenez:2019esp}.
	
	It is important, however, to distinguish \emph{dynamical equivalence} from complete physical or conceptual identity. The three formulations employ different affine structures and consequently assign different geometrical interpretations to the gravitational field. In GR, gravitational effects are represented by curvature of the Levi--Civita connection. In TEGR, the relevant connection is flat but torsionful, while in STEGR it is flat and torsion-free but non-metric. These differences become particularly relevant when considering boundary terms, gravitational energy and momentum, coupling prescriptions, modified theories of gravity, and limits in which the relativistic metric becomes degenerate or changes its geometrical character. The geometric trinity should therefore be viewed not simply as three notationally different versions of Einstein's equations, but as three distinct geometrical realizations of the same relativistic dynamics.
	
	This perspective also motivates the investigation of geometric limits of gravity. The Lorentzian structure underlying GR permits two qualitatively different singular limits associated with the contraction of the relativistic symmetry group: the non-relativistic limit $c\rightarrow\infty$ and the ultra-relativistic limit $c\rightarrow0$. The former leads to Galilean and Newton-Cartan geometries \cite{Bekaert:2015xua, Christensen:2013lma}, while the latter gives rise to Carrollian geometry \cite{levy1965nouvelle,duval2014carroll}. Importantly, these limits are not obtained merely by setting a parameter in the Lorentzian metric to a limiting value. The metric generally becomes degenerate, and the resulting geometrical structure must instead be formulated in terms of appropriate temporal and spatial tensors together with a compatible connection \cite{Henneaux:2021yzg,Ciambelli:2019lap}.
	
	The geometric trinity therefore provides a natural starting point for asking how alternative descriptions of gravitational dynamics behave when the Lorentzian structure is replaced by a non-Lorentzian one. This question has already been explored in the non-relativistic regime. In particular, recent work has shown that the geometric trinity admits a non-relativistic counterpart involving Newton-Cartan gravity and its curvature-, torsion-, and non-metricity-based descriptions \cite{Wolf:2023rad}. This result suggests that the trinity is not tied intrinsically to the Lorentzian formulation of spacetime, but may reflect a more general relation between gravitational dynamics and the possible geometrical structures of an affine spacetime.
	
	The ultra-relativistic limit is conceptually different. Carrollian geometry arises when the causal cones collapse in the limit $c\rightarrow0$, leading to a degenerate temporal structure and a geometry adapted to Carrollian rather than Lorentzian kinematics. Carrollian structures have attracted increasing attention in the context of asymptotic symmetries \cite{duval2014conformal}, flat-space holography \cite{Ruzziconi:2026bix, Nguyen:2025zhg,Bagchi:2023cen}, Carrollian field theories\cite{Henneaux:2021yzg,Ahmadi-Jahmani:2025iqc}, and ultra-relativistic limits of gravitational systems \cite{Hartong:2015xda,Campoleoni:2022ebj,Bergshoeff:2017btm,Hansen:2021fxi}. In particular, recent constructions of Carrollian gravity indicate that torsion and curvature can play non-trivial and, in some cases, independent roles in the resulting non-Lorentzian geometry \cite{Hansen:2021fxi,Musaeus:2023oyp,Ciambelli:2025unn}. These developments raise a natural question: whether the geometrical trinity of relativistic gravity possesses a meaningful Carrollian counterpart, and, if so, how curvature, torsion, and non-metricity reorganize when the underlying Lorentzian metric becomes degenerate.
	
	The purpose of the following analysis is to address this question by treating the geometric trinity as a structural framework rather than merely as an equivalence between three relativistic actions. Starting from the decomposition of a general affine connection, one can investigate how the curvature, torsion, and non-metricity sectors behave under an ultra-relativistic contraction. Such an analysis provides a systematic route from Lorentzian gravity to Carrollian geometry and allows one to determine which elements of the relativistic trinity survive the Carrollian limit, which combine into new geometrical objects, and which require a reformulation due to the degeneracy of the limiting metric. In this sense, the Carrollian geometry offers a particularly sharp setting in which to examine the geometrical content of gravitational dynamics beyond the conventional Lorentzian framework.

	\section{Carrollian Geometry and Gravity}\label{sec:CarrollGandG}
	A Carrollian manifold $\mathcal{C}$ is a differentiable manifold endowed with a degenerate metric structure formally specified by \((h_{\mu\nu}, v^\mu )\), where \(h_{\mu\nu}\) is a positive semi-definite spatial metric of rank \(d-1\) and \(v^\mu\) is a nowhere-vanishing vector field defining the time direction \cite{Dautcourt:1997hb,Hansen:2021fxi,Duval:2014uoa,Figueroa-OFarrill:2022nui}. This geometric framework naturally accommodates torsion and non-metricity, allowing for a rich classification of Carrollian structures\cite{Duval:2014uva,Figueroa-OFarrill:2020gpr,Bekaert:2015xua,Ciambelli:2025unn,Vigneron:2025vgn,Ahmadi-Jahmani:2025bls}. 
	
	The degeneracy of the Carrollian metric, however, leaves considerable freedom in how the geometry is represented and extended away from the spatial directions. In particular, the Carrollian data do not uniquely determine a spacetime connection, and additional choices are required to specify how the spatial metric and the preferred temporal direction are transported across the manifold. These freedoms are closely related to the choice of temporal one-form, the representative of the spatial inverse metric, and the connection compatible with the Carrollian structure. These choices can result in geometrically equivalent descriptions. However, when promoted to dynamical fields, they can yield entirely distinct Carrollian theories. This freedom is most obvious when Carrollian geometry is derived as a limit of, or embedded in, Lorentzian geometry. Quantities that are rigidly linked in the bulk, or before the limit is taken, can split into independent geometric structures as the metric becomes degenerate \cite{Hansen:2021fxi, Ciambelli:2025unn}. The rich structure is one of the reasons Carrollian geometry is suited to studying gravitational theories beyond the standard pseudo-Riemannian framework. In this sense, Carrollian geometry provides a systematic setting in which the familiar relationships among metricity, torsion, curvature, and affine connection can be disentangled and subsequently recombined in different ways, a perspective that is especially relevant for formulating Carrollian analogues of the geometric formulations of gravity.

	\subsection{Carrollian Geometry}
	
	Given a Carrollian structure $( h_{\mu\nu}, v^{\mu} )$, the tensorial quantities \( v^{\mu} \) and \( h_{\mu\nu} \) are utilized to represent the notions of temporal and spatial structures, respectively \cite{Duval:2014uoa}, which satisfy the fundamental orthonormality condition \cite{Hansen:2021fxi,Musaeus:2023oyp}:
	\begin{align}\label{eq:vhzero}
		v^{\mu}h_{\mu\nu}=0.
	\end{align}
	We can define inverse quantities \(\tau_{\mu}\) and \(h^{\mu\nu}\) which also satisfy the completeness relations \footnote{The existence of the Carrollian vector field \(v^{\mu}\) is ensured provided there exists a non-vanishing one-form \(\tau_{\mu}\) such that \(v^{\mu}\tau_{\mu} = -1\).}:
	\begin{align}\label{eq:carroll-conditions}
		\tau_{\mu}v^{\mu}= -1,\quad \tau_{\mu}h^{\mu\nu}=v^{\mu}h_{\mu\nu}=0,\quad \delta_{\mu}^{\nu}=- \tau_{\mu}v^{\nu}+h^{\nu\rho}h_{\rho\mu}.
	\end{align}
By definition, the vector field \(v^\mu\) and the spatial metric \(h_{\mu\nu}\) are invariant under local homogeneous Carroll transformations. In contrast, the projective inverses \(\tau_\mu\) and \(h^{\mu\nu}\) are not invariant under local Carrollian boosts, as is evident from the transformation laws \cite{Hansen:2021fxi,Baiguera:2022lsw}:
\begin{subequations}
    \begin{align}
    \delta_{C}v^{\mu}&=\mathcal{L}_{\xi}v^{\mu}\,,\\
    \delta_{C}\tau_{\mu}&=\lambda_{\mu}\,,\\
    \delta_{C}h_{\mu\nu}&=\mathcal{L}_{\xi}h_{\mu\nu}\,,\\
    \delta_{C}h^{\mu\nu}&=\mathcal{L}_{\xi}h^{\mu\nu}+2\lambda^{(\mu}\, v^{\nu)}\,.
    \end{align}
\end{subequations}
Here, \(\lambda_{\mu}\) represents a Carrollian boost, with \(\lambda^{\mu}\) defined as \(h^{\mu\nu}\lambda_{\nu}\). It has been demonstrated that the projective inverses \(\tau_{\mu}\) and \(h^{\mu\nu}\) are not invariant under local Carrollian boosts \cite{Musaeus:2023oyp,Hansen:2021fxi}; the ambiguity in defining these inverses can be understood as arising from this non-invariance. Nevertheless, at the end we can construct combinations that are invariant. To study equivalent theories, we divide our analysis into two parts based on metricity (torsionfull) and non-metricity (torsionless) conditions. To construct dynamical Carroll theories, we need to define a Carroll invariant volume form: a non-singular \(D\times D\) matrix with non-vanishing determinant,
\begin{align}
    e\equiv \det(\tau_{\mu},e_{\mu}{}^{a})=\sqrt{\det\left(\tau_{\mu}\tau_{\nu}+h_{\mu\nu} \right)}\,,
\end{align}
which is invariant under internal Carroll boost transformations:
\begin{align}
    \delta_{C}\, e
    &= \frac{1}{2} e\left(\tau^{\mu}\tau^{\nu}+h^{\mu\nu} \right)\delta_{C} \left(\tau_{\mu}\tau_{\nu}+h_{\mu\nu}\right)
    =0\,.
\end{align}
Here, the variation \(\delta_C\) denotes the internal Carroll boost (at a fixed spacetime point), under which \(\delta_C \tau_\mu = \lambda_\mu\) and \(\delta_C h_{\mu\nu}=0\). The vanishing follows because \(\lambda_\mu v^\mu = 0\) \cite{Musaeus:2023oyp,Ahmadi-Jahmani:2025bls}.

Another important geometrical Carroll entity that we will use later in our analysis is the extrinsic curvature associated with the spatial metric \(h_{\mu\nu}\). It is defined as its Lie derivative along the time-flow vector \(v^\mu\) \cite{Campoleoni:2022ebj,Musaeus:2023oyp,Hansen:2021fxi,Ahmadi-Jahmani:2025bls}:
\begin{align}\label{extrinsiccurv}
    K_{\mu\nu}\equiv-\frac{1}{2}\mathcal{L}_{v}h_{\mu\nu}
    =-\frac{1}{2}\left(v^{\rho}\partial_{\rho}h_{\mu\nu}
    +h_{\mu\rho}\partial_{\nu}v^{\rho}
    +h_{\nu\rho}\partial_{\mu}v^{\rho}\right).
\end{align}
Condition \(K_{\mu\nu}=0\) implies that the spatial metric \(h_{\mu\nu}\) does not evolve along the temporal direction defined by \(v^\mu\). In this case, the Carrollian spatial geometry is frozen in time. Conversely, a non-vanishing extrinsic curvature signals a dynamical evolution of the spatial slices and can be interpreted as a manifestation of torsional effects in Carrollian geometry \cite{Figueroa-OFarrill:2020gpr}.
	
	The natural Carrollian \textbf{metric compatibility conditions}
	are \cite{Hansen:2021fxi,Musaeus:2023oyp}:
	\begin{align}\label{eq:metric-compatibility}
		\nabla_{\rho} v^{\mu}&=0,\quad\quad
		\nabla_{\rho}h_{\mu\nu}=0.
	\end{align}
	One may then proceed along the lines of Riemannian geometry and attempt to characterize compatible connections by their torsion. However, it has been demonstrated that the compatibility with the degenerate Carroll structure implies the existence of ``\textbf{intrinsic torsion}'', which generally cannot be set to zero \cite{Figueroa-OFarrill:2020gpr}. Thus, Carroll structures with metricity conditions can be classified based on their intrinsic and extrinsic torsion, a property natural to geometry i.e.($v^{\mu}$, $h_{\mu\nu}$). Recent work has shown that conditions \eqref{eq:metric-compatibility} can arise when we take ultra-relativistic limit of a Lorentzian manifold with Levi-Civita connection \cite{Hansen:2021fxi}.
	
	We can also consider Carrollian geometry with a \textbf{non-metricity condition} that can be identified as:
	\begin{align}\label{eq:nonmetricityC}
		\nabla_{\rho} v^{\mu}&=Q_{\rho}{}^{\mu},\quad\quad
		\nabla_{\rho}h_{\mu\nu}=Q_{\rho\mu\nu} \,,
	\end{align}
	and torsion $T^{\rho}{}_{\mu\nu}=2{\Gamma}_{[\mu\nu]}^{\rho}$. It has been demonstrated that non-metricity conditions \eqref{eq:nonmetricityC} appear in Carrollian geometries embedded in a higher-dimensional spacetime with a Levi-Civita connection \cite{Ciambelli:2025unn}. 
	\subsection{Carrollian Gravity}
	
	Carrollian geometry features a degenerate metric \(h_{\mu\nu} = \delta_{ab} e_\mu^a e_\nu^b\) (with \(h_{\mu\nu} v^\nu = 0\)) and extrinsic curvature \(K_{\mu\nu} = -\frac{1}{2} \mathcal{L}_n h_{\mu\nu}\). Unlike Riemannian geometry, as stated before, no unique torsion-free, metric-compatible connection exists unless \(K_{\mu\nu} = 0\); otherwise, parallel transport is limited.
	Carrollian gravity emerges as the ultra-relativistic limit of general relativity when the speed of light approaches zero, yielding a degenerate metric structure with a clock form \(\tau_\mu\) and spatial vielbein \(e_\mu^a\), where spatial propagation freezes \cite{Guerrieri:2021cdz,Bergshoeff:2017btm}. This framework describes dynamics on null hypersurfaces, contrasting with Lorentzian geometry \cite{Campoleoni:2022ebj,Adami:2023wbe,Ruzziconi:2026bix}. Carrollian gravity naturally splits into two dynamical sectors, Electric and Magnetic \cite{Henneaux:2021yzg,Guerrieri:2021cdz,Hansen:2021fxi}.

	\paragraph{Electric Carroll Gravity:}
	Electric Carroll gravity emerges as the leading order limit in the expansion of General Relativity, corresponding to an intrinsic torsional phase in which space can be dynamic. This phase is well-known and described by the action 
    \begin{equation}
       S_{\rm ECG} = \int e (K_{\mu\nu}K^{\mu\nu} - K^2 ) \, ,
    \end{equation}
    where \(e\) denotes the determinant of the Carrollian vielbein, serving as the volume density factor \cite{Hansen:2021fxi,Hartong:2015xda}. It encodes the shear and expansion of the spatial slices along the time flow, reproducing the kinetic terms of the ADM Hamiltonian constraint, with the extrinsic curvature \(K_{\mu\nu}\) absorbed into intrinsic torsion \cite{Hansen:2021fxi,deBoer:2023fnj,Figueroa-OFarrill:2020gpr,Donnay:2019jiz,Ahmadi-Jahmani:2025bls}.
	
	\paragraph{Magnetic Carroll Gravity:}
	Magnetic Carroll gravity can be obtained either by gauging the Carroll algebra or by considering the next-to-leading order limit in the expansion of General Relativity. Its action is expressed as 
    \begin{equation}
       S_{\rm MCG} = \int e\,\tilde{R}\, ,
    \end{equation}
    and it is characterized by the conditions of vanishing torsion, \(T^\rho{}_{\mu\nu} = 0\), and zero extrinsic curvature, \(K_{\mu\nu} = 0\). The corresponding connection is both torsion-free and compatible with the metric; nevertheless, it is not uniquely specified. Instead, it is determined up to an arbitrary symmetric tensor or $X_{\mu\nu}$ that is transverse with respect to the time-flow vector, satisfying the relation \(X_{\mu\nu} v^\nu = 0\) \cite{Bergshoeff:2017dqq,Campoleoni:2022ebj,Hansen:2021fxi,Ahmadi-Jahmani:2025bls}.

    Physically, the magnetic sector describes a rigid Carrollian universe whose gravitational content is encoded in the curvature $\tilde{R}$. The two sectors, electric and magnetic, are not independent, as we shall demonstrate, they are related by a dynamical equivalence within the Carrollian Trinity.

\section{Teleparallel Equivalent of Magnetic Carrollian Gravity}\label{sec:metric-sector}
	
In this section, we systematically develop a framework for investigating equivalent theories of Magnetic Carroll gravity. Subsection (\ref{section:GMetricCC}) constructs the most general metric-compatible Carroll connection to build the geometric foundation for the later analysis. Subsection (\ref{section:McompatibleEquivalentT}) then introduces first the Magnetic Carroll connection, which serves as the baseline structure for formulating Magnetic Carroll gravity and its equivalent theories. Building on this baseline, Subsections (\ref{section:ETI-ECG}-\ref{section:ETIV-MaxwellElectricT}) present the dynamically equivalent formulations of Magnetic Carrollian gravity, in close analogy with the geometric trinity of gravity in Lorentzian spacetimes. This analysis highlights how curvature and torsion can be reorganized within Carrollian geometry to obtain distinct but equivalent dynamical descriptions.

\subsection{General Metric Compatible Carroll Connection} \label{section:GMetricCC}
A Carrollian manifold is equipped with a degenerate metric with structures \((h_{\mu\nu}, v^\mu, \tau_\mu, h^{\mu\nu})\) satisfying complementary conditions \eqref{eq:carroll-conditions}, while the geometry admits a completeness relation (projector) given by $\delta^\nu_\mu$ as defined in \eqref{eq:carroll-conditions}. Here in this section, we demand a strictly metric-compatible Carrollian connection \({\nabla}_\mu\), such that the spatial metric and the time-flow vector are covariantly constant, i.e., \eqref{eq:metric-compatibility}. To find the most general metric-compatible Carrollian connection, we can construct it algebraically by using the completeness relation (projector) and solving for the connection coefficients piece by piece. Using the completeness relation any affine connection $\Gamma^\lambda_{\mu\nu}$ can be decomposed into temporal and spatial components:
\begin{equation}\label{eq:Gamma-decomposition}
		\Gamma^\lambda_{\mu\nu} = -v^\lambda A_{\mu\nu} + h^{\lambda\sigma} B_{\sigma\mu\nu}\, ,
	\end{equation}
	where the tensors $A$ and $B$ are defined as:
	\begin{equation}\label{def:AandB}
		A_{\mu\nu} = \tau_\lambda \Gamma^\lambda_{\mu\nu}, \quad B_{\sigma\mu\nu} = h_{\sigma\lambda} \Gamma^\lambda_{\mu\nu}\, .
	\end{equation}
	By definition, $B$ is completely spatial in its first index: $v^\sigma B_{\sigma\mu\nu} = 0$.
	We require the connection to be metric-compatible with the flow of time:
	\begin{equation}
		\nabla_\mu v^\rho = \partial_\mu v^\rho + \Gamma^\rho_{\mu\nu} v^\nu = 0
	\end{equation}
	Contracting this with $\tau_\rho$, we obtain a constraint on $A_{\mu\nu}$:
	\begin{equation}\label{eq:Aconstraint}
		\tau_\rho \partial_\mu v^\rho + A_{\mu\nu} v^\nu = 0 \implies A_{\mu\nu}v^\nu = -\tau_\rho \partial_\mu v^\rho \, .
	\end{equation}
	We also require the preservation of the normalization, $\nabla_\mu(\tau_\nu v^\nu) = 0$. Because $\nabla_\mu v^\nu = 0$, this implies $v^\nu \nabla_\mu \tau_\nu = 0$. Writing out the covariant derivative of $\tau$:
	\begin{equation}
		v^\nu (\partial_\mu \tau_\nu - \Gamma^\lambda_{\mu\nu} \tau_\lambda) = v^\nu \partial_\mu \tau_\nu - A_{\mu\nu}v^\nu = 0 \, .
	\end{equation}
	Notice that because $\tau_\nu v^\nu = -1$ is a constant, $\partial_\mu(\tau_\nu v^\nu) = 0 \implies v^\nu \partial_\mu \tau_\nu = -\tau_\nu \partial_\mu v^\nu$. Hence, this is perfectly consistent. 
	We can parameterize the most general solution for $A_{\mu\nu}$ \eqref{eq:Aconstraint} as:
	\begin{equation}\label{def:Amunu}
		A_{\mu\nu} = \partial_\mu \tau_\nu + X_{\mu\nu} \, ,
	\end{equation}
	where $X_{\mu\nu}$ is an arbitrary tensor satisfying $X_{\mu\nu}v^\nu = 0$. This corresponds to the freedom
	\begin{equation}\label{eq:nablatau-M}
		-\nabla_\mu \tau_\nu = X_{\mu\nu} \, ,
	\end{equation}
    where \(X_{\mu\nu}\) parametrizes the failure of the connection to preserve the clock form. We refer to it as the \emph{clock-form ambiguity}. The tensor \(X_{\mu\nu}\) plays a central role throughout this paper, and its principal choices correspond to physically distinct regimes. 

	Contracting $\nabla_\mu v^\rho = 0$ with $h_{\sigma\rho}$ yields a constraint on $B_{\sigma\mu\nu}$:
	\begin{equation}\label{eq:secondconstraintB}
		h_{\sigma\rho} \partial_\mu v^\rho + B_{\sigma\mu\nu} v^\nu = 0 \implies B_{\sigma\mu\nu} v^\nu = -h_{\sigma\lambda} \partial_\mu v^\lambda \, .
	\end{equation}
	Then we consider the spatial metric compatibility which is given by:
	\begin{equation}
		\nabla_\mu h_{\sigma\nu} = \partial_\mu h_{\sigma\nu} - \Gamma^\alpha_{\mu\sigma} h_{\alpha\nu} - \Gamma^\alpha_{\mu\nu} h_{\sigma\alpha} = 0 \, .
	\end{equation}
	Plugging in our definition of $B$ from \eqref{def:AandB}, this mandates that the symmetric part of $B$ in the 1st and 3rd indices perfectly equals the partial derivative of $h$:
	\begin{equation}
		B_{\nu\mu\sigma} + B_{\sigma\mu\nu} = \partial_\mu h_{\sigma\nu} \, .
	\end{equation}
	To build $B_{\sigma\mu\nu}$, we start with the following tensor:
	\begin{equation}
		C_{\sigma\mu\nu} = \frac{1}{2}\left( \partial_\mu h_{\nu\sigma} + \partial_\nu h_{\mu\sigma} - \partial_\sigma h_{\mu\nu} \right) \, ,
	\end{equation}
	which is $C_{\sigma\mu\nu} + C_{\nu\mu\sigma} = \partial_\mu h_{\sigma\nu}$, by construction.

	Now we check the required condition $C_{\sigma\mu\nu} v^\nu$. Using the identity $\mathcal{L}_v h_{\mu\nu} = v^\rho \partial_\rho h_{\mu\nu} + h_{\rho\nu} \partial_\mu v^\rho + h_{\mu\rho} \partial_\nu v^\rho = -2 K_{\mu\nu}$, we find:
	\begin{equation}
		C_{\sigma\mu\nu} v^\nu = -h_{\sigma\lambda} \partial_\mu v^\lambda - K_{\mu\sigma} \, .
	\end{equation}
	Because $B_{\sigma\mu\nu}$ requires $B_{\sigma\mu\nu} v^\nu = -h_{\sigma\lambda} \partial_\mu v^\lambda$, the term $K_{\mu\sigma}$ needs to be canceled out. Furthermore, we must enforce $v^\sigma B_{\sigma\mu\nu} = 0$. Evaluating $v^\sigma C_{\sigma\mu\nu}$ yields exactly $+K_{\mu\nu}$. 
	
	Then, we can systematically correct $C_{\sigma\mu\nu}$ by adding an anti-symmetric piece $A_{\sigma\mu\nu}$ (so as not to spoil $B_{\nu\mu\sigma} + B_{\sigma\mu\nu} = \partial_\mu h_{\sigma\nu}$) using the clock form $\tau$, we obtain the general form:
	\begin{equation}\label{def:Btensor}
		B_{\sigma\mu\nu} = C_{\sigma\mu\nu} + \tau_\sigma K_{\mu\nu} - \tau_\nu K_{\mu\sigma} + A_{\sigma\mu\nu} \, .
	\end{equation}
	Now we need to verify if form \eqref{def:Btensor} satisfies the constraints for $B$ tensor. We have first-index spatiality condition $v^\sigma B_{\sigma\mu\nu} = 0$ \eqref{def:AandB}. Since $\tau_\sigma v^\sigma = -1$ and $v^\sigma K_{\mu\sigma} = 0$, in order to satisfy this condition, we need to have $v^\sigma A_{\sigma\mu\nu} = 0$. 
	For $B$ we also have the third-index contraction constraint \eqref{eq:secondconstraintB}. To meet this requirement, we also need to impose $A_{\sigma\mu\nu} v^\nu = 0$.
	
	Thus, $A_{\sigma\mu\nu}$ must be an arbitrary tensor that is purely spatial in both its 1st and 3rd indices and anti-symmetric ($A_{\sigma\mu\nu} = -A_{\nu\mu\sigma}$).
	Now we plug $A_{\mu\nu}$ and $B_{\sigma\mu\nu}$ as obtained in \eqref{def:Amunu} and \eqref{def:Btensor} back into our decomposition \eqref{eq:Gamma-decomposition}. Because $h^{\lambda\sigma}\tau_\sigma = 0$, the $\tau_\sigma K_{\mu\nu}$ term in \eqref{def:Btensor} drops out.
	Thus, the most general Carrollian affine connection satisfying all given metric compatibilities is:
	\begin{equation}\label{eq:Gconnection}
		\Gamma^\lambda_{\mu\nu} = -v^\lambda \partial_\mu \tau_\nu + \gamma^\lambda_{\mu\nu} - \tau_\nu K^\lambda_\mu - v^\lambda X_{\mu\nu} + A^\lambda_{\mu\nu} \, .
	\end{equation}
	For convenience, the geometric quantities appearing in \eqref{eq:Gconnection} are defined below and summarized in Table \ref{tab:dictionary}, which serves as a reference guide to the notation used throughout this work:
	\begin{enumerate}[label=(\roman*)]
		\item $\gamma^\lambda_{\mu\nu}$ is the spatial Carrollian Christoffel symbol: $\gamma^\lambda_{\mu\nu} = \frac{1}{2} h^{\lambda\sigma} (\partial_\mu h_{\nu\sigma} + \partial_\nu h_{\mu\sigma} - \partial_\sigma h_{\mu\nu})$.
		\item $K^\lambda_\mu$ is the raised extrinsic curvature: $K^\lambda_\mu = h^{\lambda\sigma} K_{\mu\sigma}$.
		\item $X_{\mu\nu}$ is an arbitrary tensor field dictating the ambiguity of the spatial derivative of the clock form ($-\nabla_\mu \tau_\nu$), restricted only by $X_{\mu\nu} v^\nu = 0$. Parametrizing the failure of the clock form to be covariantly constant.
		\item $A^\lambda_{\mu\nu} = h^{\lambda\sigma}A_{\sigma\mu\nu}$ is an arbitrary tensor field dictating the intrinsic spatial torsion ambiguity, restricted by being spatial in its 1st and 3rd indices ($\tau_\lambda A^\lambda_{\mu\nu} = 0$ and $A^\lambda_{\mu\nu} v^\nu = 0$) and anti-symmetric in the spatial metric ($h_{\lambda\sigma} A^\sigma_{\mu\nu} = -h_{\nu\sigma} A^\sigma_{\mu\lambda}$).
	\end{enumerate}

\begin{table}[h]
\centering
\renewcommand{\arraystretch}{1.3}
\caption{Dictionary of the key tensors of the general metric-compatible Carrollian connection.}
\label{tab:dictionary}
\begin{tabularx}{\textwidth}{@{} l l >{\raggedright\arraybackslash}X @{}}
\toprule
 \textbf{Tensor} & \textbf{Description} \\
\midrule
 $A_{\mu\nu}$ & Temporal part; governs parallel transport of the clock form. \\
\midrule
 $B_{\sigma\mu\nu}$ & Spatial part; governs parallel transport of spatial vectors. \\
\midrule
$X_{\mu\nu}$ & Clock-form ambiguity: time dilation, clock-synchronization freedom. \\
\midrule
$A^\lambda_{\mu\nu}$ & Spatial contorsion: arbitrary spatial frame-dragging. \\
\midrule
 $K_{\mu\nu}$ & Extrinsic curvature: expansion and shear of space along the time flow. \\
\midrule
$F_{\mu\nu}$ & Clock-form field strength: temporal torsion (analogue of the electromagnetic field). \\
\midrule
$\gamma^\lambda_{\mu\nu}$ & Spatial Christoffel symbol: parallel transport within a spatial slice. \\
\bottomrule
\end{tabularx}
\end{table}
	
	\paragraph{Calculating $\nabla_\mu \tau_\nu$ and $\nabla_\mu h^{\nu\rho}$}: Now we want to know how $\tau_mu$ and $h^{\mu\nu}$ behave under parallel transportation using covariant derivative $\nabla_{\mu}$.
	Using the standard definition $\nabla_\mu \tau_\nu = \partial_\mu \tau_\nu - \Gamma^\lambda_{\mu\nu} \tau_\lambda$, we contract the connection with $\tau_\lambda$. 
	We will use the known properties of our connection's components in \eqref{eq:Gconnection}:
	\begin{subequations}\label{eq:Gconnectionrels}
		\begin{align}
			\tau_\lambda \gamma^\lambda_{\mu\nu} &= \frac{1}{2}\tau_\lambda h^{\lambda\sigma}(\dots) = 0 \, ,\\
			\tau_\lambda K^\lambda_\mu &= \tau_\lambda h^{\lambda\sigma}K_{\mu\sigma} = 0 \, ,\\
			\tau_\lambda A^\lambda_{\mu\nu} &= 0 \; \text{(by definition of spatiality)} \, ,\\
			\tau_\lambda v^\lambda &= -1 \, .
		\end{align}
	\end{subequations}
	Using relations \eqref{eq:Gconnectionrels} and \eqref{eq:Gconnection}, we find:
	\begin{equation}
		\tau_\lambda \Gamma^\lambda_{\mu\nu} = \partial_\mu \tau_\nu + X_{\mu\nu} \, .
	\end{equation}
	Thus, the covariant derivative of the clock form is:
	\begin{equation}\label{eq:covtau}
		\nabla_\mu \tau_\nu = \partial_\mu \tau_\nu - (\partial_\mu \tau_\nu + X_{\mu\nu}) \implies \nabla_\mu \tau_\nu = -X_{\mu\nu} \, .
	\end{equation}
	This reveals that $X_{\mu\nu}$ parametrizes the failure of the connection to preserve the clock form.
	
	To find the covariant derivative of $h^{\nu\rho}$, we take the covariant derivative of the completeness relation $\delta_\alpha^\rho = -\tau_\alpha v^\rho + h_{\alpha\sigma}h^{\sigma\rho}$. Since $\nabla_\mu \delta_\alpha^\rho = 0$, we obtain:
	\begin{equation}
		0 = -(\nabla_\mu \tau_\alpha) v^\rho - \tau_\alpha (\nabla_\mu v^\rho) + (\nabla_\mu h_{\alpha\sigma}) h^{\sigma\rho} + h_{\alpha\sigma} (\nabla_\mu h^{\sigma\rho}) \, .
	\end{equation}
	Using metric compatibility conditions \eqref{eq:metric-compatibility} and $\nabla_\mu \tau_\alpha = -X_{\mu\alpha}$, this simplifies to:
	\begin{equation}
		h_{\alpha\sigma} \nabla_\mu h^{\sigma\rho} = -X_{\mu\alpha} v^\rho \, .
	\end{equation}
	Contract this with the inverse metric $h^{\alpha\beta}$ and use the identity $h^{\alpha\beta} h_{\alpha\sigma} = \delta_\sigma^\beta + \tau_\sigma v^\beta$, we obtain:
	\begin{align}
		-X_{\mu\alpha} h^{\alpha\beta} v^\rho = \nabla_\mu h^{\beta\rho} + v^\beta (\tau_\sigma \nabla_\mu h^{\sigma\rho}) \, . \label{eq:nabla-contrah}
	\end{align}
	To find the $\tau_\sigma \nabla_\mu h^{\sigma\rho}$ term in the above relation, we use the fact that $\tau_\sigma h^{\sigma\rho} = 0$, meaning that we have $\nabla_\mu(\tau_\sigma h^{\sigma\rho}) = (\nabla_\mu \tau_\sigma)h^{\sigma\rho} + \tau_\sigma \nabla_\mu h^{\sigma\rho} = 0$. Therefore, we obtain $\tau_\sigma \nabla_\mu h^{\sigma\rho} = X_{\mu\sigma}h^{\sigma\rho}$. Substituting this back in \eqref{eq:nabla-contrah} gives the final formula for covariant derivative of $h^{\nu\rho}$:
	\begin{equation}\label{eq:covh}
		\nabla_\mu h^{\nu\rho} = - v^\nu h^{\rho\sigma} X_{\mu\sigma} - v^\rho h^{\nu\sigma} X_{\mu\sigma} \, .
	\end{equation}
	
	As a consistency check, we use our formulas \eqref{eq:covtau} and \eqref{eq:covtau} to prove that $\tau_\mu h^{\mu\nu} = 0$ is covariantly conserved. For this objective, we start from:
	\begin{equation}
		\nabla_\rho(\tau_\mu h^{\mu\nu}) = (\nabla_\rho \tau_\mu) h^{\mu\nu} + \tau_\mu (\nabla_\rho h^{\mu\nu}) \, ,
	\end{equation}
	then, substituting our results from \eqref{eq:covtau} and \eqref{eq:covh} we get:
	\begin{equation}
		= (-X_{\rho\mu})h^{\mu\nu} + \tau_\mu \left( -v^\mu h^{\nu\sigma} X_{\rho\sigma} - v^\nu h^{\mu\sigma} X_{\rho\sigma} \right) \, .
	\end{equation}
	By using the complementary relations \eqref{eq:carroll-conditions}, we can finally show:
	\begin{equation}
		\nabla_\rho(\tau_\mu h^{\mu\nu}) = -X_{\rho\mu}h^{\mu\nu} + X_{\rho\sigma}h^{\sigma\nu} = 0 \, .
	\end{equation}
	So the condition $\tau_\mu h^{\mu\nu} = 0$ is perfectly preserved, even though $\tau_\mu$ and $h^{\mu\nu}$ individually are \textbf{not} covariantly constant (unless $X=0$). On the other hand, by the very definition of a covariant derivative as a tensor derivation, it obeys the Leibniz product rule. Because $\tau_\mu h^{\mu\nu} = 0$ is an algebraic identity everywhere on the manifold, its covariant derivative is trivially zero: $\nabla_\rho(\tau_\mu h^{\mu\nu}) = 0$.

	Now we want to calculate the torsion tensor $T^\lambda_{\mu\nu} = \Gamma^\lambda_{\mu\nu} - \Gamma^\lambda_{\nu\mu}$. 
	Using the connection \eqref{eq:Gconnection}, we subtract its index-swapped counterpart. Noting that the spatial Christoffel symbol $\gamma^\lambda_{\mu\nu}$ is symmetric in $\mu$ and $\nu$ (so it cancels out), we get:
	\begin{equation}
		T^\lambda_{\mu\nu} = -v^\lambda (\partial_\mu \tau_\nu - \partial_\nu \tau_\mu) - (\tau_\nu K^\lambda_\mu - \tau_\mu K^\lambda_\nu) - v^\lambda (X_{\mu\nu} - X_{\nu\mu}) + (A^\lambda_{\mu\nu} - A^\lambda_{\nu\mu}) \, .
	\end{equation}
	Let the exterior derivative of the clock form $\tau$ be 
	\begin{equation}\label{eq:Fmunu}
		F_{\mu\nu} = \partial_\mu \tau_\nu - \partial_\nu \tau_\mu \, ,
	\end{equation}
	and denote the antisymmetrization by brackets (e.g., $M_{[\mu\nu]} = \frac{1}{2}(M_{\mu\nu} - M_{\nu\mu})$). Then, the exact form of torsion will be:
	\begin{equation}\label{eq:gtorsion}
		T^\lambda_{\mu\nu} = -v^\lambda F_{\mu\nu} - 2\tau_{[\nu} K^\lambda_{\mu]} - 2v^\lambda X_{[\mu\nu]} + 2A^\lambda_{[\mu\nu]} \, .
	\end{equation}
	
	A remark is in order regarding the tensor $A^\lambda_{\mu\nu}$. Mathematically, $A^\lambda_{\mu\nu}$ represents the ``spatial contorsion tensor''. Because it is defined by $A_{\sigma\mu\nu} = h_{\sigma\lambda}A^\lambda_{\mu\nu} = -A_{\nu\mu\sigma}$, which means it lives entirely in the spatial subspace \footnote{it vanishes when contracted with $v$ or $\tau$}. It parametrizes the intrinsic \textit{gauge freedom} of affine connections on degenerate metrics. In standard pseudo-Riemannian geometries, metric compatibility uniquely fixes a torsion-free connection. In Carrollian geometry, the metric degeneracy leaves this purely spatial anti-symmetric ambiguity. Physically, it represents arbitrary \textit{spatial frame-dragging or twisting}. If we parallel-transport a spatial vector around a closed loop using this connection \eqref{eq:Gconnection}, $A^\lambda_{\mu\nu}$ dictates an extra, arbitrary internal rotation of our spatial frame that does not alter vector lengths (since $\nabla h = 0$) and does not mix space with time (since $\nabla v = 0$).
	We can fix $A^\lambda_{\mu\nu}$ uniquely by making a physical choice about the torsion of Carrollian spacetime. 
	
	If we project the torsion tensor $T^\lambda_{\mu\nu}$ completely onto the spatial subspace using the spatial projector $h^\alpha_\beta = \delta^\alpha_\beta + \tau_\beta v^\alpha$, all terms involving $v^\lambda$ or $\tau$ vanish. Thus, the purely spatial torsion is exactly:
	\begin{equation}\label{eq:spatialtorsion}
		T^{\perp \lambda}_{\mu\nu} = h^\lambda_\alpha h^\beta_\mu h^\gamma_\nu T^\alpha_{\beta\gamma} = A^\lambda_{\mu\nu} - A^\lambda_{\nu\mu} \, .
	\end{equation}
	If we demand that spatial torsion is zero, then $A^\lambda_{\mu\nu}$ must be symmetric in $\mu$ and $\nu$ ($A_{\sigma\mu\nu} = A_{\sigma\nu\mu}$).
	However, metric-compatibility already forces it to be anti-symmetric in its outer indices ($A_{\sigma\mu\nu} = -A_{\nu\mu\sigma}$). 
	A tensor that is symmetric in its last two indices but anti-symmetric in its first and last must be identically zero:
	\begin{equation}\label{eq:zeroST}
		A_{\sigma\mu\nu} = A_{\sigma\nu\mu} = -A_{\mu\nu\sigma} = -A_{\mu\sigma\nu} = A_{\nu\sigma\mu} = A_{\nu\mu\sigma} = -A_{\sigma\mu\nu} \implies A_{\sigma\mu\nu} = 0 \, .
	\end{equation}
	Therefore, \textit{imposing zero spatial torsion} \textbf{uniquely} fixes $A^\lambda_{\mu\nu} = 0$.

	\subsection{Magnetic Carrollian Gravity and its Dynamical Equivalent Theories}\label{section:McompatibleEquivalentT}
	
	The foundation of our analysis is Magnetic Carrollian Gravity, which serves as the baseline geometric theory of the Carrollian regime here. In this formulation, the manifold is equipped with a strictly metric-compatible and torsion-free affine connection, denoted as $\tilde{\Gamma}^\lambda_{\mu\nu}$. As established by the geometric constraints of the degenerate metric \eqref{eq:carroll-conditions}, enforcing simultaneous metricity and vanishing torsion mathematically forbids any temporal evolution of the spatial metric \cite{Campoleoni:2022ebj}, locking the spacetime into a rigid configuration where the extrinsic curvature identically vanishes ($K_{\mu\nu} = 0$). The dynamics of this base theory are governed by the magnetic Carrollian Ricci scalar, $\tilde{R}$ \cite{Hansen:2021fxi}. 
	
	Following the approach of the ``Geometrical Trinity of Gravity'' in General Relativity \cite{BeltranJimenez:2019esp,Aldrovandi:2013wha,Koivisto:2018loq,Nester:1998mp, BeltranJimenez:2017tkd, Maluf:2013gaa}, we construct dynamically equivalent theories by shifting this magnetic Carrollian curvature into alternative geometric sectors. Starting with the general metric-compatible connection \eqref{eq:Gconnection} and relaxing the zero-torsion constraint, we isolate specific components of the Carrollian torsion tensor \eqref{eq:gtorsion}. If we subsequently demand that these new torsion-bearing connections be globally flat, meaning their respective Riemann curvature tensors vanish, the magnetic Carrollian curvature $\tilde{R}$ will be algebraically traded for the kinetic energy density of the isolated torsion fields. Each specific choice of torsion activation carries its own distinct physical motivation, that we will discussed below in each subsection, generating dynamically equivalent actions governed by fundamentally different kinetic behaviors. 
	
	In Lorentzian spacetime, the Geometrical Trinity maps Einstein curvature strictly to alternative \textit{gravitational/geometric} frameworks, namely Teleparallel and Symmetric Teleparallel gravity. In contrast, here the degenerate fiber-bundle structure of Carrollian geometry dictates that some of these equivalent theories may manifest as \textit{non-gravitational} theories. Depending on which sector of the torsion tensor is isolated, we demonstrate that the gravitational curvature of Magnetic Carroll Gravity can be recast into the Electric Carroll Gravity, Hamiltonian of a scalar-charge Fracton gauge theory, the electric field energy of Maxwell electromagnetism, or an ultralocal scalar dilaton theory. This finding establishes a geometric dictionary translating Carrollian gravity into gauge field and condensed matter theories.

	\subsubsection{Magnetic Carrollian Gravity}\label{section:MCG}
	To establish our reference theory, we first isolate pure intrinsic curvature by demanding a completely torsion-free connection \((T^\lambda_{\mu\nu} = 0)\).
	
	Setting the entire torsion of a Carrollian connection to zero has profound geometric consequences. In pseudo-Riemannian geometry, metric compatibility uniquely fixes a torsion-free connection (the Levi-Civita connection). In Carrollian geometry, due to the degenerate metric, requiring zero torsion places a strict constraint on the spacetime itself, while still leaving a symmetric ambiguity in the connection.
	
	We derived the torsion previously in \eqref{eq:Gconnection}. If we demand $T^\lambda_{\mu\nu} = 0$ in Eq.~\eqref{eq:gtorsion}, we first examine the spatial projection of the torsion by contracting $T^\lambda_{\mu\nu} = 0$ with the spatial projector $h^\rho_\lambda$. From this contraction, we obtain:
	\begin{equation}
		A^\rho_{\mu\nu} - A^\rho_{\nu\mu} = \tau_\nu K^\rho_\mu - \tau_\mu K^\rho_\nu \, .
	\end{equation}
	By definition, $A_{\sigma\mu\nu}$ must be anti-symmetric in its first and third indices ($A_{\sigma\mu\nu} = -A_{\nu\mu\sigma}$) to preserve $\nabla h_{\mu\nu}=0$. Applying this cyclic symmetry requirement to the equation above forces:
	\begin{equation}
		\tau_\nu K_{\sigma\mu} = -\tau_\mu K_{\sigma\nu} \, .
	\end{equation}
	If we contract this with $v^\mu$, we get $ \implies K_{\sigma\nu} = 0$.
	So, a torsion-free, metric-compatible Carrollian connection only exists if the extrinsic curvature vanishes identically ($K_{\mu\nu} = 0$). The Carrollian spacetime must be ``rigid'' (its spatial metric does not evolve along the flow of time). Thus, we must assume $K_{\mu\nu} = 0$ to proceed.
	
	Assuming $K_{\mu\nu} = 0$, the $A^\lambda_{\mu\nu}$ tensor vanishes, as discussed in \eqref{eq:zeroST}. Then, the torsion-free condition requires 
	\begin{equation}\label{eq:antiX-magnetic}
		2X_{[\mu\nu]} = -F_{\mu\nu} \, .
	\end{equation}
	In order to write the final expression for the connection in a more compact form, let us define a symmetric combination 
	\begin{equation}\label{def:Upsilon}
		\Upsilon_{\mu\nu} := \partial_\mu \tau_\nu + X_{\mu\nu} \, ,
	\end{equation}
	with the spatiality condition $X_{\mu\nu}v^\nu = 0$, which allows us to define:
	\begin{equation}\label{def:bmu}
		\Upsilon_{\mu\nu} v^\nu = -\tau_\sigma \partial_\mu v^\sigma =: b_\mu \, ,
	\end{equation} 
	Then, from \eqref{eq:Gconnection}, the new unique torsion-free connection simplifies to:
	\begin{equation}\label{eq:magnetic-connection}
		\tilde{\Gamma}^\lambda_{\mu\nu} = \gamma^\lambda_{\mu\nu} - v^\lambda \Upsilon_{\mu\nu}\,,
	\end{equation}
	where $\gamma^\lambda_{\mu\nu} = \frac{1}{2}h^{\lambda\sigma}(\partial_\mu h_{\nu\sigma} + \partial_\nu h_{\mu\sigma} - \partial_\sigma h_{\mu\nu})$ is the spatial Christoffel symbol and $\Upsilon_{\mu\nu}$ is an arbitrary symmetric tensor satisfying $\Upsilon_{\mu\nu} v^\nu = b_\mu$ \footnote{The spatially symmetric part of $\Upsilon_{\mu\nu}$ remains an unfixed gauge degree of freedom, which will be constrained in the following sections according to the physical modifications of the problem.}.
	
	\paragraph{Calculating the Curvature}:
	Let $\hat{\nabla}$ and $\hat{R}^\rho_{\ \sigma\mu\nu}$ be respectively the covariant derivative and Riemann tensor associated purely with the spatial Christoffel symbol $\gamma^\lambda_{\mu\nu}$. Because $\tilde{\Gamma}^\lambda_{\mu\nu}$ is just $\gamma^\lambda_{\mu\nu}$ shifted by a tensor $C^\lambda_{\mu\nu} = -v^\lambda \Upsilon_{\mu\nu}$, we can compute the curvature exactly.
	Using the following identity which is obtained from the metricity condition \eqref{eq:metric-compatibility}
	\begin{equation}\label{eq:hatnablav}
		\hat{\nabla}_\mu v^\rho = v^\rho b_\mu \, ,
	\end{equation}
	the non-linear $C \times C$ terms in the Riemann tensor perfectly cancel the derivative $v^\rho$ terms, leading to a linear shift. 
	Thus, the Riemann tensor will be
	\begin{equation}\label{eq:tildeRiemann}
		\tilde{R}^\rho_{\ \sigma\mu\nu} = \hat{R}^\rho_{\ \sigma\mu\nu} - v^\rho \left( \hat{\nabla}_\mu \Upsilon_{\nu\sigma} - \hat{\nabla}_\nu \Upsilon_{\mu\sigma} \right) \, .
	\end{equation}
	As a check, we contract this with $v^\sigma$ which gives $R^\rho_{\ \sigma\mu\nu} v^\sigma = 0$, that correctly reflects $[\tilde{\nabla}_\mu, \tilde{\nabla}_\nu]v^\rho = 0$.
	We trace the Riemann tensor $R_{\sigma\nu} = R^\rho_{\ \sigma\rho\nu}$ to derive the Ricci tensor. Using the identity 
    \begin{equation}\label{eq:identityI}
        v^\rho \hat{\nabla}_\nu \Upsilon_{\rho\sigma} = \hat{\nabla}_\nu b_\sigma - b_\nu b_\sigma \, ,
    \end{equation}
    we obtain:
	\begin{equation}\label{eq:magnetic-riccis0}
		\tilde{R}_{\sigma\nu} = \hat{R}_{\sigma\nu} - v^\rho \hat{\nabla}_\rho \Upsilon_{\nu\sigma} + \hat{\nabla}_\nu b_\sigma - b_\nu b_\sigma \, ,
	\end{equation}
	where $\hat{R}_{\sigma\nu} = \hat{R}^\rho_{\ \sigma\rho\nu}$ is the spatial Ricci tensor \cite{Bergshoeff:2017btm,Hansen:2021fxi}.
	
	For the Ricci scalar, we contract the Ricci tensor \eqref{eq:magnetic-riccis0} with the spatial metric $\tilde{R} = h^{\sigma\nu} \tilde{R}_{\sigma\nu}$. Let $\Upsilon := h^{\sigma\nu}\Upsilon_{\sigma\nu}$ be the spatial trace of $\Upsilon$. 
	Using the identity 
    \begin{equation}\label{eq:identityII}
        v^\rho \hat{\nabla}_\rho h^{\sigma\nu} = (E^\nu + b^\nu)v^\sigma + (E^\sigma + b^\sigma)v^\nu \, ,
    \end{equation}
    where the Carrollian \textit{gravitoelectric} field is defined by \footnote{$E_\mu$ is the \emph{Carrollian acceleration} vector introduced in \cite{Ciambelli:2025unn,Hansen:2021fxi} and satisfies $v^\sigma \partial_\sigma \tau_\mu = E_\mu + b_\mu$ with $b_\mu$ defined in \eqref{def:bmu}.}:
    \begin{equation}\label{eq:Emu}
     E = \mathcal{L}_v \tau \, ,
    \end{equation}
     one may express the traced contribution  $h^{\sigma\nu} v^\rho \hat{\nabla}_\rho \Upsilon_{\nu\sigma}$ appearing in \eqref{eq:magnetic-riccis0} as $\mathcal{L}_v \Upsilon - 2 b^\mu b_\mu - 2 E^\nu b_{\nu}$. The motivation for referring to $E_\mu$ as a gravitoelectric field will become apparent in Sec.~\ref{section:ETIV-MaxwellElectricT}.
     Substituting this result into the trace of \eqref{eq:magnetic-riccis0} finally leads to the Carrollian Ricci scalar:
	\begin{equation}\label{eq:magnetic-riccis}
		\tilde{R} = \hat{R} - \mathcal{L}_v \Upsilon + h^{\sigma\nu} \hat{\nabla}_\nu b_\sigma + b^\mu b_\mu + 2 E^\nu b_{\nu} \, ,
	\end{equation}
	where $\mathcal{L}_v \Upsilon = v^\rho \partial_\rho \Upsilon$ is the time evolution of the trace $\Upsilon$ and we also defined $b^\mu = h^{\mu\nu}b_\nu$. For a review on gauging algebra method for deriving the action of Magnetic Carrollian gravity, see App.~\ref{app:gauging}.
	
	Thus, setting the spatial torsion to zero requires \(K_{\mu\nu} = 0\), forcing the spacetime to be rigidly non-expanding. Furthermore, the temporal torsion vanishes if \(\Upsilon_{\mu\nu} \equiv \partial_\mu \tau_\nu + X_{\mu\nu}\) is strictly symmetric. Under these constraints, we obtained a torsion-free ``Magnetic Connection'' \eqref{eq:magnetic-connection}, whose curvature \eqref{eq:magnetic-riccis} acts as the Lagrangian for \textit{Magnetic Carrollian Gravity} \cite{Hansen:2021fxi,Bergshoeff:2017btm}:
	\begin{equation}\label{eq:MCaction}
		S_{\text{Mag}} = \int d^D x \, e \, \tilde{R}\,,
	\end{equation}
	where \(e\) is the Carrollian volume measure. In the subsequent sections, we will demonstrate how shifting this curvature into the isolated torsion sectors generates dynamically equivalent theories.
	In \cite{Hansen:2021fxi}, the Magnetic Carroll action \eqref{eq:MCaction} is obtained through a limiting procedure that automatically fixes the freedom in the symmetric part of $X_{\mu\nu}$ to $X_{\mu\nu}=\tau_{(\mu}\mathcal{L}_v\tau_{\nu)}$, whereas in our construction here, the symmetric degree of freedom of $X_{\mu\nu}$ is left unrestricted.
	\subsubsection{Equivalent Theory I: Electric Carrollian Gravity}\label{section:ETI-ECG}
	
	In General Relativity, the ADM formalism splits the Hamiltonian constraint into intrinsic curvature and a kinetic term dependent on extrinsic curvature. Using the fiber bundle structure of Carrollian geometry, we can recover this structure by relaxing the rigidity constraint \((K_{\mu\nu} \neq 0)\) and letting the connection absorb extrinsic curvature as torsion.
	By keeping the extrinsic curvature term while turning off all other sources of torsion, we are constructing a very important connection in Carrollian geometry (often closely related to the intrinsic Carrollian connections studied when the spacetime is not rigid, i.e., $K_{\mu\nu} \neq 0$) \cite{Hansen:2021fxi,Musaeus:2023oyp,Bekaert:2015xua,Hartong:2015xda}.
	
	To define the new connection $\check{\Gamma}^\lambda_{\mu\nu}$,
	we start from the most general metric-compatible connection derived earlier in \eqref{eq:Gconnection} and impose three conditions: I) Setting spatial torsion to zero: $A^\lambda_{\mu\nu} = 0$, II) Keeping the extrinsic curvature torsion, where the term $-\tau_\nu K^\lambda_\mu$ stays, III) Setting the torsion of $F$ and $X$ to zero: The remaining torsion from these terms is $-v^\lambda (F_{\mu\nu} + 2X_{[\mu\nu]})$, setting this to zero requires:
	\begin{equation}
		\partial_\mu \tau_\nu - \partial_\nu \tau_\mu + X_{\mu\nu} - X_{\nu\mu} = 0 \, .
	\end{equation}
	This implies that the combination $\Upsilon_{\mu\nu} \equiv \partial_\mu \tau_\nu + X_{\mu\nu}$ is perfectly symmetric ($\Upsilon_{\mu\nu} = \Upsilon_{\nu\mu}$).
	Substituting these conditions back into \eqref{eq:Gconnection}, the new connection becomes:
	\begin{equation}
		\check{\Gamma}^\lambda_{\mu\nu} = \gamma^\lambda_{\mu\nu} - v^\lambda \Upsilon_{\mu\nu} - \tau_\nu K^\lambda_\mu \, .
	\end{equation}
	Notice that the first two terms ($\gamma^\lambda_{\mu\nu} - v^\lambda \Upsilon_{\mu\nu}$) are exactly the connection $\tilde{\Gamma}^\lambda_{\mu\nu}$ from \eqref{eq:magnetic-connection}. Because $\Upsilon_{\mu\nu}$ is symmetric, $\tilde{\Gamma}$ acts as a completely torsion-free (symmetric) reference connection. Therefore, we can write our new connection as a simple tensor shift:
	\begin{equation}\label{eq:connection-electric}
		\check{\Gamma}^\lambda_{\mu\nu} = \tilde{\Gamma}^\lambda_{\mu\nu} - \tau_\nu K^\lambda_\mu \, ,
	\end{equation}
	where the torsion of this new connection is purely made by extrinsic curvature $\check{T}^\lambda_{\mu\nu} = -2\tau_{[\nu} K^\lambda_{\mu]}$, as requested.
	
	The symmetric part of \( X_{\mu\nu} \) remains undetermined. To construct Magnetic Carroll gravity from the connection \eqref{eq:magnetic-connection}, it suffices to fix its antisymmetric component, as done in \eqref{eq:antiX-magnetic}. Here, however, we specify the symmetric part by requiring it to be proportional to the extrinsic curvature:  
	\begin{equation}\label{eq:Xchoice}
		X_{(\mu\nu)} = \alpha K_{\mu\nu} \, ,
	\end{equation}
	where \( \alpha \) is a constant. The motivation for this choice will become clear later.
	To isolate the extrinsic curvature torsion, we also require $X_{[\mu\nu]} = -\frac{1}{2}F_{\mu\nu}$, which was also imposed in \eqref{eq:antiX-magnetic} when constructing MCG. Therefore, the full tensor is forced to be:
	\begin{equation}\label{eq:Xtensor-eleccarr}
		X_{\mu\nu} = \alpha K_{\mu\nu} - \frac{1}{2}F_{\mu\nu} \, .
	\end{equation}
	
	Now we will do a consistency check. The tensor $X_{\mu\nu}$ must satisfy the spatiality condition with respect to the vector: $X_{\mu\nu}v^\nu = 0$. 
	Since $K_{\mu\nu} v^\nu = 0$, this requires $F_{\mu\nu}v^\nu = 0$. This term can be evaluated to:
	\begin{equation}
		F_{\mu\nu}v^\nu = v^\nu \partial_\mu \tau_\nu - v^\nu \partial_\nu \tau_\mu \, .
	\end{equation}
	Using the identities from $\tau_\mu v^\mu = -1$, we know $v^\nu \partial_\mu \tau_\nu = -\tau_\nu \partial_\mu v^\nu$. Thus, the second term in the above equation is just $v^\nu \partial_\nu \tau_\mu$, from which we obtain:
	\begin{equation}
		F_{\mu\nu}v^\nu = -\tau_\nu \partial_\mu v^\nu - v^\nu \partial_\nu \tau_\mu = -\mathcal{L}_v \tau_\mu \, .
	\end{equation}
	Therefore, we derive $X_{\mu\nu} v^\nu = 0 \implies \mathcal{L}_v \tau_\mu = 0$.
	This assumption has two physical meanings: 
	i) \textit{Stationary time}: The assumption is mathematically allowed if and only if $\mathcal{L}_v \tau_\mu = 0$. Physically, this means the clock form $\tau_\mu$ is invariant under the flow of time. There is no ``Carrollian gravitoelectric field'' \footnote{Much like the gravitoelectric field in general relativity, the ‘electric‑like’ component of gravity in the weak‑field approximation, or the lapse function \(N\) in the ADM and 1+3 decompositions of spacetime, which is tied to the clock one‑form (the covector field aligned with the time direction that governs proper time intervals), the gravitoelectric field can be interpreted as the acceleration of the congruence defined by the clock one‑form \cite{Costa:2012cw, nuastase2025general, Ruggiero:2021uag}}; the local synchronization of clocks does not drift or evolve as time progresses.
	ii) \textit{Expansion-driven time dilation}: We note that $X_{\mu\nu}$ controls the spatial gradient of time ($\nabla_\mu \tau_\nu = -X_{\mu\nu}$). Setting its symmetric part to $\alpha K_{\mu\nu}$ means that the spatial variation of clock rates (time dilation) is rigidly locked to the physical shearing and expansion ($K_{\mu\nu}$) of the space itself.
	
	By using tensor \eqref{eq:Xtensor-eleccarr}, we construct the new connection $\check{\Gamma}$ in \eqref{eq:connection-electric} with a specific choice \eqref{eq:Xchoice} for $X$, where $\tilde{\Gamma}$ is the fully torsion-free connection. Note that the specific choice of $X$ is hidden inside the definition of $\tilde{\Gamma}^\lambda_{\mu\nu} = \gamma^\lambda_{\mu\nu} - v^\lambda (\partial_\mu \tau_\nu + X_{\mu\nu})$, ensuring that $\tilde{\Gamma}$ remains totally symmetric.
	
	\paragraph{Calculating the Curvatures}:
	Let the shift tensor be $C^\lambda_{\mu\nu} = -\tau_\nu K^\lambda_\mu$. Because $\tilde{\Gamma}$ is torsion-free, the Riemann tensor of $\check{\Gamma}$ relates to the Riemann tensor of $\tilde{\Gamma}$ via the standard formula:
	\begin{equation}\label{eq:check-riemann1}
		\check{R}^\rho_{\ \sigma\mu\nu} = \tilde{R}^\rho_{\ \sigma\mu\nu} + \tilde{\nabla}_\mu C^\rho_{\nu\sigma} - \tilde{\nabla}_\nu C^\rho_{\mu\sigma} + C^\rho_{\mu\lambda} C^\lambda_{\nu\sigma} - C^\rho_{\nu\lambda} C^\lambda_{\mu\sigma}  \, .
	\end{equation}
	
	In the following we will evaluate each piece. The non-linear (quadratic) terms vanish as follows,
	\begin{equation}
		C^\rho_{\mu\lambda} C^\lambda_{\nu\sigma} = (-\tau_\lambda K^\rho_\mu) (-\tau_\sigma K^\lambda_\nu) = \tau_\sigma K^\rho_\mu (\tau_\lambda K^\lambda_\nu) 
	\end{equation}
	since the extrinsic curvature is purely spatial ($K^\lambda_\nu = h^{\lambda\alpha}K_{\alpha\nu}$), we have $\tau_\lambda K^\lambda_\nu = 0$. Thus, $C \times C = 0$ identically. Thus, the shift in curvature is purely linear.
	
	We also need $\tilde{\nabla}_\mu C^\rho_{\nu\sigma} = -(\tilde{\nabla}_\mu \tau_\sigma) K^\rho_\nu - \tau_\sigma \tilde{\nabla}_\mu K^\rho_\nu$. For this, we need the covariant derivative of the clock form with respect to the reference connection $\tilde{\Gamma}$, which we will calculate in the following, 
	\begin{equation}
		\tilde{\nabla}_\mu \tau_\sigma = \partial_\mu \tau_\sigma - \tilde{\Gamma}^\lambda_{\mu\sigma} \tau_\lambda = \partial_\mu \tau_\sigma - (\gamma^\lambda_{\mu\sigma} - v^\lambda \Upsilon_{\mu\sigma})\tau_\lambda \, ,
	\end{equation}
	using $\tau_\lambda \gamma^\lambda_{\mu\sigma} = 0$ and $\tau_\lambda v^\lambda = -1$, we get:
	\begin{equation}
		\tilde{\nabla}_\mu \tau_\sigma = \partial_\mu \tau_\sigma - \Upsilon_{\mu\sigma} \, .
	\end{equation}
	Recalling that $\Upsilon_{\mu\sigma}$ \eqref{def:Upsilon}, the above relation implies: $\tilde{\nabla}_\mu \tau_\sigma = -X_{\mu\sigma} $. Therefore, the derivative of the shift tensor will be:
	\begin{equation}
		\tilde{\nabla}_\mu C^\rho_{\nu\sigma} = X_{\mu\sigma} K^\rho_\nu - \tau_\sigma \tilde{\nabla}_\mu K^\rho_\nu \, .
	\end{equation}
	Substituting the evaluated pieces back into the definitions of Riemann \eqref{eq:check-riemann1}, we obtain the following exact relation, 
	\begin{equation}\label{eq:riemann-electric2}
		\check{R}^\rho_{\ \sigma\mu\nu} = \tilde{R}^\rho_{\ \sigma\mu\nu} + X_{\mu\sigma} K^\rho_\nu - X_{\nu\sigma} K^\rho_\mu - \tau_\sigma \big( \tilde{\nabla}_\mu K^\rho_\nu - \tilde{\nabla}_\nu K^\rho_\mu \big) \, .
	\end{equation}
	We are now simply substituting \eqref{eq:Xtensor-eleccarr} into \eqref{eq:riemann-electric2}, then we find:
	\begin{equation}\label{eq:riemann-electric3}
		\check{R}^\rho_{\ \sigma\mu\nu} = \tilde{R}^\rho_{\ \sigma\mu\nu} + \alpha(K_{\mu\sigma} K^\rho_\nu - K_{\nu\sigma} K^\rho_\mu) - \frac{1}{2}(F_{\mu\sigma} K^\rho_\nu - F_{\nu\sigma} K^\rho_\mu) - \tau_\sigma ( \tilde{\nabla}_\mu K^\rho_\nu - \tilde{\nabla}_\nu K^\rho_\mu ) \, .
	\end{equation}
	Notice the term $\alpha(K_{\mu\sigma} K^\rho_\nu - K_{\nu\sigma} K^\rho_\mu)$. This is exactly the standard Gauss equation from the ADM formulation of General Relativity.
	Tracing the Riemann tensor \eqref{eq:riemann-electric3} over $\rho$ and $\mu$ and defining $K = K^\rho_\rho$, we derive:
	\begin{equation}
		\check{R}_{\sigma\nu} = \tilde{R}_{\sigma\nu} + \alpha(K_{\rho\sigma} K^\rho_\nu - K_{\nu\sigma} K) - \frac{1}{2}(F_{\rho\sigma} K^\rho_\nu - F_{\nu\sigma} K) - \tau_\sigma ( \tilde{\nabla}_\rho K^\rho_\nu - \tilde{\nabla}_\nu K ) \, .
	\end{equation}
	
	Then, the Ricci scalar will be obtained by tracing with the spatial metric: $\check{R} = h^{\sigma\nu} \check{R}_{\sigma\nu}$.
	As established earlier, the $\tau_\sigma$ derivative term vanishes because $h^{\sigma\nu}\tau_\sigma = 0$. Then, we get:
	\begin{equation}
		\check{R} = \tilde{R} + \alpha(K_{\rho\sigma} K^{\rho\sigma} - K^2) - \frac{1}{2}(h^{\sigma\nu}F_{\rho\sigma} K^\rho_\nu - h^{\sigma\nu}F_{\nu\sigma} K) \, .
	\end{equation}
	Let us look closely at the $F$ terms in the above relation: First, this term $h^{\sigma\nu}F_{\rho\sigma} K^\rho_\nu = F_{\rho\sigma} K^{\rho\sigma}$. Because $F_{\rho\sigma}$ is anti-symmetric and $K^{\rho\sigma}$ is symmetric, their full contraction is exactly zero. For the second term, we will have $h^{\sigma\nu}F_{\nu\sigma} K = F^\sigma_{\ \sigma} K = 0$ (the trace of an anti-symmetric tensor is zero).
	Since $F_{\mu\nu}$ terms completely vanish from the scalar curvature, thus, the final expression will have a very compact form:
	\begin{equation}\label{eq:ricci-electric}
		\check{R} = \tilde{R} + \alpha \left( K_{\mu\nu} K^{\mu\nu} - K^2 \right)
	\end{equation}
	
	$\tilde{R}$ in \eqref{eq:ricci-electric} is the scalar curvature when the geometry is completely torsion-free (which mathematically forced $K_{\mu\nu} = 0$, meaning rigid space). While, $\check{R}$ allows the space to expand/shear dynamically over time ($K_{\mu\nu} \neq 0$). By letting the connection possess exactly one piece of torsion (the extrinsic curvature term), and linking the spatial time-dilation to this expansion ($X \propto K$), the Ricci scalar undergoes a shift precisely equal to $K_{\mu\nu} K^{\mu\nu} - K^2$. If we set $\alpha = 1$ (or $\alpha = -1$ depending on the signature convention), this formula perfectly replicates the ``Hamiltonian constraint of General Relativity'' in the Carrollian limit.
	Here, the specific intrinsic torsion (linked to extrinsic curvature) perfectly replicates the kinetic momentum terms of the gravitational Hamiltonian constraint.
	
	Now, we use the connection $\check{\Gamma}$ and impose the flatness condition $\textbf{Riemann}(\check{\Gamma}) = 0 \implies \check{R} = 0$.
	Then, from \eqref{eq:ricci-electric}, we find the dynamical equivalence: 
	\begin{equation}
		\int  d^D x \, e \, \tilde{R} \quad \equiv - \quad \int   d^D x \, e \, \alpha \left( K_{\mu\nu}K^{\mu\nu} - K^2 \right) \, .
	\end{equation}
	This is an interesting equivalence. Usually, Magnetic Carrollian gravity (curved, rigid spacetime) and Electric Carrollian gravity (ECG) are considered two entirely different regimes or limits of General Relativity. 
	However, this equivalence establishes an interesting duality within the Carrollian regime. A universe characterized by a fully curved spacetime but a rigid, non-evolving spatial metric ($K_{\mu\nu}=0$, $\tilde{R} \neq 0$) is dynamically indistinguishable from a universe that is globally flat ($\check{R}=0$) but whose spatial geometry dynamically expands and shears over time ($K_{\mu\nu} \neq 0$). The intrinsic curvature of the Magnetic sector has been perfectly exchanged for the kinetic momentum of the Electric sector.

	\subsubsection{Equivalent Theory II: Spatial Torsion and Tensor Gauge Theories}\label{section:ETII-FractonGT}
	
	Next, to explore more sectors of equivalent theories, we return to the rigid case \((K_{\mu\nu}=0)\) but turn on the purely spatial contorsion \(A^\lambda_{\mu\nu}\), defining a teleparallel-like connection $\bar{\Gamma}^\lambda_{\mu\nu}$. To define this new connection, we must first understand what it physically and mathematically takes to ``turn off'' all torsion terms except the spatial contorsion $A^\lambda_{\mu\nu}$.
	
	Recall the full torsion tensor of the general connection \eqref{eq:gtorsion}. If we demand that the new torsion is purely spatial ($\bar{T}^\lambda_{\mu\nu} = 2A^\lambda_{[\mu\nu]}$), the rest of the terms must vanish:
	\begin{equation}
		-v^\lambda (F_{\mu\nu} + 2X_{[\mu\nu]}) - 2\tau_{[\nu} K^\lambda_{\mu]} = 0 \, .
	\end{equation}
	Because the $\tau$ and $v$ components are linearly independent, we must have: i) $K_{\mu\nu} = 0$: The extrinsic curvature must vanish. Keeping only the spatial torsion forces the Carrollian spacetime to be perfectly rigid (the spatial metric does not flow with time). ii) $F_{\mu\nu} + 2X_{[\mu\nu]} = 0$: The tensor $\Upsilon_{\mu\nu}$ can have the combination \eqref{def:Upsilon} and must be perfectly symmetric ($\Upsilon_{\mu\nu} = \Upsilon_{\nu\mu}$).
	
	Imposing these conditions on our general connection \eqref{eq:Gconnection} leaves us with:
	\begin{equation}
		\bar{\Gamma}^\lambda_{\mu\nu} = \gamma^\lambda_{\mu\nu} - v^\lambda \Upsilon_{\mu\nu} + A^\lambda_{\mu\nu} \, .
	\end{equation}
	Notice that the first two terms ($\gamma^\lambda_{\mu\nu} - v^\lambda \Upsilon_{\mu\nu}$) perfectly constitute the fully torsion-free connection $\tilde{\Gamma}^\lambda_{\mu\nu}$ that we derived earlier for constructing MCG. Thus, the new connection is just a pure tensor shift of the fully torsion-free connection by the spatial torsion tensor:
	\begin{equation}\label{eq:fractonconnection}
		\bar{\Gamma}^\lambda_{\mu\nu} = \tilde{\Gamma}^\lambda_{\mu\nu} + A^\lambda_{\mu\nu} \, ,
	\end{equation}
	where $A^\lambda_{\mu\nu}$ has a few properties. It is purely spatial ($\tau_\lambda A^\lambda_{\mu\nu} = A^\lambda_{\mu\nu}v^\nu = 0$) and its fully lowered version is antisymmetric in its outer indices: $A_{\sigma\mu\nu} = -A_{\nu\mu\sigma}$.
	
	\paragraph{Calculating the Curvatures}:
	Because $\tilde{\Gamma}$ is completely torsion-free, we can compute the Riemann tensor of $\bar{\Gamma}$ using the standard relation, as in Eq.~\eqref{eq:check-riemann1} for shifting an affine connection by a tensor $A^\lambda_{\mu\nu}$.
	Then, the Riemann tensor for this configuration will be
	\begin{equation}
		\bar{R}^\rho_{\ \sigma\mu\nu} = \tilde{R}^\rho_{\ \sigma\mu\nu} + \tilde{\nabla}_\mu A^\rho_{\nu\sigma} - \tilde{\nabla}_\nu A^\rho_{\mu\sigma} + A^\rho_{\mu\lambda} A^\lambda_{\nu\sigma} - A^\rho_{\nu\lambda} A^\lambda_{\mu\sigma} \, .
	\end{equation}
	Here, $\tilde{\nabla}$ is the covariant derivative with respect to the torsion-free connection $\tilde{\Gamma}$.
	We obtain the Ricci tensor by tracing the first and third indices ($\bar{R}_{\sigma\nu} = \bar{R}^\rho_{\ \sigma\rho\nu}$). 
	Let us define the spatial torsion trace vector as $A_\sigma \equiv A^\rho_{\rho\sigma}$. 
	Tracing the Riemann tensor yields:
	\begin{equation}
		\bar{R}_{\sigma\nu} = \tilde{R}_{\sigma\nu} + \tilde{\nabla}_\rho A^\rho_{\nu\sigma} - \tilde{\nabla}_\nu A_\sigma + A_\lambda A^\lambda_{\nu\sigma} - A^\rho_{\nu\lambda} A^\lambda_{\rho\sigma} \, .
	\end{equation}
	To compute the Ricci scalar, we trace the Ricci tensor with the spatial metric: $\bar{R} = h^{\sigma\nu} \bar{R}_{\sigma\nu}$. 
	Then, several non-trivial geometric cancellations occur, which are:
	\begin{enumerate}[label=(\roman*)]
		\item Because $A_{\alpha\beta\gamma} = -A_{\gamma\beta\alpha}$, tracing the outer indices with the symmetric spatial metric yields zero: $h^{\sigma\nu} A^\rho_{\nu\sigma} = h^{\sigma\nu} h^{\rho\alpha} A_{\alpha\nu\sigma} = 0$.
		\item We have terms like this $h^{\sigma\nu} \tilde{\nabla}_\rho A^\rho_{\nu\sigma} = \tilde{\nabla}_\rho(h^{\sigma\nu} A^\rho_{\nu\sigma}) - (\tilde{\nabla}_\rho h^{\sigma\nu})A^\rho_{\nu\sigma}$. According to the previous item, the first term is zero. After using \eqref{eq:covh}, the second term involves terms like $v^\sigma A^\rho_{\nu\sigma} = 0$, then, the entire term evaluates to exactly zero.
		\item Similarly, $h^{\sigma\nu} \tilde{\nabla}_\nu A_\sigma$ term simplifies to $\tilde{\nabla}_\nu A^\nu$ (the divergence of the torsion vector) because the derivative acting on the metric drops out when contracted with the spatial $A_\sigma$.
		\item We also have a quadratic cancellations in the terms like $h^{\sigma\nu} A_\lambda A^\lambda_{\nu\sigma}$ which vanishes because we have $h^{\sigma\nu}A^\lambda_{\nu\sigma} = 0$. 
	\end{enumerate}
	Then, what remains is very compact. Let spatial indices be raised/lowered freely with $h$, for example $A^\mu = h^{\mu\nu}A_\nu$. Then we obtain:
	\begin{equation}\label{eq:ricci-fractonic}
		\bar{R} = \tilde{R} - \tilde{\nabla}_\mu A^\mu - A_{\alpha\beta\gamma} A^{\gamma\alpha\beta} \, .
	\end{equation}
	Keeping the \textbf{spatial torsion} $A$ introduces a pure ``frame-dragging'' or twist distortion. It alters the Ricci scalar solely through the divergence of the torsion vector ($\tilde{\nabla}_\mu A^\mu$) and the spatial square of the torsion tensor ($A_{\alpha\beta\gamma} A^{\gamma\alpha\beta}$).
	
	If we use the spatial teleparallel connection $\bar{\Gamma}$ and impose the flatness condition $\textbf{Riemann}(\bar{\Gamma}) = 0 \implies \bar{R} = 0$. Then, using \eqref{eq:ricci-fractonic}, we obtain the following equivalence: 
	\begin{equation} \label{eq:equi-gaugefield}
		\int \tilde{R} \quad \equiv \quad \int A_{\alpha\beta\gamma}A^{\gamma\alpha\beta} \, .
	\end{equation}
	Under the integral sign, the boundary term \(\tilde{\nabla}_\mu A^\mu\) drops out, establishing an exact equivalence between the Magnetic curvature and spatial torsion. Equivalence \eqref{eq:equi-gaugefield} can be regarded as one of the Teleparallel Equivalences of Magnetic Carroll Gravity (TEMCG). We declare that spacetime is fundamentally flat. To explain gravitational attraction, we say that the flat spacetime is filled with a spatial torsion field $A_{\sigma\mu\nu}$. 
	
	It would be illuminating to study the geometric nature and degrees of freedom of spatial contorsion $A_{\sigma\mu\nu}$, which is a rank-3 purely spatial tensor and anti-symmetric in its outer indices ($A_{\sigma\mu\nu} = -A_{\nu\mu\sigma}$).
	Under the $SO(d)$ rotation group, this splits into three fundamental representations: i) \textbf{Vector trace} $V_\mu$: $V_\mu = h^{\sigma\nu} A_{\sigma\mu\nu}$. This is the trace of the torsion.
	ii) \textbf{Totally anti-symmetric axial sector} $H_{\sigma\mu\nu}$: $H_{\sigma\mu\nu} = A_{[\sigma\mu\nu]}$. In 3D space, this is dual to a pseudo-vector (like the magnetic field).
	iii) \textbf{Traceless tensor} $q_{\sigma\mu\nu}$: The remaining part, which satisfies $h^{\sigma\nu} q_{\sigma\mu\nu} = 0$ and $q_{[\sigma\mu\nu]} = 0$.
	
	We can rewrite the spatial Ricci scalar \eqref{eq:ricci-fractonic} by inserting the orthogonal decomposition $\mathbf{A} = \mathbf{V} \oplus \mathbf{H} \oplus \mathbf{q}$ . Because the irreducible representations are orthogonal, cross-terms vanish, leaving a sum of independent squared fields:
	\begin{equation}\label{eq:ASOd}
		\bar{R} = \tilde{R} - \tilde{\nabla}_\mu V^\mu + c_1 V_\mu V^\mu + c_2 H_{\alpha\beta\gamma} H^{\alpha\beta\gamma} + c_3 q_{\alpha\beta\gamma} q^{\alpha\beta\gamma} \; ,
	\end{equation}
	$c_1, c_2, c_3$ are specific dimensional constants. 
	
	Let's take a closer look at the equivalence theories emerging from relations \eqref{eq:ASOd} and \eqref{eq:equi-gaugefield} to understand each term more precisely:

	\textbf{1. Axial sector $\mathbf{H}^2$}: It resembles the Kalb-Ramond field from string theory \cite{Kalb:1974yc, Polchinski:1998rq}. The structure of Kalb-Ramond requires a 3-form field strength $\mathbf{H} = d\mathbf{B}$ generated by a 2-form potential $B_{\mu\nu}$, which has a strict gauge symmetry $\delta \mathbf{B} = d\mathbf{\Lambda} \implies \delta \mathbf{H} = 0$.
	Our axial torsion $H_{\mu\nu\rho}$ is built from the totally anti-symmetric spatial torsion $A_{[\mu\nu\rho]}$ with Lagrangian $+c_2 H_{\alpha\beta\gamma} H^{\alpha\beta\gamma}$, which exactly matches the Kalb-Ramond energy Hamiltonian.
	However, it lacks the required gauge symmetry. Unless we explicitly define the connection such that $A_{[\mu\nu\rho]} = \partial_{[\mu} B_{\nu\rho]}$, the geometric axial torsion is just an algebraic 3-form. It does not possess the $\delta \mathbf{B} = d\mathbf{\Lambda}$ gauge freedom. Like $V_\mu$, it acts as an auxiliary (non-propagating) field. On the other hand, in string theory, the Kalb-Ramond field is known to identically source the axial torsion of spacetime. So while the equivalence is not identical at the level of free gauge symmetries, they are intimately close manifestation; the axial torsion is the geometric manifestation of the string fluid.
	
	\textbf{2. Fractonic sector}: In scalar-charge Fracton theory, M. Pretko defines the magnetic field as \cite{Pretko:2017fbf, Pretko:2016lgv}:
	\begin{equation}
		B^i_{\ j} = \epsilon^{ikl} \partial_k A_{lj} \, ,
	\end{equation}
	with fracton Lagrangian $\mathcal{L}_{fract} \propto B^j_{\ i}B^i_{\ j}$ \cite{Bertolini:2022ijb}. If we multiply both sides by the fully covariant Levi-Civita tensor $\epsilon_{imn}$ (contracting on the index $i$), we obtain:
	\begin{equation}
		\epsilon_{imn} B^i_{\ j} = \epsilon_{imn} \epsilon^{ikl} \partial_k A_{lj} \, .
	\end{equation}
	Then, using the fundamental Levi-Civita identity for contracting one index, the result will be:
	\begin{equation}\label{def:3fractonM}
		\mathcal{B}_{mnj} := \epsilon_{imn} B^i_{\ j} = \partial_m A_{nj} - \partial_n A_{mj} \, .
	\end{equation}
	Now we can see that $\mathcal{B}_{mnj}$ has the same symmetric structure as our tensor field $A_{\sigma\mu\nu}$, but to establish a strict algebraic equivalence with the Fracton magnetic field \eqref{def:3fractonM}, we must restrict the geometric degrees of freedom of the spatial torsion. In 3 spatial dimensions, the general torsion $A_{\sigma\mu\nu}$ possesses 9 degrees of freedom. However, Pretko's rank-2 magnetic matrix $B^i_{\ j}$, generated from a symmetric tensor potential, is traceless, possessing only 8 degrees of freedom. These 8 degrees of freedom can be reconstructed by merging the vector trace ($V_\mu$, 3 d.o.f.) and the traceless tensor ($q_{\sigma\mu\nu}$, 5 d.o.f.). Mathematically, this is achieved by defining a restricted connection where the completely anti-symmetric axial torsion vanishes ($A_{[\sigma\mu\nu]} = 0$). Physically, the combined tensor $\boldsymbol{\mathcal{A}} = \mathbf{V} \oplus \,\mathbf{q}$ can encapsulate the complete spectrum of standard magnetic sector of the scalar-charge Fracton gauge theory. It should be emphasized that the Fracton field $\mathcal{B}$ and the torsion field $\mathcal{A}$ share the same kinetic term and algebraic structure. However, $\mathcal{A}$ still lacks the gauge symmetry required for a complete realization of the scalar‑charge Fracton theory.
	
	\subsubsection{Equivalent Theory III: Temporal Torsion and Dilatonic Scaling}\label{section:ETII-DilatonicS}
	
	To explore another sector of equivalent theories, in the following we will isolate the torsion terms originating from the clock form $F_{\mu\nu}$ and the tensor $X_{\mu\nu}$ while eliminating all others. For this purpose, we must once again impose strict geometric constraints. We assume a torsion-free spatial sector \((A=K=0)\), but allow the time-dilation tensor to contain an anti-symmetric piece. 
	
	Recall the most general torsion tensor \eqref{eq:gtorsion},
	to keep only the $X$ and $F$ terms, we must set $A^\lambda_{\mu\nu} = 0$ (zero spatial torsion) and $K_{\mu\nu} = 0$ (zero extrinsic curvature), which means the Carrollian spacetime must be rigid.
	Then, the surviving torsion is purely along the time direction:
	\begin{equation}\label{eq:Sconnection}
		\mathring{T}^\lambda_{\mu\nu} = -v^\lambda (F_{\mu\nu} + X_{\mu\nu} - X_{\nu\mu}) \, .
	\end{equation}
	Let us define the \textit{temporal torsion} as $S_{\mu\nu} \equiv \frac{1}{2}(F_{\mu\nu} + X_{\mu\nu} - X_{\nu\mu})$. 
	Notice that $S_{\mu\nu}$ is exactly the anti-symmetric part of the tensor $\Upsilon_{\mu\nu}$ that we defined earlier in \eqref{def:Upsilon} 
	\begin{equation}\label{eq:Smunu}
		S_{\mu\nu} = \Upsilon_{[\mu\nu]} \, .
	\end{equation}
	Then, the new connection \eqref{eq:Sconnection}, after imposing $A=0$ and $K=0$ on our general connection, will be:
	\begin{equation} \label{eq:delatonconnection}
		\mathring{\Gamma}^\lambda_{\mu\nu} = \gamma^\lambda_{\mu\nu} - v^\lambda \Upsilon_{\mu\nu} \, .
	\end{equation}
	We can split $\Upsilon_{\mu\nu}$ into its symmetric part $\tilde{\Upsilon}_{\mu\nu} = \Upsilon_{(\mu\nu)}$ and its anti-symmetric part $S_{\mu\nu} = \Upsilon_{[\mu\nu]}$. 
	Recall that the fully torsion-free connection $\tilde{\Gamma}$ was defined exactly using the symmetric part of  $\Upsilon_{\mu\nu}$. Therefore, our new connection is simply the torsion-free connection \eqref{eq:magnetic-connection} shifted by the temporal torsion:
	\begin{equation} \label{eq:delatonconnection2}
		\mathring{\Gamma}^\lambda_{\mu\nu} = \tilde{\Gamma}^\lambda_{\mu\nu} - v^\lambda S_{\mu\nu} \, .
	\end{equation}
	The physical meaning of the new connection $\mathring{\Gamma}$ is that it preserves metric compatibility but allows for ``temporal torsion''. If we parallel-transport a vector around a closed loop, the spatial lengths and angles are preserved, but the time-coordinate of the vector experiences an anholonomic shift dictated by $S_{\mu\nu}$.
	
	\paragraph{Calculating the Curvatures} We now calculate the curvature using the shift tensor $C^\lambda_{\mu\nu} = -v^\lambda S_{\mu\nu}$. Because $v^\lambda$ is covariantly constant under the reference connection ($\tilde{\nabla} v = 0$), the derivatives are very clean.
	Let us define the ``\textit{Carrollian torsion vector}'' as
	\begin{equation}\label{eq:smu}
		s_\mu \equiv v^\lambda S_{\lambda\mu} \, ,
	\end{equation}
	since $S_{\mu\nu}$ is anti-symmetric, we have $s_\mu v^\mu = 0$, meaning $s_\mu$ is purely spatial.
	
	Using the standard formula \eqref{eq:check-riemann1} for our new connection \eqref{eq:delatonconnection2}, we obtain $ \mathring{R} = \tilde{R} + \tilde{\nabla} C - \tilde{\nabla} C + C \times C - C \times C$ for the Riemann tensor. The quadratic term in Riemann tensor is:
	\begin{equation}
		C^\rho_{\mu\lambda} C^\lambda_{\nu\sigma}  = -v^\rho s_\mu S_{\nu\sigma} \, .
	\end{equation}
	Subtracting the $\mu \leftrightarrow \nu$ term gives the non-linear contribution. Then, the Riemann tensor evaluates to:
	\begin{equation}\label{eq:riemann-smu}
		\mathring{R}^\rho_{\ \sigma\mu\nu} = \tilde{R}^\rho_{\ \sigma\mu\nu} - v^\rho \big( \tilde{\nabla}_\mu S_{\nu\sigma} - \tilde{\nabla}_\nu S_{\mu\sigma} \big) + v^\rho \big( s_\mu S_{\nu\sigma} - s_\nu S_{\mu\sigma} \big) \, .
	\end{equation}
	We trace over $\rho$ and $\mu$ ($\mathring{R}_{\sigma\nu} = \mathring{R}^\rho_{\ \sigma\rho\nu}$) to build the Ricci tensor as usual.
	Thus, the first shift term becomes $-v^\rho \tilde{\nabla}_\rho S_{\nu\sigma} + v^\rho \tilde{\nabla}_\nu S_{\rho\sigma}$. Because $\tilde{\nabla} v = 0$, the first part is exactly the Lie derivative along the time flow: $-\mathcal{L}_v S_{\nu\sigma}$. The second part can be written as $\tilde{\nabla}_\nu (v^\rho S_{\rho\sigma}) = \tilde{\nabla}_\nu s_\sigma$. The third term in \eqref{eq:riemann-smu} dies, since $s_\mu$ is spatial ($v^\mu s_\mu = 0$). 
	Putting all these together, we obtain clean relation:
	\begin{equation}
		\mathring{R}_{\sigma\nu} = \tilde{R}_{\sigma\nu} - \mathcal{L}_v S_{\nu\sigma} + \tilde{\nabla}_\nu s_\sigma + s_\nu s_\sigma \, .
	\end{equation}
	
	Then, we trace the Ricci tensor with the spatial metric: $\mathring{R} = h^{\sigma\nu} \mathring{R}_{\sigma\nu}$.
	Since $h^{\sigma\nu}$ is symmetric and $S_{\nu\sigma}$ is purely anti-symmetric, their contraction vanishes completely: $h^{\sigma\nu} \mathcal{L}_v S_{\nu\sigma} = 0$. The time-derivative term drops out entirely.
	If we define $s^\mu = h^{\mu\nu}s_\nu$, $h^{\sigma\nu} \tilde{\nabla}_\nu s_\sigma$ term simply becomes the spatial divergence $\tilde{\nabla}_\mu s^\mu$, while the final term is just the spatial norm squared of $s_\mu$.
	Then, the Ricci scalar takes an exceptionally compact form:
	
	\begin{equation}\label{eq:ricci-smu}
		\mathring{R} = \tilde{R} + \tilde{\nabla}_\mu s^\mu + s_\mu s^\mu \, .
	\end{equation}
	
	In order to find the equivalent theory, we use the temporal torsion connection $\mathring{\Gamma}$, as defined in \eqref{eq:Sconnection}, and impose the flatness condition $\textbf{Riemann}(\mathring{\Gamma}) = 0 \implies \mathring{R} = 0$.
	We, thus, find the following equivalence:
	\begin{equation}\label{eq:ssquared}
		\int \tilde{R} \quad \equiv \quad \int -s_\mu s^\mu \, .
	\end{equation}
	We again restate that in this section, instead of spatial defects like $A^\lambda_{\mu\nu}$, we allow the ``clock form'' $\tau_\mu$ to possess a spatial and temporal gradient. 
	
	In differential geometry, combination \eqref{eq:ricci-smu} is the famous signature of \textit{Vector Torsion} (also known as the trace of torsion). In Einstein-Cartan gravity, if we take the full torsion tensor $T^\lambda_{\mu\nu}$ and trace over it, we get a vector: $T_\mu \equiv T^\lambda_{\ \lambda\mu}$. 
	When we compute the Ricci scalar of a connection with vector torsion, it shifts by exactly $\nabla_\mu T^\mu + T_\mu T^\mu$. Physically, it acts as a geometric vector potential governing the local synchronization of the Carrollian clock. Furthermore, if the geometry is restricted such that this vector field is irrotational ($\partial_{[\mu} s_{\nu]} = 0$), it can be integrated into a scalar potential ($s_\mu = \partial_\mu \phi$). In this irrotational limit, the term $\tilde{\nabla}_\mu s^\mu + s_\mu s^\mu$ identically reproduces the transformation of the Ricci scalar under a conformal Weyl rescaling, rendering the theory equivalent to a Dilaton or a Newtonian scalar gravity model. In the following, we will show how this can be done. 

	Let us first evaluate \(s_\mu\) explicitly from its definition, in order to clarify its independent degrees of freedom. 
	Based on the definitions \eqref{def:Upsilon}, \eqref{eq:Smunu} and \eqref{eq:smu}, we obtain $s_\mu$ as follows:
	\begin{equation}\label{eq:smu2}
		s_\mu = \frac{1}{2} v^\lambda \left( \partial_\lambda \tau_\mu - \partial_\mu \tau_\lambda + X_{\lambda\mu} - X_{\mu\lambda} \right) \, .
	\end{equation}
	For evaluating the derivative of the clock form $\tau$, we start from the definition of the Lie derivative
	\begin{equation}\label{eq:partialtau1}
		v^\lambda \partial_\lambda \tau_\mu = \mathcal{L}_v \tau_\mu - \tau_\lambda \partial_\mu v^\lambda \, ,
	\end{equation}
	and using $v^\lambda \partial_\mu \tau_\lambda = - \tau_\lambda \partial_\mu v^\lambda$, which was obtained from $\tau_\lambda v^\lambda = -1$. Then, equation \eqref{eq:partialtau1} can be written as:
	\begin{equation}
		v^\lambda (\partial_\lambda \tau_\mu - \partial_\mu \tau_\lambda) = \mathcal{L}_v \tau_\mu \, .
	\end{equation} 
	Recall that $X$ represents the spatial gradient of the clock form ($\nabla_\mu \tau_\nu = -X_{\mu\nu}$), and its only algebraic restriction is $X_{\mu\lambda}v^\lambda = 0$, which means the other contraction $v^\lambda X_{\lambda\mu}$ is completely unconstrained. It can be any arbitrary spatial vector.
	Putting it all together, we find:
	\begin{equation}\label{eq:smu-expression}
		s_\mu = \frac{1}{2} \Big( \mathcal{L}_v \tau_\mu + v^\lambda X_{\lambda\mu} \Big)\, .
	\end{equation}
    
	The shift term \(\tilde{\nabla}_\mu s^\mu + s_\mu s^\mu\) appearing in \eqref{eq:ricci-smu} is mathematically equivalent to the transformation of the Ricci scalar under a conformal Weyl rescaling \(\tilde{g}_{\mu\nu} = e^{2\phi}g_{\mu\nu}\), where \(s_\mu \sim \partial_\mu \phi\). Its resemblance to a Weyl gauge field is not accidental; rather, it reflects the anisotropic scaling symmetries intrinsic to Carrollian geometry. In fact, under a local Carrollian scaling transformation, the clock form and the time‑flow vector transform with opposite dynamical exponents: $\tau_\mu \to e^{z\phi}\tau_\mu$ and $v^\mu \to e^{-z\phi}v^\mu$. 
	The temporal torsion $s_\mu$ is defined as \eqref{eq:smu-expression}.
	Explicitly evaluating the Lie and covariant derivative of the scaled clock form reveals that the exponential factors cancel and leave behind a gauge shift: $\mathcal{L}_{\hat{v}} \hat{\tau}_\mu = \mathcal{L}_v \tau_\mu + z \bar{\partial}_\mu \phi$ and $ \hat{v}^\lambda \hat{X}_{\lambda\mu} = {v}^\lambda {X}_{\lambda\mu} + z \bar{\partial}_\mu \phi$, where $\bar{\partial}_\mu$ is the spatial gradient: $\bar{\partial}_\mu \phi = \partial_\mu \phi + (v^\alpha \partial_\alpha \phi) \tau_\mu = \partial_\mu \phi + (\mathcal{L}_v \phi)\tau_\mu$.
	Combining these results, we can see that the temporal torsion $\hat{s}_\mu$ transforms as a Weyl gauge connection:
	\begin{equation}\label{eq:weylgauge}
	\hat{s}_\mu = s_\mu + {z} \, \bar{\partial}_\mu \phi \, .
	\end{equation}
	This confirms that the scaling freedom of the clock form natively generates the gauge transformation of $s_\mu$. It means, if we restrict the geometry to a pure conformal gauge where the background torsion vanishes, $s_\mu$ becomes exactly proportional to the spatial gradient of the scaling field $\bar{\partial}_\mu \phi$, validating the mapping of the equivalent action $\int -s^2$ to the kinetic Lagrangian of the Carrollian Dilaton.
    
	Because $X_{\lambda\mu}$ is arbitrary in its first index, $s_\mu$ is an arbitrary spatial vector field. In the following we will show its potential connection to \textit{Newtonian-like} potential theory, once an appropriate gauge fixing of the clock form is imposed.
	By Poincaré’s Lemma, an arbitrary vector field can only be written as the gradient of a scalar ($s_\mu = \partial_\mu \phi$) if and only if its curl vanishes ($\partial_{[\nu} s_{\mu]} = 0$). Since $X$ is unrestricted, $s_\mu$ will generally have a non-zero curl. Therefore, it is generally a vector field, not a scalar field.
	
	To find the exact condition under which $s_\mu$ becomes a pure scalar gradient, we must calculate its curl, $\partial_{[\mu} s_{\nu]}$, and set it to zero. 
	In the following, we will show that it leads us to one of the interesting physical regimes in Carrollian gravity, the stationary Newtonian-like limit.
	Recalling our formula \eqref{eq:smu-expression} for the temporal vector torsion, we want to calculate the anti-symmetric derivative (the curl): $\partial_{[\mu} s_{\nu]} = \frac{1}{2}(\partial_\mu s_\nu - \partial_\nu s_\mu)$. 
	In the language of differential forms, we are taking the exterior derivative $ds$.
	Using the properties of the exterior derivative ($d$) and Cartan's magic formula for Lie derivatives ($\mathcal{L}_v \tau = \iota_v d\tau + d(\iota_v \tau)$), we can evaluate the required curl of $s_\mu$. 
	For the first term in \eqref{eq:smu-expression} we obtain:
	\begin{equation}
		d(\mathcal{L}_v \tau) = \mathcal{L}_v (d\tau) = \mathcal{L}_v F \, .
	\end{equation}
	To calculate the second term of \eqref{eq:smu-expression}, we introduce the vector $D_\mu \equiv v^\lambda X_{\lambda\mu}$ (which represents the covariant drift of the clock form along the time flow, $-\nabla_v \tau_\mu$). Its curl is simply $\partial_{[\mu} D_{\nu]}$.
	Putting them together, the curl of $s_\mu$ is:
	\begin{equation}
		\partial_{[\mu} s_{\nu]} = \frac{1}{2} (\mathcal{L}_v F)_{\mu\nu} + \frac{1}{2} \partial_{[\mu} D_{\nu]} \, .
	\end{equation}
	For $s_\mu$ to be written as the gradient of a scalar ($\partial_\mu \phi$), its curl must vanish:
	\begin{equation}\label{eq:scalarcondition}
		(\mathcal{L}_v F)_{\mu\nu} + \partial_{[\mu} D_{\nu]} = 0 \, .
	\end{equation}
	How do we satisfy this condition in a physically meaningful way? One perfect configuration is a stationary Carrollian spacetime. To answer this question, let us write the condition \eqref{eq:scalarcondition} in the language of differential forms. The anti-symmetric derivative is the exterior derivative, so we have $dD = -\mathcal{L}_v F$. We know that the 2-form $F$ is the exterior derivative of the clock form: $F = d\tau$. Because the Lie derivative commutes with the exterior derivative ($\mathcal{L}_v d = d \mathcal{L}_v$), we can rewrite the right-hand side as:
    \begin{equation}
        \mathcal{L}_v F = d(\mathcal{L}_v \tau) \, .
    \end{equation}
    Using the Carrollian \textit{gravitoelectric} field $E_\mu$ \eqref{eq:Emu}, the condition \eqref{eq:scalarcondition} can be written as:
    \begin{equation}
        dD = -dE \implies d(D + E) = 0 
    \end{equation}
    By Poincaré’s Lemma, if the exterior derivative of a 1-form is zero, that 1-form must (locally) be the exact gradient of a scalar function. Let’s call this arbitrary scalar function $2\phi$ (the factor of 2 is just for convenience): 
    \begin{equation} \label{eq:dilatonsolution}
        D_\mu + E_\mu = 2\partial_\mu \phi \, ,
    \end{equation}
    which can be solved for the drift vector $D_\mu$. If we plug solution \eqref{eq:dilatonsolution} back into the definition \eqref{eq:smu-expression} of the temporal torsion $s_\mu$, we find:
    \begin{equation} \label{eq:phi}
        s_\mu = \partial_\mu \phi \, .
    \end{equation}
    Since we defined $D_\mu \equiv v^\lambda X_{\lambda\mu}$, we must check if solution \eqref{eq:dilatonsolution} violates any rules for $X_{\mu\nu}$.
    The only strict constraint on $X_{\mu\nu}$ for metric-compatibility is $X_{\mu\nu} v^\nu = 0$. 
    This implies that $D_\mu$ must be a purely spatial vector:
    \begin{equation}
        D_\mu v^\mu = (v^\lambda X_{\lambda\mu}) v^\mu = 0 \, .
    \end{equation}
    So we must enforce $v^\mu D_\mu = 0$ on our solution \eqref{eq:dilatonsolution}. Since $E_\mu$ is spatial $v^\mu E_\mu = v^\lambda \partial_\lambda (v^\mu \tau_\mu) = 0$, the constraint \eqref{eq:scalarcondition} reduces to the stationary condition on the scalar field $\phi$:
    \begin{equation}
        2 v^\mu \partial_\mu \phi = 0 \implies \mathbf{\mathcal{L}_v \phi = 0} \, .
    \end{equation}
    This means that while the spacetime geometry can be evolving in time, the equivalent scalar field $\phi(x)$ must be constant along the time fibers. This result is closely related to \eqref{eq:weylgauge}, which arises from a different choice of geometric ambiguity and gauge fixing within Carrollian geometry. This is analogous to the Newtonian potential in Carrollian geometry. 
    We do not need to restrict the Carrollian geometry to a stationary spacetime to recover a scalar equivalent theory. By exploiting the internal gauge freedom of the connection $X_{\mu\nu}$, the geometric drift can dynamically absorb the non-stationary evolution of the clock form. Provided the emergent Dilaton field $\phi$ is constant along the time fibers, any general Magnetic Carrollian Gravity can be mapped equivalently to the Dilatonic kinetic action $S_{\rm Newton~P.} = \int - \partial_\mu \phi \partial^\mu \phi$.
	In other words, under the given conditions, \textit{temporal torsion} is dynamically equivalent to embedding a Newtonian gravitational potential into the Carrollian geometry.
	Physically, meaning that a curved rigid spacetime is indistinguishable from a flat spacetime where the local \textit{rate of time} varies from point to point, generating a gradient field that mimics gravity.
	Because \(s_\mu\) fundamentally measures the spatial and temporal gradient of the local clock synchronization, it can act as a scalar potential (or a Dilaton). 
	
	\textbf{D.o.f decomposition of $S_{\mu\nu}$}:
	Since anti-symmetric tensor $S_{\mu\nu}$ is not a purely spatial tensor, in the Carrollian framework, we decompose this tensor using the time-flow vector $v^\nu$ and the spatial metric $h^{\mu\nu}$: i) Temporal vector ($s_\mu$): Its contraction with time, $s_\mu = S_{\mu\nu}v^\nu$ can contain the scalar field \eqref{eq:phi}, ii) Spatial tensor ($\mathcal{S}_{\mu\nu}$): The purely spatial projection, $\mathcal{S}_{\mu\nu} = h^\alpha_\mu h^\beta_\nu S_{\alpha\beta}$, which acts as a spatial tensor field.
	
	When we computed the temporal shift in Curvature \eqref{eq:ricci-smu}, the shift tensor $C^\lambda_{\mu\nu} \propto -v^\lambda S_{\mu\nu}$ forced all spatial components to drop out during the metric contraction, which means the Ricci scalar evaluates to:
	\begin{equation}
		\mathring{R} = \tilde{R} + \tilde{\nabla}_\mu s^\mu + s_\mu s^\mu + 0 \cdot \mathcal{S}_{\mu\nu} \mathcal{S}^{\mu\nu} \, ,
	\end{equation}
	with no contribution from the spatial sector $\mathcal{S}_{\mu\nu}$ in the equivalent theory.

	\subsubsection{Equivalent Theory IV: Carrollian Clock Torsion and Maxwell Electric Theory}\label{section:ETIV-MaxwellElectricT}
	Another sector of this Teleparallel equivalence emerges by manipulating the different contributions of the torsion tensor. By setting all torsion terms proportional to \( A^\lambda_{\mu\nu} \), \( X_{\mu\nu} \), and \( K_{\mu\nu} \) to zero, we isolate the torsion that is intrinsically generated by the Carrollian clock form \( \tau_\mu \). This bridges gravity directly to Maxwell Electromagnetism (specifically, the Carrollian gravitoelectric limit).
	
	We begin with the most general Carrollian connection \eqref{eq:Gconnection} and impose the following three constraints: 1)  $A^\lambda_{\mu\nu} = 0$, zero spatial torsion, 2)  $K_{\mu\nu} = 0$, rigid space; no extrinsic curvature, 3)  $X_{[\mu\nu]} = 0$, no arbitrary anti-symmetric spatial gradient of the clock form.
	Substituting these into the general connection \eqref{eq:Gconnection} leaves us with a streamlined form:
	\begin{equation}
		\breve{\Gamma}^\lambda_{\mu\nu} = \gamma^\lambda_{\mu\nu} - v^\lambda \big[ \partial_{(\mu} \tau_{\nu)} + X_{(\mu\nu)} \big] - \frac{1}{2}v^\lambda F_{\mu\nu} \, .
	\end{equation}
	We want to compare $\breve{\Gamma}^\lambda_{\mu\nu}$ with the fully torsion-free case $\tilde{\Gamma}$ adapted for MCG. We recall that $\tilde{\Gamma}$ required the symmetric part of $\partial_\mu \tau_\nu + X_{\mu\nu}$, thus, we can freely set $X_{[\mu\nu]} = 0$. 
	Therefore, our new connection is just the torsion-free connection shifted by the anti-symmetric part of $\partial_\mu \tau_\nu$:
	\begin{equation}\label{def:connection-maxwell}
		\breve{\Gamma}^\lambda_{\mu\nu} = \tilde{\Gamma}^\lambda_{\mu\nu} - \frac{1}{2} v^\lambda F_{\mu\nu}\,,
	\end{equation}
	where $F_{\mu\nu}$ is the exterior derivative of the clock form \eqref{eq:Fmunu}.
	The torsion of new connection \eqref{def:connection-maxwell} is purely $\breve{T}^\lambda_{\mu\nu} = -v^\lambda F_{\mu\nu}$.
	
	It is useful to examine the shift tensor \(C^\lambda_{\mu\nu} = -\tfrac{1}{2} v^\lambda F_{\mu\nu}\), and study its properties. In our analysis, a physically relevant vector naturally emerges from contracting \(F_{\mu\nu} = \partial_\mu \tau_\nu - \partial_\nu \tau_\mu\) with the time‑flow vector \(v\). In analogy with electromagnetism and gravitoelectromagnetism \cite{Costa:2012cw, Lynden-Bell:1996dpw}, we define this new vector as:
	\begin{equation}
		E_\mu \equiv v^\lambda F_{\lambda\mu}\,.
	\end{equation}
	which is the Carrollian \textit{gravitoelectric} field \(E_\mu\) introduced earlier in \eqref{eq:Emu}.
	If we treat $\tau_\mu$ as a $U(1)$ gauge potential like in electromagnetism, $v^\lambda F_{\lambda\mu}$ is the definition of an Electric Field vector.
	
	Because $\breve{\Gamma}$ has the exact same mathematical structure as the temporal torsion case $\mathring{\Gamma}$ \eqref{eq:delatonconnection} (when substituting $S_{\mu\nu} \to \frac{1}{2}F_{\mu\nu}$ and $s_\mu \to \frac{1}{2}E_\mu$), we can perfectly map the curvature equations calculated previously for this new connection \eqref{def:connection-maxwell}.
	Substituting $S_{\mu\nu} = \frac{1}{2}F_{\mu\nu}$ into our previously computed Riemann formula (for example \eqref{eq:riemann-smu}), we get:
	\begin{equation}
		\breve{R}^\rho_{\ \sigma\mu\nu} = \tilde{R}^\rho_{\ \sigma\mu\nu} - \frac{1}{2} v^\rho \big( \tilde{\nabla}_\mu F_{\nu\sigma} - \tilde{\nabla}_\nu F_{\mu\sigma} \big) + \frac{1}{4} v^\rho \big( E_\mu F_{\nu\sigma} - E_\nu F_{\mu\sigma} \big) \, ,
	\end{equation}
	and tracing the Riemann tensor yields:
	\begin{equation}
		\breve{R}_{\sigma\nu} = \tilde{R}_{\sigma\nu} - \frac{1}{2} \mathcal{L}_v F_{\nu\sigma} + \frac{1}{2} \tilde{\nabla}_\nu E_\sigma + \frac{1}{4} E_\nu E_\sigma \, .
	\end{equation}
	Tracing again with the spatial metric $h^{\sigma\nu}$, the $\mathcal{L}_v F_{\nu\sigma}$ term vanishes completely because $F$ is anti-symmetric and $h$ is symmetric. Thus, we are left with a simple and compact scalar:
	\begin{equation}
		\breve{R} = \tilde{R} + \frac{1}{2} \tilde{\nabla}_\mu E^\mu + \frac{1}{4} E_\mu E^\mu \, ,
	\end{equation}
	where we defined $E^\mu = h^{\mu\nu} E_\nu$ and $E^2 = E_\mu E^\mu$ is the spatial norm squared of the Carrollian gravitoelectric field. 
	By integrating the Ricci scalar to form an action (and dropping the total boundary divergence $\frac{1}{2}\tilde{\nabla}_\mu E^\mu$), we establish the following algebraic equivalence:
	\begin{equation}
		\int d^D x \, e \, \tilde{R} \quad \equiv \quad \int d^D x \, e \, \Big[ \breve{R} - \frac{1}{4} E_\mu E^\mu \Big] \, .
	\end{equation}
	
	Following the Teleparallel logic, if we assume the universe is globally flat, under this connection ($\textbf{Riemann}(\breve{\Gamma}) = 0 \implies \breve{R} = 0$), the action for Carroll Magnetic Gravity becomes equivalent to:
	\begin{equation}\label{eq:action-gravitoelectric}
		S_{\text{Equivalent}} = \int d^D x \, e \, \left( - \frac{1}{4} E_\mu E^\mu \right) \, .
	\end{equation}
	
	If $\tau_\mu$ is a gauge potential, then $F_{\mu\nu}$ is the electromagnetic field strength tensor. In standard Lorentzian spacetime, the Maxwell action is:
	\begin{equation}
		\mathcal{L}_{\text{Maxwell}} = -\frac{1}{4} F_{\mu\nu} F^{\mu\nu} \, ,
	\end{equation}
	where it can then be decomposed into electric and magnetic parts: $\frac{1}{2}(E^2 - B^2)$.
	
	Let's take a closer look at our equivalent action \eqref{eq:action-gravitoelectric}.
	Where did the spatial magnetic field ($B^2$, where $B_{\mu \nu} = h^\alpha_\mu h^\beta_\nu F_{\alpha\beta}$ \footnote{Similarly to the decomposition of \(S_{\alpha \beta }\), the degrees of freedom of \(F_{\alpha \beta }\) can be decomposed by projecting it along the time-flow vector \(v\) and its spatial components, which we call Magnetic part.}) go? 
	Because the Carrollian spatial metric $h^{\mu\nu}$ is degenerate along the time direction, the connection shift vector was aligned with $v^\lambda$. When the shift tensor is squared ($C \times C$), $v^\lambda$ vectors will contract with $F_{\mu\nu}$ and project out only the Electric sector, rendering the spatial magnetic part completely invisible to the Ricci scalar.
	
	On the nature of the $U(1)$ Equivalence:
	The identification of the clock form $\tau_\mu$ as a $U(1)$ gauge potential is not a mathematical artifact, but a reflection of the underlying diffeomorphism algebra of the Carrollian manifold. 
	In a Carrollian universe, the speed of light is zero, meaning light cones have collapsed. Because light cannot travel between spatial points to synchronize clocks, every point in space has its own completely independent clock.
	If we decide to redefine the zero-point of our clock at every position in space by an arbitrary function $\alpha(x)$:
	\begin{equation}
		t' = t + \alpha(x) \, ,
	\end{equation}
	the differential of time transforms as $dt' = dt + \partial_\mu \alpha dx^\mu$. 
	Since $\tau_\mu$ is the 1-form that measures the passage of local time (locally $\tau \sim dt$), this coordinate transformation forces the clock form to transform exactly as:
	\begin{equation}
		\tau_\mu \longrightarrow \tau_\mu + \partial_\mu \alpha \, .
	\end{equation}
	This is the mathematical definition of a $U(1)$ gauge symmetry. 
	Therefore, $\tau_\mu$ is the gauge field that ensures physics remains invariant under our arbitrary choice of local clock synchronization.
	Because $\tau_\mu$ is the gauge field of time translations, the electric charge of the Carrollian universe is energy (or mass). 
	When we write the Lagrangian $-\frac{1}{4} E_\mu E^\mu$, we are writing the gravitational kinetic energy of the time-synchronization field. This is why it looks like electromagnetism, but acts as gravity.

	There is one mathematical restriction we must be aware of. 
	A true electromagnetic field $A_\mu$ has no algebraic constraints. However, the clock form must satisfy the normalization condition: $ \tau_\mu v^\mu = -1 $.
	Because of this constraint, we cannot shift $\tau_\mu$ by any arbitrary $\partial_\mu \alpha$. It can only be shifted by functions where $v^\mu \partial_\mu \alpha = 0$ (meaning the clock shift $\alpha$ can vary from place to place, but cannot evolve in time). 
	Therefore, $\tau_\mu$ is a \textbf{constrained} $U(1)$ gauge field.

	\section{Symmetric Teleparallel Equivalent of Magnetic Carrollian Gravity}\label{sec:STEMCG}
	In this section, we construct a minimally non‑metric, torsion‑free Carroll connection, providing the foundation for analyzing departures from strict metric compatibility. In the subsequent part, we introduce a dynamically equivalent formulation of Magnetic Carroll gravity, following the same methodological pattern established in the preceding sections for metric connections in the teleparallel Carroll gravity framework, Sections (\ref{section:ETI-ECG}-\ref{section:ETIV-MaxwellElectricT}). This analysis falls within the category of the symmetric teleparallel equivalent of Magnetic Carroll gravity, not in its most general form, but specifically under the assumption of minimal non‑metricity motivated by physical considerations. These motivations will be elaborated upon in the following subsections.

	\subsection{Non-Metric Carroll Connection}
	Here, we consider Carrollian geometry with structures $(h_{\mu\nu}, v^\mu, \tau_\mu, h^{\mu\nu})$ and the orthonormality conditions \eqref{eq:carroll-conditions}.
	We want to find the general torsion-free Carrollian connection with the following specified non-metricity,
	\begin{align}\label{eq:minimalNM}
		\mathring{\bar{\nabla}}_{\rho}h_{\mu\nu} &= Q_{\rho\mu\nu}, \\
		v^{\rho}Q_{\rho\mu\nu} &= 0, \qquad h^{\mu}_{\alpha} h^{\nu}_{\beta} Q_{\rho\mu\nu} = 0, \qquad v^{\nu}Q_{[\rho\mu]\nu} =0 \, .
	\end{align}
	This minimal non‑metricity is motivated by scenarios in which Carrollian geometry is embedded within a higher‑dimensional spacetime equipped with a Levi‑Civita connection \cite{Ciambelli:2025unn}. As shown by Ciambelli et al., the conditions \eqref{eq:minimalNM} single out a Carrollian connection that is fixed by requiring it to coincide with the projection of the bulk Levi‑Civita connection.
	$[\rho\mu]$ means anti-symmetrization over indices $\rho$ and $\mu$.
	
	By definition, the spatial metric is symmetric, so the non-metricity tensor is symmetric in its last two indices: $Q_{\rho\mu\nu} = Q_{\rho\nu\mu}$. 
	We are given the constraint $h^\alpha_\mu h^\beta_\nu Q_{\rho\alpha\beta} = 0$. This implies that $Q_{\rho\mu\nu}$ has no purely spatial components. Therefore, every term in $Q_{\rho\mu\nu}$ must contain at least one clock form $\tau$. We can parameterize the non-metricity as:
	\begin{equation}
		Q_{\rho\mu\nu} = \tau_\mu W_{\rho\nu} + \tau_\nu W_{\rho\mu} + \tau_\mu \tau_\nu Z_\rho \, .
	\end{equation}
	To preserve the condition $v^\mu h_{\mu\nu} = 0$, its covariant derivative must vanish:
	\begin{equation}
		\mathring{\bar{\nabla}}_\rho(v^\mu h_{\mu\nu}) = 0 \implies v^\mu Q_{\rho\mu\nu} = - h_{\mu\nu} \mathring{\bar{\nabla}}_\rho v^\mu \, .
	\end{equation}
	Contracting this relation with $v^\nu$ gives $v^\mu v^\nu Q_{\rho\mu\nu} = -h_{\mu\nu} v^\nu \mathring{\bar{\nabla}}_\rho v^\mu = 0$. 
	Evaluating $v^\mu v^\nu Q_{\rho\mu\nu}$ using our parameterization forces $Z_\rho = 0$. Without loss of generality, we can absorb the remaining degrees of freedom into a tensor $G_{\rho\mu}$ that is spatial in its second index ($G_{\rho\mu} v^\mu = 0$):
	\begin{equation}\label{eq:Gmunu}
		Q_{\rho\mu\nu} = -\tau_\mu G_{\rho\nu} - \tau_\nu G_{\rho\mu} \, ,
	\end{equation}
	which yields $v^\mu Q_{\rho\mu\nu} = G_{\rho\nu}$. Comparing this with the preservation condition above, we find:
	\begin{equation}
		G_{\rho\nu} = -h_{\mu\nu} \mathring{\bar{\nabla}}_\rho v^\mu \, .
	\end{equation}
	Now we apply the constraint $v^\nu Q_{[\rho\mu]\nu} = 0$:
	\begin{equation}
		v^\nu Q_{[\rho\mu]\nu} = v^\nu (-\tau_{[\rho} G_{\mu]\nu} - \tau_\nu G_{[\rho\mu]}) = 0 \implies G_{[\rho\mu]} = 0 \, ,
	\end{equation}
	which means, $G_{\rho\mu}$ is a symmetric tensor ($G_{\rho\mu} = G_{\mu\rho}$).
	Using the standard identity linking an arbitrary connection to the Christoffel symbols of the first kind $C_{\rho\mu\nu} = \frac{1}{2}(\partial_\rho h_{\mu\nu} + \partial_\mu h_{\nu\rho} - \partial_\nu h_{\rho\mu})$, we have:
	\begin{equation}
		\mathring{\bar{\Gamma}}^\sigma_{\rho\mu}h_{\sigma\nu} = C_{\rho\mu\nu} - \frac{1}{2}\left(Q_{\rho\mu\nu} + Q_{\mu\nu\rho} - Q_{\nu\rho\mu}\right) \, .
	\end{equation}
	If we substitute our expression for $Q_{\rho\mu\nu}$ \eqref{eq:Gmunu} into the above non-metricity connection, yields:
	\begin{equation}
		Q_{\rho\mu\nu} + Q_{\mu\nu\rho} - Q_{\nu\rho\mu} = -2\tau_\nu G_{\rho\mu} \, ,
	\end{equation}
	because $G_{\rho\mu}$ is symmetric. This simplifies the spatial projection of the connection to:
	\begin{equation}\label{eq:Qconnection-SP}
		\mathring{\bar{\Gamma}}^\sigma_{\rho\mu}h_{\sigma\nu} = C_{\rho\mu\nu} + \tau_\nu G_{\rho\mu} \, .
	\end{equation}
	Now, we evaluate the contraction $v^\nu C_{\rho\mu\nu}$:
	\begin{equation}
		v^\nu C_{\rho\mu\nu} = \frac{1}{2} \big[ -h_{\mu\nu}\partial_\rho v^\nu - h_{\rho\nu}\partial_\mu v^\nu - (\mathcal{L}_v h_{\rho\mu} - h_{\alpha\mu}\partial_\rho v^\alpha - h_{\rho\alpha}\partial_\mu v^\alpha) \big] \, ,
	\end{equation}
	upon using $\mathcal{L}_v h_{\rho\mu} = -2K_{\rho\mu}$, the derivative terms perfectly cancel out, leaving exactly:
	\begin{equation}
		v^\nu C_{\rho\mu\nu} = K_{\rho\mu} \, .
	\end{equation}
	Now, we multiply the connection \eqref{eq:Qconnection-SP} by $v^\nu$. Since $v^\nu h_{\sigma\nu} = 0$, the left-hand side vanishes. Thus, we obtain:
	\begin{equation}
		0 = v^\nu C_{\rho\mu\nu} + v^\nu \tau_\nu G_{\rho\mu} = K_{\rho\mu} - G_{\rho\mu} \, ,
	\end{equation}
	which gives us 
	\begin{equation}\label{eq:GK}
		G_{\rho\mu} = K_{\rho\mu} \, .
	\end{equation}
	Hence, the non-metricity tensor is completely and uniquely fixed by the extrinsic curvature:
	\begin{equation}
		Q_{\rho\mu\nu} = -\tau_\mu K_{\rho\nu} - \tau_\nu K_{\rho\mu} \, .
	\end{equation}
	Notice that the constraint $v^\rho Q_{\rho\mu\nu} = 0$ is naturally satisfied because $v^\rho K_{\rho\nu} = 0$.
	
	Now, we can build the full connection $\mathring{\bar{\Gamma}}^\lambda_{\mu\nu}$ by inserting the completeness relation \eqref{eq:carroll-conditions}:
	\begin{equation}
		\mathring{\bar{\Gamma}}^\lambda_{\mu\nu} = \delta^\lambda_\sigma \mathring{\bar{\Gamma}}^\sigma_{\mu\nu} = -v^\lambda (\tau_\sigma \mathring{\bar{\Gamma}}^\sigma_{\mu\nu}) + h^{\lambda\alpha} (\mathring{\bar{\Gamma}}^\sigma_{\mu\nu}h_{\sigma\alpha}) \, ,
	\end{equation}
	where we have $\mathring{\bar{\Gamma}}^\sigma_{\mu\nu}h_{\sigma\alpha} = C_{\mu\nu\alpha} + \tau_\alpha K_{\mu\nu}$. Substituting from \eqref{eq:Qconnection-SP} and \eqref{eq:GK} into the second term of the above equation, we find:
	\begin{equation}
		h^{\lambda\alpha} (\mathring{\bar{\Gamma}}^\sigma_{\mu\nu}h_{\sigma\alpha})  = h^{\lambda\alpha} C_{\mu\nu\alpha} = \gamma^\lambda_{\mu\nu} \, ,
	\end{equation}
	where $\gamma^\lambda_{\mu\nu}$ is the standard spatial Christoffel symbol.
	If we define $\mathring{\bar{A}}_{\mu\nu} \equiv \tau_\sigma \mathring{\bar{\Gamma}}^\sigma_{\mu\nu}$, the connection then takes the simple form:
	\begin{equation}\label{def:gconnection-NM}
		\mathring{\bar{\Gamma}}^\lambda_{\mu\nu} = \gamma^\lambda_{\mu\nu} - v^\lambda \mathring{\bar{A}}_{\mu\nu} \, .
	\end{equation}
	We require the connection to be \textbf{torsion-free}, which means $\mathring{\bar{\Gamma}}^\lambda_{\mu\nu} = \mathring{\bar{\Gamma}}^\lambda_{\nu\mu}$. Because $\gamma^\lambda_{\mu\nu}$ is symmetric in $\mu, \nu$, $\mathring{\bar{A}}_{\mu\nu}$ must also be a completely symmetric tensor ($\mathring{\bar{A}}_{\mu\nu} = \mathring{\bar{A}}_{\nu\mu}$).
	
	Finally, we ensure that the normalization $\tau_\mu v^\mu = -1$ is preserved by connection \eqref{def:gconnection-NM}. We start from:
	\begin{equation}
		\mathring{\bar{\nabla}}_\rho(\tau_\mu v^\mu) = v^\mu \mathring{\bar{\nabla}}_\rho \tau_\mu + \tau_\mu \mathring{\bar{\nabla}}_\rho v^\mu \, ,
	\end{equation}
	then by using connection \eqref{def:gconnection-NM}, we obtain 
    \begin{equation}
        v^\mu(\partial_\rho \tau_\mu - \mathring{\bar{A}}_{\rho\mu}) + \tau_\mu(\partial_\rho v^\mu + \gamma^\mu_{\rho\sigma}v^\sigma - v^\mu \mathring{\bar{A}}_{\rho\sigma}v^\sigma) \, ,
    \end{equation}
    which collapses to $\partial_\rho(v^\mu \tau_\mu) = 0$. Which means it puts no further constraints on $\mathring{\bar{A}}_{\mu\nu}$.
	
	Therefore, by imposing the specific non‑metricity $Q_{\rho\mu\nu} = -2\tau_{(\mu} K_{\nu)\rho}$,  one finds that the spatial non‑metricity effectively absorbs the extrinsic curvature. This mechanism ensures that the spacetime admits a torsion‑free connection even in situations where the spatial volume is undergoing shear or expansion. In some sense, this shows that we are essentially deriving the \textbf{Symmetric Teleparallel Equivalent of Magnetic Carrollian Gravity} (STEMCG), where non-metricity replaces torsion.

	Now we express \(\mathring{\bar{A}}_{\mu\nu}\) \eqref{def:gconnection-NM} in terms of geometrical quantities with clear physical meaning, since clarifying these structures is essential for studying the equivalent theories and for better understanding the geometric dynamics arising from them. In what follows, after computing the covariant derivatives of \(\tau\) and \(v\), we will obtain an alternative geometrical representation of \(\mathring{\bar{A}}_{\mu\nu}\), which will then allow us to recast \eqref{def:gconnection-NM} in a more transparent form. We now proceed to carry out this in detail, demonstrating each step explicitly. Although parts of these calculations can already be found in \cite{Ciambelli:2025unn}, our treatment follows a different route and emphasizes a distinct geometric perspective. 
	
	\paragraph{Calculating $\mathring{\bar{\nabla}}_\mu v^\nu$}: 
	Using the newly derived non-metric connection \eqref{def:gconnection-NM}, the covariant derivative of the time-flow vector is:
	\begin{equation}\label{eq:nablav-NM0}
		\mathring{\bar{\nabla}}_\mu v^\nu = \partial_\mu v^\nu + \mathring{\bar{\Gamma}}^\nu_{\mu\rho} v^\rho = \partial_\mu v^\nu + \gamma^\nu_{\mu\rho} v^\rho - v^\nu (\mathring{\bar{A}}_{\mu\rho} v^\rho) \, .
	\end{equation}
	To better understand the different geometrical contributions in the parallel transport of $v^\mu$, we must first evaluate the term $\gamma^\nu_{\mu\rho} v^\rho$ in the above relation. 
	Using the definition of the spatial Christoffel symbol $\gamma^\nu_{\mu\rho} = h^{\nu\sigma} C_{\mu\rho\sigma}$ and evaluating the contraction $C_{\mu\rho\sigma} v^\rho$ via the product rule and Lie derivative $\mathcal{L}_v h_{\mu\sigma} = -2K_{\mu\sigma}$, one finds the exact identity:
	\begin{equation}
		C_{\mu\rho\sigma} v^\rho = - h_{\alpha\sigma} \partial_\mu v^\alpha - K_{\mu\sigma} \, ,
	\end{equation}
	multiplying this by the inverse metric $h^{\nu\sigma}$ and using the completeness relation, we find:
	\begin{equation}
		\gamma^\nu_{\mu\rho} v^\rho = -\partial_\mu v^\nu - (\tau_\alpha \partial_\mu v^\alpha)v^\nu - K^\nu_\mu \, .
	\end{equation}
	Substituting this expression back into \eqref{eq:nablav-NM0}, two terms cancel, yielding the simplified relation
	\begin{equation}
		\mathring{\bar{\nabla}}_\mu v^\nu = -K^\nu_\mu - (\tau_\alpha \partial_\mu v^\alpha) v^\nu - (\mathring{\bar{A}}_{\mu\rho} v^\rho) v^\nu \, .
	\end{equation}
	Upon defining the vector $\tau_\nu \mathring{\bar{\nabla}}_\mu v^\nu = a_\mu$, the covariant derivative takes the compact form:
	\begin{equation}\label{eq:nablav-NM}
		\mathring{\bar{\nabla}}_\mu v^\nu = -K^\nu_\mu - a_\mu v^\nu
	\end{equation}
	This relation demonstrates that the spatial non‑metricity is entirely governed by the extrinsic curvature \(K\), while the temporal non‑metricity is encoded in \(a_\mu\).
	
	\paragraph{Calculating $\mathring{\bar{\nabla}}_\mu \tau_\nu$}:
	It is also instructive to examine how the clock form \(\tau\) behaves under parallel transport when adopting the specific choice of non‑metricity \eqref{eq:minimalNM} within Carrollian geometry. Using the connection \eqref{def:gconnection-NM} for the clock form, we find:
	\begin{equation}
		\mathring{\bar{\nabla}}_\mu \tau_\nu = \partial_\mu \tau_\nu - (\gamma^\rho_{\mu\nu} - v^\rho \mathring{\bar{A}}_{\mu\nu}) \tau_\rho \, ,
	\end{equation}
	which simplifies to:
	\begin{equation}\label{eq:nablatau-NM}
		\mathring{\bar{\nabla}}_\mu \tau_\nu = \partial_\mu \tau_\nu - \mathring{\bar{A}}_{\mu\nu} \, . 
	\end{equation}
	
	To preserve the normalization, the covariant derivative of the scalar $v^\nu \tau_\nu$ must be zero:
	\begin{equation}
		\mathring{\bar{\nabla}}_\mu(v^\nu \tau_\nu) = (\mathring{\bar{\nabla}}_\mu v^\nu)\tau_\nu + v^\nu \mathring{\bar{\nabla}}_\mu \tau_\nu = 0 \, .
	\end{equation}
	The non-metricity of $v$ induces terms in $\tau$. From our calculation of \(v\) in \eqref{eq:nablav-NM}, and in order to preserve the constraint, we are forced to impose $v^\nu \mathring{\bar{\nabla}}_\mu \tau_\nu = -a_\mu$.
	It is also necessary to verify its consistency using the relation \eqref{eq:nablatau-NM}, which yields: 
	\begin{equation}
		v^\nu \mathring{\bar{\nabla}}_\mu \tau_\nu = v^\nu \partial_\mu \tau_\nu - v^\nu \mathring{\bar{A}}_{\mu\nu} \, ,
	\end{equation}
	which can be recast to:
	\begin{equation}
		v^\nu \mathring{\bar{\nabla}}_\mu \tau_\nu = -(\tau_\nu \partial_\mu v^\nu + \mathring{\bar{A}}_{\mu\nu} v^\nu) = -a_\mu \, .
	\end{equation}
	This completes our consistency check. 
	One can also decompose the vector $a_\mu$ into vertical and horizontal components using the projector \eqref{eq:carroll-conditions}, which gives us:
	\begin{equation}\label{def:amu}
	a_\mu = -(v^\nu a_\nu)\tau_\mu + \pi_\mu \, ,
	\end{equation}
	where we defined $\pi_\mu \equiv h^\nu_\mu a_\nu$.
	
	As promised, in order to express $\mathring{\bar{A}}_{\mu\nu}$ in terms of geometrical quantities, in the following, we adopt a general parametrization for \(\mathring{\bar{\nabla}}_\mu \tau_\nu\), motivated by \cite{Ciambelli:2025unn}, and establish its relation to the connection \eqref{eq:magnetic-connection}. For this purpose, we make use of the decomposition of the clock form field strength, 
	\begin{equation}\label{def:Ehresmann}
		d\tau = -(\tau \wedge E + F) \, ,
	\end{equation}
    together with the parametrization $\mathring{\bar{\nabla}}_\mu \tau_\nu = a_\mu \tau_\nu + \tau_\mu(\pi_\nu - E_\nu) - \bar{X}_{\mu\nu}$ to find $\mathring{\bar{A}}_{\mu\nu}$. We start with:
	\begin{equation}
		\mathring{\bar{A}}_{\mu\nu} = \partial_\mu \tau_\nu - \mathring{\bar{\nabla}}_\mu \tau_\nu \, ,
	\end{equation}
	from which, we explicitly obtain,
	\begin{equation}\label{def:mathringAmunu}
		\mathring{\bar{A}}_{\mu\nu} = \partial_\mu \tau_\nu - a_\mu \tau_\nu - \tau_\mu(\pi_\nu - E_\nu) + \bar{X}_{\mu\nu} \, .
	\end{equation}
	Since we want to have a torsion-free connection \eqref{def:gconnection-NM}, the tensor $\mathring{\bar{A}}_{\mu\nu}$ in \eqref{def:mathringAmunu} should satisfy condition $\mathring{\bar{A}}_{\mu\nu}v^\nu = \mathring{\bar{A}}_{\nu\mu}v^\nu$. In order to check this, we contract both sides with $v^\nu$ to ensure symmetry. (Recalling that $v^\nu \pi_\nu = v^\nu E_\nu = v^\nu \bar{X}_{\mu\nu} = 0$).
	For the Left-Hand Side, we derive:
	\begin{equation}
		\mathring{\bar{A}}_{\mu\nu} v^\nu = (\partial_\mu \tau_\nu)v^\nu + a_\mu \, ,
	\end{equation}
	while using $v^\nu \partial_\mu \tau_\nu = -\tau_\alpha \partial_\mu v^\alpha$, we get:
	\begin{equation}
		LHS = -\tau_\alpha \partial_\mu v^\alpha + a_\mu \, .
	\end{equation}
	For the Right‑Hand Side we have:
	\begin{equation}
		\mathring{\bar{A}}_{\nu\mu} v^\nu = v^\nu \partial_\nu \tau_\mu - (v^\nu a_\nu)\tau_\mu + \pi_\mu - E_\mu \, .
	\end{equation}
	Using the definition of the Carrollian gravitoelectric field \eqref{eq:Emu}, one can rewrite the above expression as:
	\begin{equation}
		RHS = -\tau_\alpha \partial_\mu v^\alpha - (v^\nu a_\nu)\tau_\mu + \pi_\mu \, .
	\end{equation}
	Now equating LHS and RHS gives:
	\begin{equation}
		a_\mu = -(v^\nu a_\nu)\tau_\mu + \pi_\mu \, ,
	\end{equation}
	which is the same definition we had for $a_{\mu}$ in \eqref{def:amu}.
	For the next step, we want to fix the anti-symmetric part $\mathring{\bar{A}}_{[\mu\nu]} = 0$, which is
	\begin{equation}\label{eq:Aantisymm0}
		\mathring{\bar{A}}_{[\mu\nu]} = \partial_{[\mu} \tau_{\nu]} - a_{[\mu} \tau_{\nu]} - \tau_{[\mu}\pi_{\nu]} + \tau_{[\mu}E_{\nu]} + \bar{X}_{[\mu\nu]} \, .
	\end{equation}
	Starting from the field strength definition \eqref{def:Ehresmann}, one obtains
	\begin{equation}
		\partial_{[\mu} \tau_{\nu]} = \frac{1}{2}d\tau_{\mu\nu}  = -\tau_{[\mu}E_{\nu]} - \frac{1}{2}F_{\mu\nu} \, ,
	\end{equation}
	when this relation is inserted into \eqref{eq:Aantisymm0}, the antisymmetric part of the connection takes the form
	\begin{equation}\label{eq:Aantisymm1}
		\mathring{\bar{A}}_{[\mu\nu]} = -\frac{1}{2}F_{\mu\nu} - a_{[\mu} \tau_{\nu]} - \tau_{[\mu}\pi_{\nu]} + \bar{X}_{[\mu\nu]} \, .
	\end{equation}
	At this stage, the mixed term $a_{[\mu}\tau_{\nu]}$ can be simplified using \eqref{def:amu}, which shows that:
	\begin{equation}
		a_{[\mu} \tau_{\nu]}  = -\tau_{[\mu}\pi_{\nu]} \, .
	\end{equation}
	With this simplification in place, the expression for \eqref{eq:Aantisymm1}, reduces to:
	\begin{equation}
		\mathring{\bar{A}}_{[\mu\nu]} = -\frac{1}{2}F_{\mu\nu} + \bar{X}_{[\mu\nu]} = 0 \, ,
	\end{equation}
	and the condition that follows is
	\begin{equation}
		\bar{X}_{[\mu\nu]} = \frac{1}{2}F_{\mu\nu} \, .
	\end{equation}
	Hence, according to \eqref{def:gconnection-NM} and \eqref{def:mathringAmunu} for $\mathring{\bar{\Gamma}}^\lambda_{\mu\nu}$, we find
	\begin{equation}\label{def:gconnection-NM2}
		\mathring{\bar{\Gamma}}^\lambda_{\mu\nu} = \gamma^\lambda_{\mu\nu} - v^\lambda(\partial_\mu \tau_\nu + \tau_\mu E_\nu + \bar{X}_{\mu\nu}) + v^\lambda(a_\mu \tau_\nu + \tau_\mu \pi_\nu) \, .
	\end{equation}
	In order to compare \eqref{def:gconnection-NM2} to the metric-compatible torsion-free connection $\tilde{\Gamma}^\lambda_{\mu\nu}$ \eqref{eq:magnetic-connection}, we make the following identification:
	\begin{equation}
		X_{\mu\nu} \equiv \tau_\mu E_\nu + \bar{X}_{\mu\nu} \, ,
	\end{equation}
	in which $X_{\mu\nu}$ is the tensor introduced previously in the definition of $\tilde{\Gamma}^\lambda_{\mu\nu}$.
	Then the final relation will be:
	\begin{equation}\label{def:gconnection-NM3}
		\mathring{\bar{\Gamma}}^\lambda_{\mu\nu} = \tilde{\Gamma}^\lambda_{\mu\nu} + v^\lambda(a_\mu \tau_\nu + \tau_\mu \pi_\nu) \, .
	\end{equation}
	
	If we restrict the general non-metric connection $\mathring{\bar{\Gamma}}^\lambda_{\mu\nu}$ by extinguishing the excess temporal non-metricity ($a_\mu = 0$), the shift tensor vanishes, and the connection collapses algebraically to $\tilde{\Gamma}^\lambda_{\mu\nu}$. This apparent paradox, that a non-metric connection reduces to the ``metric'' torsion-free connection, highlights an intriguing structural property in Carrollian geometry.
	The connection $\tilde{\Gamma}^\lambda_{\mu\nu}$ is only metric-compatible if the spacetime is rigid ($K_{\mu\nu} = 0$). In the presence of spatial expansion ($K_{\mu\nu} \neq 0$), the zero-torsion constraint breaks metric compatibility\footnote{A similar observation has been made in \cite{Ciambelli:2025unn}.}. In this regime, applying the algebraic formula for $\tilde{\Gamma}^\lambda_{\mu\nu}$ natively yields $\tilde{\nabla}_\mu v^\nu = -K^\nu_\mu$ and $\tilde{\nabla}_\rho h_{\mu\nu} = -2\tau_{(\mu}K_{\nu)\rho}$. Thus, we conclude that $\tilde{\Gamma}^\lambda_{\mu\nu}$ serves a dual purpose: in rigid spacetimes, it is the unique metric-compatible connection, while in dynamical spacetimes, it acts as the \textbf{Minimal Symmetric Teleparallel} connection, absorbing the extrinsic curvature directly into the non-metricity tensor without requiring the excess deformations parameterized by $a_\mu$.

	\subsection{Magnetic Carrollian Gravity and its Dynamical Equivalent Theories}
	
	While the Sec.~\ref{sec:metric-sector} established dynamically equivalent theories by relaxing the zero-torsion constraint within a strictly metric-compatible framework, we now turn to explore the complementary geometric sector: Carrollian connections that are torsionless but possess non-metricity. We continue to utilize Magnetic Carrollian gravity, characterized by a metric-compatible, torsion-free connection and a rigid spatial geometry, as our baseline reference theory (as constructed in Sec.~\ref{section:MCG}). However, we now seek to construct its dynamical equivalents by activating a specific minimal non-metricity. This choice of minimal non-metricity is not mathematically arbitrary; rather, it is rooted in physical geometric considerations. When a Carrollian manifold is physically realized by embedding it as a null hypersurface within an ambient Lorentzian spacetime equipped with a standard Levi-Civita connection, the induced Carrollian connection naturally inherits a specific non-metricity directly proportional to the extrinsic curvature of the embedding \cite{Ciambelli:2025unn}. 
	
	The mathematical freedom to systematically construct and explore these alternative connections stems fundamentally from the degenerate structure of the Carrollian spacetime. In standard pseudo-Riemannian geometry, the non-degenerate metric uniquely isolates the Levi-Civita connection via the fundamental theorem of Riemannian geometry. In the Carrollian manifold, however, the degeneracy of the spatial metric and the independence of the clock form $\tau_\mu$ destroy this uniqueness. This degeneracy introduces an inherent ambiguity into the definition of the covariant derivative. In a completely general Carrollian connection, these ambiguities manifest as arbitrary tensor fields that dictate the temporal and spatial evolution of the geometric structures, such as the spatial gradients of the clock form \eqref{eq:nablatau-M}, the horizontal drift of the time-flow vector $\pi_\mu$ and the unconstrained temporal non-metricity $N = - v^{\mu} a_{\mu}$ in \eqref{def:amu}. By establishing a consistent Ehresmann parameterization, we showed that the torsion-free constraint imposes no algebraic restrictions on the drift vector. $\pi_\mu$ remains a completely free, tunable geometric ambiguity, entirely decoupled from the dynamical evolution of the clock form $\varphi_\mu$.
	
	In the present section, we apply the symmetric teleparallel methodology to the non-metric sector. By anchoring the connection’s ambiguity to the minimal non-metricity induced by an ambient Lorentzian embedding, we construct the Carrollian analogue to Symmetric Teleparallel Equivalent of General Relativity. We will demonstrate how fixing the residual geometric ambiguities in this non-metric connection allows us to ``flatten'' the spacetime globally, trading the curvature of a rigid Magnetic Carrollian universe for a dynamically equivalent, flat universe governed by trace-squared volume expansion and non-metric interactions.
	
	\subsubsection{Equivalent Theory V: Isotropic Electric Carrollian Gravity}
	The non-metricity generated by $K_{\mu\nu}$ is \textbf{unavoidable} if we want zero torsion. 
    Notice that $K^\nu_\mu$ is purely spatial: $v^\mu K^\nu_\mu = 0$. 
	The vector $a_\mu$ represents \textbf{excess non-metricity}. 
	By setting $a_\mu \neq 0$, we are allowing the time-vector $v^\nu$ to stretch along its own direction (since $a_\mu$ contains the vertical component $v \cdot a$) and drift arbitrarily in space (the horizontal component $\pi_\mu$). 
	
	Because setting $a_\mu = 0$ kills the entire shift tensor ($v^\lambda a_\mu \tau_\nu + v^\lambda \tau_\mu \pi_\nu = 0$) in connection \eqref{def:gconnection-NM3}, the new connection $\mathring{\tilde{\Gamma}}^\lambda_{\mu\nu}$ is algebraically identical to $\tilde{\Gamma}^\lambda_{\mu\nu}$:
	\begin{equation} 
		\mathring{\tilde{\Gamma}}^\lambda_{\mu\nu} = \gamma^\lambda_{\mu\nu} - v^\lambda \Upsilon_{\mu\nu} \, ,
	\end{equation}
	where $\Upsilon_{\mu\nu} = \partial_\mu \tau_\nu + X_{\mu\nu}$.
	
	The key physical difference is that when we calculated the curvatures for $\tilde{\Gamma}$ previously, we forced the spacetime to be rigid ($K_{\mu\nu} = 0$) to preserve metric compatibility. Now, we are evaluating the exact same algebraic connection in an expanding spacetime ($K_{\mu\nu} \neq 0$), which is what causes the non-metricity. 
	In the special case where $K_{\mu\nu} = 0$, we had the simple identity $\hat{\nabla}_\mu v^\rho = v^\rho b_\mu$ as shown in \eqref{eq:hatnablav}. Here \(b_\mu\), defined in \eqref{def:bmu}, plays the role of the acceleration one‑form, while \(\hat{\nabla}\) denotes the spatial covariant derivative associated with the connection \(\gamma^\lambda_{\mu\nu}\). 
	However, when $K_{\mu\nu} \neq 0$, the identity picks up the extrinsic curvature:
	\begin{equation} \label{eq:minimal-nonM}
    \hat{\nabla}_\mu v^\rho = v^\rho b_\mu - K^\rho_\mu \, .
    \end{equation}
	This extra $-K^\rho_\mu$ term acts as the non-metric source that propagates through the entire curvature calculation.
	
	\paragraph{Calculating the Riemann Curvature $\mathring{\tilde{R}}^\rho_{\ \sigma\mu\nu}$}:
	Let the shift be $C^\rho_{\mu\nu} = -v^\rho \Upsilon_{\mu\nu}$. To account for the subtleties arising from non-metricity, it is better to use $\hat{\nabla}$ as the base connection and extract the extra terms directly. The Riemann tensor is formed by:
	\begin{equation} \label{eq:mathringtildeR}
		\mathring{\tilde{R}}^\rho_{\ \sigma\mu\nu} = \hat{R}^\rho_{\ \sigma\mu\nu} + \hat{\nabla}_\mu C^\rho_{\nu\sigma} - \hat{\nabla}_\nu C^\rho_{\mu\sigma} + C^\rho_{\mu\lambda} C^\lambda_{\nu\sigma} - C^\rho_{\nu\lambda} C^\lambda_{\mu\sigma} \, .
	\end{equation}
    which contains the following term:
	\begin{align} \label{eq:nablaC-NM}
		\begin{split}
		\hat{\nabla}_\mu (-v^\rho \Upsilon_{\nu\sigma}) &= -(\hat{\nabla}_\mu v^\rho)\Upsilon_{\nu\sigma} - v^\rho \hat{\nabla}_\mu \Upsilon_{\nu\sigma} \\
		&= -v^\rho(b_\mu \Upsilon_{\nu\sigma} + \hat{\nabla}_\mu \Upsilon_{\nu\sigma}) + K^\rho_\mu \Upsilon_{\nu\sigma} \, .
			\end{split}
	\end{align}
	We now turn to the evaluation of the non‑linear terms $C \times C$. Making use of the definition \eqref{def:bmu}, one finds:
	\begin{equation} \label{eq:C2-NM}
		C^\rho_{\mu\lambda} C^\lambda_{\nu\sigma} = (-v^\rho \Upsilon_{\mu\lambda})(-v^\lambda \Upsilon_{\nu\sigma}) = v^\rho (b_\mu) \Upsilon_{\nu\sigma} \, ,
	\end{equation}
	Alongside this, recall the contribution from $\hat{\nabla}C$ 
	in \eqref{eq:nablaC-NM}. When both pieces are combined, two terms cancel, leaving a simplified expression that will serve as the starting point for the next step:
	\begin{equation} 
		\hat{\nabla}_\mu C^\rho_{\nu\sigma} + C^\rho_{\mu\lambda} C^\lambda_{\nu\sigma} = -v^\rho \hat{\nabla}_\mu \Upsilon_{\nu\sigma} + K^\rho_\mu \Upsilon_{\nu\sigma} \, .
	\end{equation}
	Subtracting the $\mu \leftrightarrow \nu$ terms, we get the final Riemann tensor \eqref{eq:mathringtildeR}. Let $\tilde{R}$ denote the curvature evaluated in the rigid $K=0$ spacetime as defined in \eqref{eq:tildeRiemann}. Then, the new curvature is a linear shift, which is:
	\begin{equation} 
		\mathring{\tilde{R}}^\rho_{\ \sigma\mu\nu} = \tilde{R}^{\rho}_{\ \sigma\mu\nu} + K^\rho_\mu \Upsilon_{\nu\sigma} - K^\rho_\nu \Upsilon_{\mu\sigma} \, .
	\end{equation}
	We trace the Riemann tensor over $\rho$ and $\mu$ to obtain the Ricci tensor ($\mathring{\tilde{R}}_{\sigma\nu} = \mathring{\tilde{R}}^\rho_{\ \sigma\rho\nu}$); thus,
	tracing the shift terms gives:
	\begin{equation} \text{Shift} = K^\rho_\rho \Upsilon_{\nu\sigma} - K^\rho_\nu \Upsilon_{\rho\sigma} = K \Upsilon_{\nu\sigma} - K^\rho_\nu \Upsilon_{\rho\sigma} \, . \end{equation}
    where $K \equiv K^\rho_\rho$ is the scalar expansion trace. 
	But there is a second hidden cancellation. The term $\tilde{R}_{\sigma\nu}$ contained the contraction $-v^\rho \hat{\nabla}_\nu \Upsilon_{\rho\sigma}$. 
	Because $\hat{\nabla}_\nu v^\rho$ shifted by $-K^\rho_\nu$, this term shifts by exactly $+K^\rho_\nu \Upsilon_{\rho\sigma}$. 
	When added together, the $K^\rho_\nu \Upsilon_{\rho\sigma}$ terms cancel each other out.
	The final Ricci tensor picks up only the scalar trace of the extrinsic curvature:
	\begin{equation} 
		\mathring{\tilde{R}}_{\sigma\nu} = \tilde{R}_{\sigma\nu} + K \Upsilon_{\nu\sigma} \, .
	\end{equation}
	Finally, we trace the Ricci tensor with the spatial metric: $\mathring{\tilde{R}} = h^{\sigma\nu} \mathring{\tilde{R}}_{\sigma\nu}$. Let $\Upsilon \equiv h^{\sigma\nu}\Upsilon_{\sigma\nu}$, then the shift will be the simple expression:
	\begin{equation} 
		\mathring{\tilde{R}} = \tilde{R} + K \Upsilon \, .
	\end{equation}
	
	This result is physically interesting. 
	If we use the formula for $\tilde{\Gamma}$ in an expanding universe, the non-metricity it generates ($K_{\mu\nu}$) does not cause the curvature formulas to break down or spawn endless non-linear derivative terms. 
	Instead, the non-metricity shifts the Riemann tensor by a simple algebraic coupling ($K \times \Upsilon$), and shifts the final scalar curvature Lagrangian by exactly $K \Upsilon$. 
	This establishes that in a Symmetric Teleparallel setup, the expansion of the universe ($K$) directly couples to the field ($\Upsilon$), acting as an interaction term $K \Upsilon$ that sources gravity without needing torsion.
	
	By relaxing metric-compatibility while maintaining zero torsion, the connection $\mathring{\tilde{\Gamma}}$ algebraically separates from its rigid metric counterpart. Following the Teleparallel logic, if we assume the universe is globally flat under this condition $\textbf{Riemann}(\mathring{\tilde{\Gamma}}) = 0 \implies \mathring{\tilde{R}}  = 0$, the action for Carroll Magnetic Gravity becomes:
	\begin{equation}\label{eq:action-IsoElecCarroll}
		\tilde{R} \equiv -K \Upsilon \, .
	\end{equation}
	Within the fiber bundle structure of Carrollian geometry, $K = -\frac{1}{2}h^{\mu\nu}\mathcal{L}_v h_{\mu\nu}$ represents the expansion scalar of the spatial base metric along the time fibers. Conversely, $\Upsilon = h^{\mu\nu}(\partial_\mu \tau_\nu + X_{\mu\nu})$ measures the spatial trace of the temporal connection. 
    
    Because the spatial trace of the connection ambiguity $X_{\mu\nu}$ is an independent, unconstrained degree of freedom, we are free to impose a dynamic constraint linking these two quantities.
	We postulate a gauge wherein the spatial divergence of the temporal connection is strictly driven by the expansion of the spatial metric: 
	\begin{equation} \label{eq:UpsilonK}
		\Upsilon = \lambda K \, ,
	\end{equation}
	In Carrollian framework, the arbitrary tensor $X_{\mu\nu}$ represents the intrinsic ambiguity of the Carrollian connection, essentially, the freedom to choose how local clocks are synchronized across space. By writing \eqref{eq:UpsilonK}, we are removing this ambiguity and manually restricting the geometry so that the variation of clock synchronization ($\Upsilon$) is locked to the physical expansion of the universe ($K$). For this purpose, only one scalar component of symmetric part of $X_{\mu\nu}$ is fixed and all traceless components remain free, unlike gauge-fixing \eqref{eq:Xchoice} for equivalent theory I. Substituting this ansatz yields the equivalent action:
	\begin{equation}\label{eq:action-IsoElecCarroll2}
		S_{\text{STECG}} = \int d^D x \, e \, \Big( -\lambda K^2 \Big)
	\end{equation}
	 This Physically establishes a Carrollian analogue to STEGR. Unlike the Torsion equivalence (which generated the full kinetic term $K_{\mu\nu}K^{\mu\nu} - K^2$), the non-metric equivalence suppresses the anisotropic shear ($K_{\mu\nu}K^{\mu\nu}$), isolating the isotropic volume expansion. Consequently, MCG is dynamically equivalent to an isotropic, shear-free Electric Carroll gravity model. This means the curvature $\tilde{R}$ is mathematically indistinguishable from the pure trace-squared kinetic expansion of the Carrollian spacetime with minimal non-metricity \eqref{eq:minimal-nonM} and gauge-fixing \eqref{eq:UpsilonK}.

	\section{Discussion and Conclusions}\label{sec:conclusion}
	
	The motivation of this work has been to explore whether the ``Geometrical Trinity of Gravity'', the property of General Relativity that allows curvature to be dynamically traded for torsion or non-metricity, possesses an analogue in the Carrollian spacetime. In standard Lorentzian manifolds, the non-degenerate nature of the pseudo-Riemannian metric guarantees the existence of a \textit{unique}, metric-compatible, and torsion-free affine connection via the fundamental theorem of Riemannian geometry. This uniqueness locks the dynamics of the gravitational field to the Levi-Civita connection. However, the transition to Carrollian geometry fundamentally changes this paradigm. Because the Carrollian metric is degenerate \eqref{eq:vhzero}, the spacetime structure natively behaves as a fiber bundle rather than as a Riemannian manifold. This degeneracy prevents the unique inversion of the metric compatibility equations, giving rise to an inherent ambiguity in the definition of the Carrollian covariant derivative. 
	
	In Carrollian connection, these ambiguities manifest as arbitrary tensor fields that dictate the temporal and spatial evolution of the geometric structures. Specifically, the connection possesses unconstrained degrees of freedom governing the spatial gradients of the clock form $X_{\mu\nu}$ \eqref{eq:nablatau-M}, the horizontal spatial drift of the time-flow vector $\pi_\mu$, the arbitrary temporal non-metricity $a_\mu$ \eqref{def:amu}, and the spatial frame-twisting torsion $A^\rho_{\mu\nu}$ \eqref{eq:Gconnection}. Throughout our analysis, we treated these geometric ambiguities as freedoms to be exploited. Rather than setting these tensors to zero, we connected them to physical motivations. For instance, we locked the arbitrary clock gradient to \eqref{eq:dilatonsolution} in order to probe the Newton-like potential theory. Freedom in fixing $X_{\mu\nu}$ reflects the physical fact that, in a Carrollian universe where the speed of light vanishes, there is no causal mechanism to enforce a global synchronization of clocks. The freedom to resynchronize them locally is therefore a genuine gauge symmetry, and \(X_{\mu\nu}\) is its geometric manifestation. Fixing $X_{\mu\nu}$ has precedents in the Carrollian gravity literature. An example is Newton-Carroll gravity, which is obtained by imposing $X_{\mu\nu}=0$. Under this condition, the clock form is covariantly constant, yielding an absolute notion of time and reducing the connection to the Newton-Carroll connection \cite{Bekaert:2015xua}. The symmetric part of $X_{\mu\nu}$ is often represented by the notation $S_{\mu\nu}$ \cite{Bergshoeff:2017btm,Campoleoni:2022ebj}.
  
    We established Magnetic Carroll Gravity (MCG), characterized by a metric-compatible, torsion-free connection and a rigid spatial geometry ($K_{\mu\nu}=0$), as our baseline theory for Carroll gravity with Lagrangian $\tilde{R}$ in Sec.~\ref{section:MCG}. By taking advantage of the physically motivated choices for the ambiguities, we obtained dynamically equivalent theories for MCG.
    Following the teleparallel methodology \cite{Aldrovandi:2013wha,Koivisto:2018loq, Maluf:2013gaa}, we systematically relaxed the zero-torsion and metric-compatibility constraints, activating isolated sectors of the geometric ambiguities in torsion. By subsequently demanding that these new connections be globally flat (Riemann $= 0$), we mapped the intrinsic curvature of the magnetic sector $\tilde{R}$ \eqref{eq:tildeRiemann} into entirely new dynamical equivalent theories in Sec.~\ref{section:ETI-ECG}-Sec.~\ref{section:ETIV-MaxwellElectricT}. Our analysis shows that the teleparallel sector of Geometrical Trinity not only exists at the Carrollian manifolds but also expands into a broader framework, revealing that Magnetic Carroll curvature $\tilde{R}$ can be dynamically traded for fundamentally distinct kinetic interactions.
	
	By relaxing spatial rigidity and absorbing the extrinsic curvature into the torsion tensor ($K_{\mu\nu} \neq 0$), we proved that the Lagrangian of a rigid Magnetic Carroll universe is dynamically indistinguishable from the kinetic, ultralocal time-evolution of Electric Carroll Gravity Sec.~\ref{section:ETI-ECG}. Furthermore, by isolating the spatial torsion ($A^\rho_{\mu\nu}$), we uncovered an interesting geometric equivalence linking Magnetic Carroll gravity to condensed matter physics, Sec.~\ref{section:ETII-FractonGT}. We demonstrated that the kinetic penalty of having spatial torsion \eqref{eq:equi-gaugefield} mirrors the magnetic Hamiltonian of a scalar-charge Fracton gauge theory \eqref{def:3fractonM}. 
	
	Our analysis of the temporal torsion sector also yielded interesting results. We showed that when the torsion arises solely due to the exact exterior derivative of clock form ($F_{\mu\nu}$) \eqref{def:connection-maxwell}, the equivalent teleparallel theory identically reproduces the electric sector of Maxwell’s electromagnetism in Sec.~\ref{section:ETIV-MaxwellElectricT}, casting the clock form as a $U(1)$ Carrollian \textit{gravitoelectric} gauge potential \eqref{eq:Emu}. 
	In addition, allowing an arbitrary temporal vector torsion \(s_\mu\) \eqref{eq:smu} brings a specific equivalent teleparallel theory \eqref{eq:ssquared} which opens two distinct interpretations. Exploiting the scale symmetry of the clock form \eqref{eq:weylgauge}, it becomes equivalent to a massless scalar dilaton. Alternatively, fixing the freedom associated with the spatial gradient of the clock form through the condition \eqref{eq:dilatonsolution} leads to a dynamically equivalent formulation of MCG, which may be viewed as a Carrollian analogue of Newtonian potential theory.

\begin{table}[htbp]
    \centering
    \renewcommand{\arraystretch}{1.5}
    \caption{Dynamically equivalent theories of Magnetic Carrollian gravity and their local Carroll boost invariance. Under a Carroll boost, $\delta \tau_\mu = \lambda_\mu$ and $\delta h^{\mu\nu} = -2\lambda^{(\mu} v^{\nu)}$. Theories constructed from purely transverse tensors ($v^\mu T_{\mu\dots} = 0$) achieve native invariance without auxiliary multipliers.}
    \label{tab:boost_invariance}
    
    \begin{tabular}{l l l}
        \toprule
        \textbf{Equivalent Theory} & \textbf{Kinetic Field} & \textbf{Boost Invariant?} \\
        \midrule
        
        \textbf{Magnetic Carrollian Gravity (MCG)} & 
        $\tilde{R} = h^{\mu\nu}\tilde{R}_{\mu\nu}$ & 
        \textbf{No} ($\delta \tilde{R} \neq 0$)  \\
        
        \textbf{Electric Carrollian Gravity (TEMCG)} & 
        $K_{\mu\nu}$ & 
        \textbf{Yes} ($\delta K_{\mu\nu} = 0$)  \\
        
        \textbf{STEMCG Cosmology} & 
        $K^2$ & 
        \textbf{Yes} ($\delta K = 0$) \\
        
        \textbf{Spatial Torsion} & 
        $A^\rho_{\mu\nu}$ & 
        \textbf{Yes} ($\delta A^2 = 0$)  \\
        
        \textbf{Maxwell Electric Theory} & 
        $E_\mu = \mathcal{L}_v \tau_\mu$ & 
        \textbf{No} ($\delta E_\mu = \mathcal{L}_v \lambda_\mu$)  \\
        
        \textbf{Dilaton Scaling} & 
        $s_\mu \sim \mathcal{L}_v \tau_\mu$ & 
        \textbf{No} \\
        
        \bottomrule
    \end{tabular}
\end{table}
    
\begin{figure}[htbp]
			\centering
			\begin{tikzpicture}[
			node distance=2.5cm and 3cm,
			box/.style={rectangle, draw=black, thick, rounded corners, align=center, minimum width=3.5cm, minimum height=1cm, fill=blue!5},
			subbox/.style={rectangle, draw=black, thick, rounded corners, align=center, minimum width=2.5cm, minimum height=0.8cm, fill=red!5},
			arrow/.style={-{Latex[length=3mm, width=2mm]}, thick},
			dashedarrow/.style={-{Latex[length=3mm, width=2mm]}, thick, dashed}
			]
			
			\node[box, fill=green!5] (Curvature) at (0, 4) {\textbf{Curvature} \\ Magnetic Carroll Gravity \\ \textbf{(MCG)}\\ $\int \tilde{R}$};
			
			\node[box] (NonMetricity) at (-4, -1.3) {\textbf{Minimal Non-Metricity} \\\textbf{(STEMCG)} \\ Isotropic ECG \\ $\int -\lambda K^2$};

			\node[subbox] (Torsion) at (4, -1.3) {\textbf{Intrinsic} ($K$) \\ ECG};
			\node[subbox] (Electric) at (4, -3.3) {\textbf{Temporal} ($S$) \\ Dilaton Scaling};
			\node[subbox] (Fracton) at (8, -1.3) {\textbf{Spatial} ($A$) \\ Tensor G. Theory};
			\node[subbox] (Maxwell) at (8, -3.3) {\textbf{Clock T.} ($F$) \\ Elec. Maxwell};
			
			\draw[arrow, <->] (Curvature) -- node[above left, align=center] {Riemann $= 0$ \\ $T=0, \nabla g \neq 0$} (NonMetricity);
			\draw[arrow, <->] (Curvature) -- node[above right, align=center] {Riemann $= 0$ \\ $\nabla g = 0, T \neq 0$} ([xshift=0.5cm,yshift=1.5cm]Torsion.west);
			\draw[arrow, <->] (NonMetricity) -- node[below, align=center] {} ([xshift=-0.4cm]Torsion.west);
			
			
			\begin{pgfonlayer}{background}
            
			\draw[rounded corners, fill=gray!10, draw=gray, dashed] 
			(2.2, 0.2) rectangle (10, -4.2);
			\node[text=gray, anchor=north west] at (2.2, 0.1) {\textit{Torsion Sectors (TEMCG)}};

			\draw[dashed] (Torsion) -- (Electric);
			\draw[dashed] (Torsion) -- (Fracton);
			\draw[dashed] (Torsion) |- (Maxwell);
			\draw[dashed] (Fracton) |- (Electric);
            
			\end{pgfonlayer}
			
			\end{tikzpicture}
			\caption{The Extended Geometrical Trinity of Carroll Gravity. The baseline Magnetic Carroll Gravity (top) can be dynamically mapped to flat-space equivalent theories by shifting the curvature into Non-Metricity (left) or Torsion (right). Unlike Lorentzian GR, the degenerate Carrollian geometry splits the Torsion sector into three distinct physical theories, explicitly linking gravity to scalar, tensor and gauge theories.}
			\label{fig:carroll_trinity}
		\end{figure}
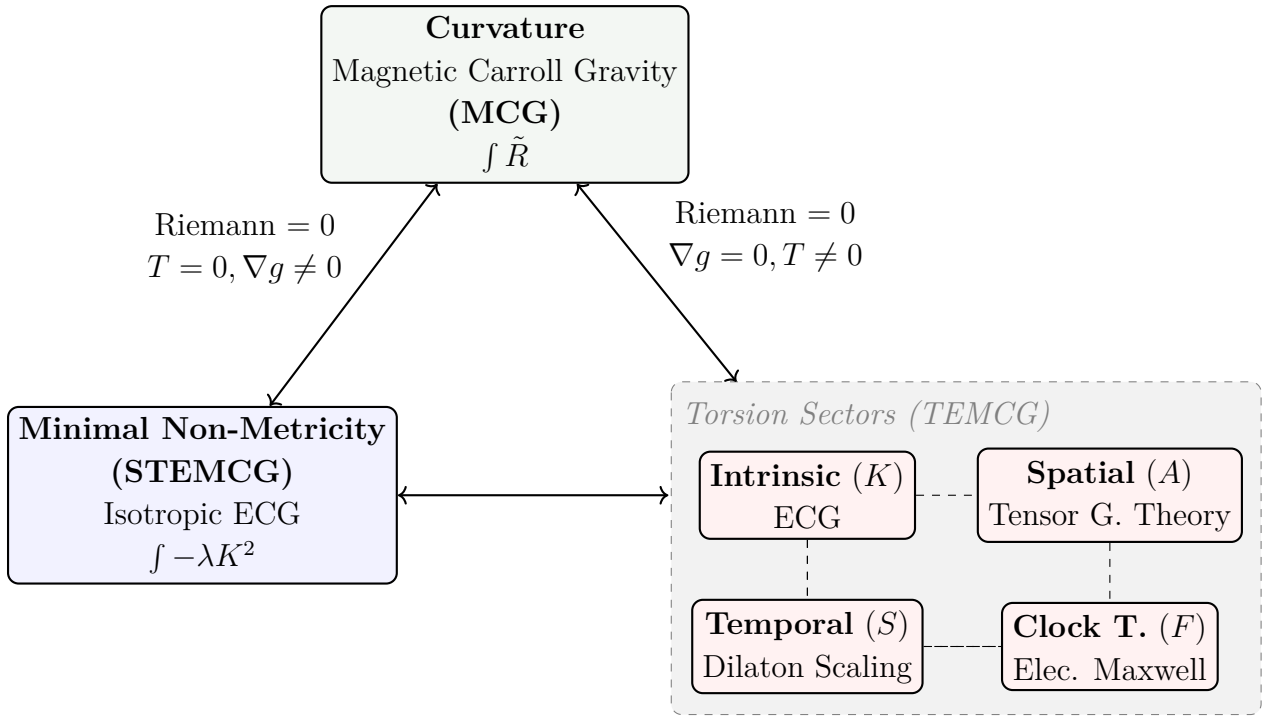

	Finally, we constructed the Symmetric Teleparallel Equivalent of Magnetic Carroll Gravity (STEMCG) by enforcing a torsion-free connection with minimal non-metricity in Sec.~\ref{sec:STEMCG}. By adopting a gauge choice that ties the spatial divergence of the temporal connection to the volumetric expansion of the universe ($\Upsilon \propto K$) \eqref{eq:UpsilonK}, we proved that MCG is dynamically equivalent to an ultralocal, isotropic Electric Carroll gravity governed exclusively by a trace-squared kinetic Lagrangian \eqref{eq:action-IsoElecCarroll2}, $\int -K^2$. A schematic illustration of our main results is presented in Fig.~\ref{fig:carroll_trinity}, while Table~\ref{tab:boost_invariance} summarizes the transformation properties of the kinetic fields associated with the different theories under local Carroll boosts. Recent classifications by Vigneron et al. \cite{Vigneron:2025vgn} have established the existence of affine connections permitting both torsion and non-metricity on general Carrollian structures. While their work identifies the structural existence of an ultra-relativistic Geometric Trinity by taking the limits of Lorentzian spacetimes, the explicit dynamical Lagrangians governing these equivalent theories were left as an open challenge. In this article, we investigated this challenge. Operating intrinsically within the Carrollian fiber bundle, we explicitly computed the scalar curvature shifts induced by these non-metric and torsionful connection ambiguities. In doing so, we demonstrate that the Carrollian Trinity is far richer than its Lorentzian counterpart.

    It is important to emphasize the variational mechanics required to formally establish these dynamical equivalences. The equivalent actions derived in this work (e.g., $S \propto \int -K^2$, $\int A^2$, $\int -E^2$) are algebraically evaluated by substituting specific geometric constraints into the baseline scalar curvature. However, to ensure that these configurations emerge as true dynamical solutions of the underlying theory, the constraints that define each specific equivalence, namely the global flatness of the modified connection, the precise conditions imposed on the torsion or non-metricity sectors, and the physical fixings of the connection ambiguities (such as \eqref{eq:Xchoice} or \eqref{eq:UpsilonK}) must be dynamically enforced. It can be achieved by introducing these constraints into the action using Lagrange multipliers.

		
	Our results give more perspective to the standard Geometrical Trinity of General Relativity. In the Lorentzian manifold, trading curvature for torsion or non-metricity yields theories that remain strictly gravitational in nature. However, our findings demonstrate that due to the causal structure of the Carrollian manifold, the equivalent theories frequently cross the boundary into non-gravitational domains. Electric Carroll gravity, tensor gauge theories, Maxwell electromagnetism, and scalar field theories, can be equivalently mapped to Magnetic Carroll gravity. This establishes a geometric dictionary that translates Carrollian gravitational dynamics into the language of gauge theories and condensed matter physics. 
    
    This framework allows us to exploit the inherent ambiguities of degenerate geometry, turning them into a feature rather than a limitation, where they become a resource that can be systematically exploited within the framework. The importance of our findings lies in their potential to serve as a framework for researchers working across different fields of theoretical physics, such as the gravity, holography and condensed matter communities.

    This study may open several directions for future research. While we focused on minimal non-metricity motivated by ambient embeddings, exploring the full parameter space of Carrollian non-metricity may yield equivalents to more exotic gravity or gauge theories. Additionally, a next step would be to introduce matter couplings into this framework, specifically, studying how Carrollian fermions couple to the spatial torsion ($A^\rho_{\mu\nu}$) via the spin connection, which could reveal new equivalence in Carrollian gravity.

\acknowledgments
We are grateful to M.M. Sheikh‑Jabbari for valuable discussions and insightful comments. MAJ would also like to thank Bahram Shakerin and Hamid Afshar for stimulating discussions on Carrollian gravity. This work was supported by the Iran National Science Foundation (INSF) under Project No. 40504679.

\appendix
\section{Review on Gauging Carroll Algebra: First-Order Formalism}\label{app:gauging}

Carroll gravity can be formulated as a gauge theory of the Carroll algebra \cite{Hartong:2015xda,Bergshoeff:2017dqq,Campoleoni:2022ebj}, whose non-vanishing commutators are
\begin{align}
    [J_{ab},P_c]&=2\delta_{c[a}P_{b]}, \qquad [J_{ab},G_c]=2\delta_{c[a}G_{b]}, \nonumber \\
    [J_{ab},J_{cd}]&=4\delta_{[a[d}J_{c]b]}, \qquad [P_a,G_b]=\delta_{ab}H.
\end{align}
Here \(J_{ab}\), \(P_a\), \(G_a\), and \(H\) generate spatial rotations, spatial translations, Carroll boosts, and time translations, respectively.

Introducing a connection \(A_\mu\) valued in the Carroll algebra,
\begin{align}\label{eq:connection}
    A_\mu = H\,\tau_\mu + P_a\,e_\mu{}^a + G_a\,\omega_\mu{}^a + \tfrac{1}{2}J_{ab}\,\omega_\mu{}^{ab},
\end{align}
where the index \(\mu\) runs over \(d+1\) values. The connection transforms as
\begin{align}
    \delta A_\mu = \partial_\mu\Pi + [A_\mu,\Pi], \qquad
    \Pi = \xi^\mu A_\mu + \Sigma, \qquad
    \Sigma = \tfrac{1}{2}J_{ab}\,\lambda^{ab} + G_a\,\lambda^a,
\end{align}
with \(\xi^\mu\) the spacetime parameter and \(\lambda^{ab}\), \(\lambda^a\) the local rotation and Carroll boost parameters.

To obtain a proper internal transformation, we use the curvature 2-form
\begin{align}
    F_{\mu\nu} = \partial_\mu A_\nu - \partial_\nu A_\mu + [A_\mu,A_\nu],
\end{align}
and define
\begin{align}\label{eq:proper-transformation}
    \bar{\delta} A_\mu = \delta A_\mu - \xi^\nu F_{\mu\nu} = \mathcal{L}_\xi A_\mu + \partial_\mu\Sigma + [A_\mu,\Sigma].
\end{align}
The vielbeins transform as
\begin{subequations}\label{eq:vielbein-transformations}
\begin{align}
    \bar{\delta}\tau_\mu &= \mathcal{L}_\xi\tau_\mu + \lambda_a e_\mu{}^a, &
    \bar{\delta}e_\mu{}^a &= \mathcal{L}_\xi e_\mu{}^a + \lambda_b{}^a e_\mu{}^b, \\
    \bar{\delta}\omega_\mu{}^a &= \mathcal{L}_\xi\omega_\mu{}^a + \partial_\mu\lambda^a + \lambda_b{}^a\omega_\mu{}^b - \lambda_b\omega_\mu{}^{ab}, &
    \bar{\delta}\omega_\mu{}^{ab} &= \mathcal{L}_\xi\omega_\mu{}^{ab} + \partial_\mu\lambda^{ab} + \lambda^a{}_c\omega_\mu{}^{cb} - \lambda^b{}_c\omega_\mu{}^{ca}.
\end{align}
\end{subequations}

The curvature 2-form decomposes as
\begin{align}
    F_{\mu\nu} = H\,R_{\mu\nu}(H) + P_a\,R_{\mu\nu}{}^a(P) + \tfrac{1}{2}J_{ab}\,R_{\mu\nu}{}^{ab}(J) + G_a\,R_{\mu\nu}{}^a(G),
\end{align}
with components
\begin{subequations}\label{eq:curvature-components}
\begin{align}
    R_{\mu\nu}(H) &= 2\partial_{[\mu}\tau_{\nu]} - 2\omega_{[\mu}{}^a e_{\nu]a}, &
    R_{\mu\nu}{}^a(P) &= 2\partial_{[\mu}e_{\nu]}{}^a - 2\omega_{[\mu}{}^{ab}e_{\nu]b}, \\
    R_{\mu\nu}{}^a(G) &= 2\partial_{[\mu}\omega_{\nu]}{}^a - 2\omega_{[\mu}{}^{ab}\omega_{\nu]b}, &
    R_{\mu\nu}{}^{ab}(J) &= 2\partial_{[\mu}\omega_{\nu]}{}^{ab} - 2\omega_{[\mu}{}^{ca}\omega_{\nu]}{}^b{}_c.
\end{align}
\end{subequations}

A frame \((v^\mu, e^\mu{}_a)\) dual to \((\tau_\mu, e^a{}_\mu)\) is defined by
\begin{align}\label{eq:dualframe}
    v^\mu\tau_\mu = -1, \quad v^\mu e^a{}_\mu = e^\mu{}_a\tau_\mu = 0, \quad e^a{}_\mu e^\mu{}_b = \delta^a_b, \quad \delta^\nu_\mu = -v^\nu\tau_\mu + e^a{}_\mu e^\nu{}_a.
\end{align}
The transformations \(\bar{\delta}v^\mu = 0\), \(\bar{\delta}e^\mu{}_a = \lambda_av^\mu + \lambda_a{}^b e^\mu{}_b\) show that only \(v^\mu\) is Carroll-invariant, allowing the definition of the degenerate metric
\begin{align}
    h_{\mu\nu} \equiv \delta_{ab}\,e_\mu{}^a e_\nu{}^b.
\end{align}

\subsection{Torsionless Carroll Connection}\label{app:torsionless}

The vielbein postulates \(\mathcal{D}_\mu\tau_\nu = \mathcal{D}_\mu e_\nu{}^a = \mathcal{D}_\mu v^\nu = \mathcal{D}_\mu e^\nu{}_a = 0\) yield the most general metric-compatible Carrollian affine connection
\begin{align}\label{eq:connection-solution}
    \Gamma^\sigma_{\mu\nu} = -v^\sigma(\partial_\mu\tau_\nu - \omega_{\mu a}e_\nu{}^a) + e^\sigma{}_a(\partial_\mu e_\nu{}^a - \omega_{\mu}{}^a{}_b e_\nu{}^b),
\end{align}
and express the torsion as
\begin{align}\label{eq:torsion-gauged}
    2\Gamma^\rho_{[\mu\nu]} = -v^\rho R_{\mu\nu}(H) + e^\rho{}_a R_{\mu\nu}{}^a(P).
\end{align}
Here \(R_{\mu\nu}(H)\) and \(R_{\mu\nu}{}^a(P)\) are the temporal and spatial components of torsion. The Riemann and Ricci tensors take the form
\begin{align}
    R_{\mu\nu\sigma}{}^\rho &= -v^\rho e_{\sigma a} R_{\mu\nu}{}^a(G) - e^\rho{}_b e_{\sigma a} R_{\mu\nu}{}^{ab}(J), \\
    R_{\mu\nu} &= -v^\rho e_{\mu a} R_{\rho\nu}{}^a(G) - e^\rho{}_a e_{\mu b} R_{\rho\nu}{}^{ab}(J).
\end{align}

Imposing the torsion-free conditions \(R_{\mu\nu}(H) = R_{\mu\nu}{}^a(P) = 0\) gives the solutions
\begin{align}
    \omega_\mu{}^a
    &=\tau_\mu e^{\nu a}v^\rho\partial_{[\rho}\tau_{\nu]}+e^{\nu a}\partial_{[\mu}\tau_{\nu]}+S^{ab}e_{\mu b},
    \\[4pt]
    \omega_\mu{}^{ab}
    &=
    -2e^{\nu[a}\partial_{[\mu}e_{\nu]}{}^{b]}
    +
    e_{\mu c}e^{\rho a}e^{\sigma b}\partial_{[\sigma}e_{\rho]}{}^c,
\end{align}
where \(S^{ab}\) is an independent symmetric, transverse, spatial tensor.

The spatial part of the connection is
\begin{align}
    \Gamma^\sigma_{\mu\nu}\big|_{\rm spatial}
    =
    e^\sigma{}_a\left(\partial_\mu e_\nu{}^a - \omega_\mu{}^a{}_b e_\nu{}^b\right).
\end{align}
Using the vielbein postulate
\begin{align}
    \partial_\mu e_\nu{}^a - \gamma^\sigma_{\mu\nu}e_\sigma{}^a - \omega_\mu{}^a{}_b e_\nu{}^b = 0,
\end{align}
and contracting with \(e^\sigma{}_a\), we obtain
\begin{align}
    \Gamma^\sigma_{\mu\nu}\big|_{\rm spatial}
    =
    \gamma^\sigma_{\mu\nu}.
\end{align}
This is exactly the spatial part of the second-order Magnetic Carroll connection.

For the temporal part, we use the explicit form of \(\omega_\mu{}^a\). Contracting with \(e_\nu{}^a\) and using \(e^{\rho a}e_{\nu a} = \delta^\rho_\nu + v^\rho\tau_\nu\),
\begin{align}
    \omega_{\mu a}e_\nu{}^a&=\tau_\mu v^\sigma\partial_{[\sigma}\tau_{\nu]}+\partial_{[\mu}\tau_{\nu]}+\tau_\nu v^\sigma\partial_{[\sigma}\tau_{\mu]}+S_{\mu\nu}.
\end{align}
Using the Lie-derivative identities
\begin{align}
    v^\sigma\partial_{[\sigma}\tau_{\nu]} = \tfrac{1}{2}\left(\mathcal{L}_v\tau\right)_\nu,
    \qquad
    v^\sigma\partial_{[\sigma}\tau_{\mu]} = \,\tfrac{1}{2}\left(\mathcal{L}_v\tau\right)_\mu,
\end{align}
this becomes
\begin{align}
    \omega_{\mu a}e_\nu{}^a
    =
    \frac{1}{2}\tau_\mu\left(\mathcal{L}_v\tau\right)_\nu +\partial_{[\mu}\tau_{\nu]}+\frac{1}{2}\tau_\nu\left(\mathcal{L}_v\tau\right)_\mu + S_{\mu\nu}= \partial_{\mu}\tau_{\nu}+\Upsilon_{\mu\nu},
\end{align}
Substituting into the temporal part,
\begin{align}
    \Gamma^\sigma_{\mu\nu}\big|_{\rm temporal}
    &=
  -  v^\sigma\left[
        \partial_\mu\tau_\nu
        -
        \frac{1}{2}\tau_\mu\left(\mathcal{L}_v\tau\right)_\nu
        -
        \partial_{[\mu}\tau_{\nu]}
        -
        \frac{1}{2}\tau_\nu\left(\mathcal{L}_v\tau\right)_\mu
        -
        S_{\mu\nu}
    \right].
\end{align}
Using \(\partial_\mu\tau_\nu - \partial_{[\mu}\tau_{\nu]} = \partial_{(\mu}\tau_{\nu)}\),
\begin{align}
    \Gamma^\sigma_{\mu\nu}\big|_{\rm temporal}
    =
    -v^\sigma\left[
        \partial_{(\mu}\tau_{\nu)} -\frac{1}{2}\tau_\mu\left(\mathcal{L}_v\tau\right)_\nu-\frac{1}{2}\tau_\nu\left(\mathcal{L}_v\tau\right)_\mu- S_{\mu\nu} \right].
\end{align}

Within the context of second-order formalisms, we have established
\begin{align}
	\tilde{\Gamma}^\sigma_{\mu\nu}\big|_{\rm temporal}
	=
	-\,v^\sigma\Upsilon_{\mu\nu},
	\qquad
	\Upsilon_{\mu\nu}
	=\partial_{(\mu}\tau_{\nu)}+X_{(\mu\nu)}.
\end{align}
Matching this with the temporal part of the first-order connection,
\begin{align}
	\Gamma^\sigma_{\mu\nu}\big|_{\rm temporal}
	=
	-v^\sigma\bigl(\partial_\mu\tau_\nu - \omega_{\mu a}e_\nu{}^a\bigr),
\end{align}
gives the relation
\begin{align}
	\omega_{\mu a}e_\nu{}^a
	=
	\partial_\mu\tau_\nu
	-
	\Upsilon_{\mu\nu}.
\end{align}
Multiplying by $e^{\nu a}$ and using $\hat{\nabla}_\mu\tau_\rho = \partial_\mu\tau_\rho$ (since the spatial Christoffel symbol is orthogonal to the clock form), one obtains
\begin{align}\label{eq:newform}
		\omega_\mu{}^a
		=
		\bigl( \hat{\nabla}_\mu\tau_\rho -\Upsilon_{\mu\rho} \bigr)e^{\rho a}.
\end{align}
Here $\hat{\nabla}_\mu$ is the spatial covariant derivative built from the spatial spin connection $\omega_\mu{}^{ab}$.

\subsection{Ricci Scalar and MCG Action}\label{app:ricci-first-order}

We start from the first-order definition of the scalar curvature \cite{Campoleoni:2022ebj},
\begin{align}
    \mathcal{R}
    &=
    2v^\mu e^\nu{}_a \mathcal{R}_{\mu\nu}{}^a
    +
    e^\mu{}_a e^\nu{}_b \mathcal{R}_{\mu\nu}{}^{ab},
\end{align}
where $v^\mu$ is the time-flow vector and the curvature components are
\begin{align}
    \mathcal{R}_{\mu\nu}{}^a
    &\equiv
    2\partial_{[\mu}\omega_{\nu]}{}^a
    -
    2\omega_{[\mu}{}^{ab}\omega_{\nu]b},
    \\
    \mathcal{R}_{\mu\nu}{}^{ab}
    &\equiv
    2\partial_{[\mu}\omega_{\nu]}{}^{ab}
    -
    2\omega_{[\mu}{}^{ca}\omega_{\nu]}{}^b{}_c.
\end{align}

\paragraph{Spatial curvature term.}
The spatial spin connection $\omega_\mu{}^{ab}$ is identically the spin connection of the spatial Christoffel symbol $\gamma^\sigma_{\mu\nu}$. Therefore, $\mathcal{R}_{\mu\nu}{}^{ab}$ is related to the Riemann tensor of $\gamma$ through
\begin{align}
    \mathcal{R}_{\mu\nu}{}^{ab}
    &=
    e^{\sigma b}e_\rho{}^a \hat{R}^\rho{}_{\sigma\mu\nu}.
\end{align}
Contracting with the vielbeins and using $e^\mu{}_a e_\rho{}^a e^\nu{}_b e^{\sigma b} = (\delta^\mu_\rho + v^\mu\tau_\rho)h^{\nu\sigma}$, we obtain
\begin{align}
    e^\mu{}_a e^\nu{}_b \mathcal{R}_{\mu\nu}{}^{ab}
    &=
    \hat{R}
    +
    v^\mu\tau_\rho h^{\nu\sigma}\hat{R}^\rho{}_{\sigma\mu\nu},
\end{align}
where $\hat{R}$ is the spatial Ricci scalar of $\gamma$. The second term vanishes because the spatial Riemann tensor is orthogonal to the clock form,
\begin{align}
    \tau_\rho\hat{R}^\rho{}_{\sigma\mu\nu} = 0.
\end{align}

\paragraph{Boost curvature term.}
The boost curvature is
\begin{align}
    \mathcal{R}_{\mu\nu}{}^a
    &=
    2\partial_{[\mu}\omega_{\nu]}{}^a
    -
    2\omega_{[\mu}{}^{ab}\omega_{\nu]b}.
\end{align}
Because $\omega_\mu{}^{ab}$ is the spin connection of $\gamma$, this expression acts exactly as the covariant exterior derivative $\hat{D}$. Using the correct relation between $\omega_\mu{}^a$ and $\Upsilon_{\mu\nu}$ (see \eqref{eq:newform}),
\begin{align*}
    \omega_\mu{}^a
    &=
    \bigl( \hat{\nabla}_\mu\tau_\rho -\Upsilon_{\mu\rho}\bigr)e^{\rho a},
\end{align*}
and noting that the covariant derivative of $e^{\rho a}$ vanishes, we obtain
\begin{align}
    \mathcal{R}_{\mu\nu}{}^a
    &=
    2e^{\rho a}\hat{\nabla}_{[\mu}\bigl( \hat{\nabla}_{\nu]}\tau_\rho -\Upsilon_{\nu]\rho} \bigr)
    \nonumber \\[4pt]
    &=
    -2e^{\rho a}\hat{\nabla}_{[\mu}\Upsilon_{\nu]\rho}
    -
    e^{\rho a}\hat{R}^\sigma{}_{\rho\mu\nu}\tau_\sigma.
\end{align}
The second term vanishes when contracted with $v^\mu e^\nu{}_a$, because $\tau_\sigma\hat{R}^\sigma{}_{\rho\mu\nu}=0$.

\paragraph{Scalar curvature.}
Substituting the torsion-free solutions and using the identities
\begin{subequations}
\begin{align}
    v^\rho\hat{\nabla}_\nu\Upsilon_{\rho\sigma} 
    &=
    \hat{\nabla}_\nu b_\sigma - b_\nu b_\sigma \, ,\\
    h^{\sigma\nu} v^\rho \hat{\nabla}_\rho \Upsilon_{\nu\sigma}&=\mathcal{L}_v \Upsilon - 2 b^\mu b_\mu - 2 E^\nu b_{\nu}\,,
\end{align}
\end{subequations}
one finds, 
\begin{align}
    \mathcal{R}
    =
    \hat{R}
    -
    2\mathcal{L}_v\Upsilon
    +2
    h^{\mu\nu}\hat{\nabla}_\mu b_\nu
    +2
    b_\mu b^\mu+4 E^\nu b_{\nu}
\end{align}
Here $\hat{R}$ is the spatial Ricci scalar of $\gamma^\sigma_{\mu\nu}$, $\Upsilon = h^{\mu\nu}\Upsilon_{\mu\nu}$, $b_\mu = \Upsilon_{\mu\nu}v^\nu$, and $E_\mu = \mathcal{L}_v\tau_\mu$ is the gravitoelectric field. The orthogonality condition $v^\mu\tau_\sigma h^{\nu\rho}\hat{R}^\sigma{}_{\rho\mu\nu}=0$ has been used. The resulting expression closely aligns with the second-order Magnetic Carroll Ricci scalar as computed in Eq.~\eqref{eq:magnetic-riccis}, differing by a factor 2. It is important to clarify a structural discrepancy between the first-order and second-order formalisms regarding how the scalar curvature is computed. In the standard second-order affine framework, the Magnetic Carrollian Ricci scalar is conventionally defined via the direct spatial trace of the Ricci tensor, $\tilde{R} = h^{\mu\nu} \tilde{R}_{\mu\nu}$. This simple contraction equally weights the temporal (boost) and purely spatial curvature components. Conversely, in the first-order formalism, the invariant action derived from gauging the Carroll algebra inherently assigns a relative factor of 2 to the Carroll boost curvature sector. Consequently, deriving the scalar curvature merely by taking the spatial trace $h^{\mu\nu} \tilde{R}_{\mu\nu}$ does not natively reproduce this relative weight, leading to an overall factor of 2 difference in certain dynamical interaction terms between the two formalisms.

	
	\addcontentsline{toc}{section}{References}
	\bibliographystyle{fullsort.bst}
	\bibliography{biblio}

\end{document}